\documentclass[12pt]{article}
\usepackage[top=5cm,bottom=3.5cm,left=3.23cm,right=3.23cm]{geometry}
\usepackage{amsmath}
\usepackage{graphicx}
\usepackage{enumerate}
\usepackage{natbib}
\usepackage{url} % not crucial - just used below for the URL 
\usepackage{lineno}
\usepackage{amssymb}
\usepackage{amsthm}
\usepackage{authblk}
\usepackage{epsfig}
\usepackage{graphics}
\usepackage{float}
\usepackage{subfigure}
\usepackage{multirow}
\usepackage{xcolor}
\usepackage{makeidx}
\usepackage{xspace}
\usepackage{wrapfig}
\usepackage{blindtext}
\usepackage{setspace}
\usepackage{algorithm}
\usepackage{algorithmic}
\usepackage[colorlinks=true,citecolor=blue,linkcolor=blue,urlcolor=blue]{hyperref}
\usepackage{booktabs}
\usepackage{bm}
\makeindex

\renewcommand{\algorithmicrequire}{ \textbf{Input:}} 
\renewcommand{\algorithmicensure}{ \textbf{Output:}} 

\definecolor{darkred}{rgb}{1, 0.1, 0.3}
\definecolor{darkblue}{rgb}{0.1, 0.1, 1}
\definecolor{darkgreen}{rgb}{0,0.6,0.5}

\newcommand{\blind}{1}

\begin{document}

\def\spacingset#1{\renewcommand{\baselinestretch}%
{#1}\small\normalsize} \spacingset{1}

%%%%%%%%%%%%%%%%%%%%%%%%%%%%%%%%%%%%%%%%%%%%%%%%%%%%%%%%%%%%%%%%%%%%%%%%%%%%%%

\if1\blind
{
  \title{\bf Efficient stochastic generators with spherical harmonic transformation for high-resolution global climate simulations from CESM2-LENS2}
  \author{Yan Song\thanks{
    Corresponding author; email: yan.song@kaust.edu.sa}\hspace{.2cm}\\
    \vspace{-10pt}
    Statistics Program, King Abdullah University of Science and Technology, Saudi Arabia\\
    \vspace{5pt}
     Zubair Khalid \\
    School of Science and Engineering, Lahore University of Management Sciences, Pakistan\\
    \vspace{5pt}
    Marc G. Genton \\
     Statistics Program, King Abdullah University of Science and Technology, Saudi Arabia}
  \maketitle
} \fi

\if0\blind
{
\title{\bf Efficient stochastic generators with spherical harmonic transformation for high-resolution global climate simulations from CESM2-LENS2}
  \author{ 
    }
  \maketitle
}\fi

% \if0\blind
% {
%   \bigskip
%   \bigskip
%   \bigskip
%   \begin{center}
%     {\LARGE\bf Supplementary Material for ``Large-Scale
    
%     Low-Rank Gaussian Process Prediction with 
    
%     Support Points"}
% \end{center}
%   \medskip
% } \fi
%\bigskip
\vspace{-1cm}
\begin{abstract} \spacingset{1.1}
Earth system models (ESMs) are fundamental for understanding Earth's complex climate system. However, the computational demands and storage requirements of ESM simulations limit their utility. For the newly published CESM2-LENS2 data, which suffer from this issue, we propose a novel stochastic generator (SG) as a practical complement to the CESM2, capable of rapidly producing emulations closely mirroring training simulations. Our SG leverages the spherical harmonic transformation (SHT) to shift from spatial to spectral domains, enabling efficient low-rank approximations that significantly reduce computational and storage costs. By accounting for axial symmetry and retaining distinct ranks for land and ocean regions, our SG captures intricate non-stationary spatial dependencies. Additionally, a modified Tukey g-and-h (TGH) transformation accommodates non-Gaussianity in high-temporal-resolution data. We apply the proposed SG to generate emulations for surface temperature simulations from the CESM2-LENS2 data across various scales, marking the first attempt of reproducing daily data. These emulations are then meticulously validated against training simulations. This work offers a promising complementary pathway for efficient climate modeling and analysis while overcoming computational and storage limitations. 
\end{abstract}
\bigskip

\noindent%
\vfill

\newpage
\spacingset{1.8} % DON'T change the spacing!
\section{Introduction}
Earth system models (ESMs) are complex mathematical equations that describe and simulate the transformation and interaction of energy and materials within the Earth's climate system. Their development is grounded in the fundamental laws of physics, fluid dynamics, and chemistry, necessitating the identification and quantification of physical processes represented through mathematical equations \citep{climateNOAA}. ESMs are indispensable tools in both scientific research and government policymaking, enabling a deeper understanding and more accurate predictions of Earth's systems. For example, the Intergovernmental Panel on Climate Change (\href{https://www.ipcc.ch}{IPCC}) utilized datasets from the Coupled Model Intercomparison Project's fifth (CMIP5) and sixth (CMIP6) phases \citep{CMIP5,CMIP6} to compile their fifth and sixth synthesis assessment reports on climate change.

Despite their significance, ESMs have limitations as approximations of the Earth's complex system. They are sensitive to changes in model inputs, such as physics parameters and emission scenarios, which need being assessed. Additionally, distinguishing between model errors and the effects of internal variability, which arises from atmospheric, oceanic, land, and cryospheric processes and their interactions, is challenging with a limited number of simulations \citep{kay2015community}. This underscores the need for multiple climate simulations to comprehensively characterize Earth's system. The Community Earth System Model \citep[CESM,][]{kay2015community} is one such model designed to study climate change while accounting for internal climate variability.

Running ESMs is a time-consuming process, even with the support of powerful supercomputers. Scientists partition the planet into thousands of $3$-dimensional grids. Within each grid cell, they specify variables and conditions, employ computers to solve equations, propagate results to neighboring cells, and then iterate the above procedure with the updated variables and conditions. The whole process is repeated through time steps of different scales, which is the temporal resolution. The spatial resolution of the model is determined by the size of the grid cell, with smaller sizes yielding higher resolution but requiring more time. Generating multiple climate simulations demands weeks and even months of computational resources, exclusively available to several research institutes globally \citep{Huang'sEmulator}. Furthermore, while modern computational power allows for the generation of climate simulations, storing them is constrained by technological limitations, resource availability, and cost considerations. For instance, the storage requirement for CMIP5 data exceeds $2.5$ Petabyte, incurring significant expenses for institutions like the National Center for Atmospheric Research (NCAR), a prominent U.S. research center for Earth Systems science. Consequently, budget constraints may necessitate the abandonment of partial data or higher resolutions \citep{guinness2018compression}.

Emulators play a crucial role in providing rapid approximations for computationally demanding model outputs and serve as effective surrogates for the original model. Specifically, emulators are statistical models that undergo proper training using a set of existing simulations and are then employed to approximate simulations for unexplored inputs. Over the past few decades, emulators have found widespread application, including assessing the sensitivity of various physical parameters \citep{Oakley2002BayesianIF,Oakley2004Probabilistic} and performing model calibration \citep{Kennedy2001,chang2014}. Within the climate research community, emulators have been extensively developed to evaluate the influence of physics parameters \citep{Sacks1989}, emission scenarios \citep{Castruccio2013,castruccio2014statistical}, and the internal variability \citep{jeong2019stochastic,Hu2021Approximating} on climate model outputs. These emulators enable efficient exploration of model behavior and facilitate a deeper understanding of complex climate systems.

Stochastic generators \citep[SGs,][]{jeong2018reducing}, as a type of emulator for quantifying internal variability uncertainty, align closely with our requirements. 
SGs are spatio-temporal models adeptly tailored with a limited number of climate simulations, capable of generating numerous annual and monthly simulations across extensive and high-resolution domains at any moment \citep{jeong2018reducing,jeong2019stochastic,castruccio2019reproducing,Tagle2020}, yet daily scale remains elusive to this date. Their implementation is affordable and even efficient, albeit at the expense of detailed physical mechanisms. SGs require the storage of only model parameters, enabling them to generate simulations that capture the general behavior of the original climate simulations. It is essential to distinguish SGs from compression methods \citep{baker2017toward,Underwood2022Understanding}, which effectively reduce data volume using compression algorithms. While compression methods store and recover compressed data, they are designed for reconstructing individual data rather than generating new simulations. Moreover, the storage requirements for compression methods tend to increase with additional simulations.

SGs typically utilize Gaussian processes (GPs) and autoregressive models to capture the intricate spatio-temporal dependencies among climate simulations driven by underlying physical mechanisms. While SGs offer several advantages, they face notable challenges when emulating ESMs. Firstly, global climate simulations, which need the technique of SGs, involve extensive high-resolution data. Inference with GPs on such data can be computationally prohibitive, often requiring time on the order of data size cubed. Addressing this computational challenge can draw inspiration from existing techniques such as low-rank approximations \citep{FRK2008,PP}, composite likelihoods \citep{vecchia1988estimation,guinness2019gaussian}, stochastic partial differential equations \citep{Rue2009,bolin2023covariancebased}, covariance tapering \citep{taper2008} and their combinations \citep{datta2016hierarchical}. Secondly, global climate simulations frequently exhibit non-stationary spatial dependence. The covariance structure can vary significantly with latitude while may remain relatively stationary along the longitude, which is called axial symmetry \citep{jones1963stochastic} and has been noticed and discussed by \citet{stein2007spatial}, \citet{STEIN20083} and \citet{Jun2008Nonstationary}. Additionally, the smoothness and variance of data can change between land and ocean at the same latitude \citep{castruccio2017evolutionary,jeong2018reducing}. A well-defined covariance function based on geodesic distance rather than chordal distance is needed to model the non-stationary spatial dependence structure. The use of chordal distance may cause physically unrealistic distortions \citep{Jeong2017Spherical}. Lastly, global climate simulations with high temporal resolutions, such as monthly and daily scales, may deviate from Gaussian assumptions when studying temporal evolution. To address this, the Tukey g-and-h (TGH) transformation has been employed to Gaussianize the data, striking a balance between model flexibility and simplicity \citep{Yan2019Non-Gaussian,jeong2019stochastic}. 

For global surface temperature simulations from the CESM2-LENS2 data, which are newly published in the climate community and face the above-mentioned challenges, we propose a novel SG. Our SG, serves as a pragmatic complement to CESM2, offering the capability to rapidly generate an unlimited number of emulations closely resembling simulations. Central to our method is the utilization of the spherical harmonic transformation (SHT) -- the well-known spherical counterpart of the Fourier transformation, allowing us to efficiently transition global simulations from the spatial to the spectral domain. Moreover, it unlocks a practical low-rank approximation strategy, using spherical harmonics as available multi-resolution basis functions that are naturally suitable for global data. The modeling in the spectral domain ushers in significant reductions in both computational and storage costs. By judiciously retaining distinct ranks for land and ocean regions and applying a general assumption of axial symmetry, our SG captures the intricate non-stationary spatial dependence among different latitudes and between land and ocean regions. Furthermore, the employment of a modified TGH transformation equips our SG to effectively model the temporal dependence of simulations, accommodating the non-Gaussianity inherent in high-temporal-resolution data. In our case study, we apply these principles to construct SGs and generate emulations for the surface temperature simulations, spanning annual, monthly, and even daily scales. We meticulously illustrate and compare the generated emulations against the training simulations, both through visual inspection and quantitative metrics.  Notably, our work marks a significant advancement by being the first to generate, display, and evaluate emulations at a daily scale. 

The remainder of the paper is structured as follows. In Section~\ref{sec:datadescription}, we provide an overview of temperature simulations across annual, monthly, and daily scales. Section~\ref{sec:methodology} delves into the intricate procedures involved in constructing an SG using the SHT and generating emulations. The generated emulations, along with comparisons to the training simulations, are showcased in Section~\ref{sec:casestudy}. Finally, Section~\ref{sec:discussion} encapsulates our conclusions and outlines avenues for future research.

\section{CESM2-LENS2 Data}
\label{sec:datadescription}
The Community Earth System Model version 2 Large Ensembles (CESM2-LENS2), as detailed in \citet{Rogers2021}, is a recently released extensive ensemble of climate change projections. It originates from the CESM2 and serves as a valuable resource for studying the sensitivity of internal climate fluctuations to greenhouse warming. This ensemble comprises $100$ individual members, operating at a spatial resolution of  $0.9375^\circ\times 1.25^\circ$ (latitude$\times$longitude), spanning the temporal domain from 1850 to 2100, and generated through a combination of diverse oceanic and atmospheric initial states. CESM2-LENS2 draws upon CMIP6 historical forcing, spanning the period from 1850 to 2014, and integrates the Shared Socioeconomic Pathways 370 (SSP370) future radiative forcing (RF, an influential factor in climate system describing the difference between incoming and outgoing sun radiation) from 2015 to 2100. The SSP scenarios, introduced in response to the sixth IPCC report, encompass a range of socioeconomic developments and atmospheric greenhouse gas concentration pathways \citep{Kriegler2014,RIAHI2017}. Notably, SSP370 represents a medium-to-high emission scenario, projecting an additional RF of $7$ $\mathrm{Wm^{-2}}$ by the year 2100 under the social development pathway of regional rivalry. 

The choice of CESM2-LENS2 in our study is motivated by several factors. First, it provides comprehensive and credible global climate simulations at high resolutions, which often pose storage challenges. Second, its ample ensemble size allows for effective training and evaluation of our SG. Third, despite generating considerable interest within the climate community \citep{Kay2022,Munoz2023,Cameron2023}, CESM2-LENS2 has yet to gain widespread attention in the field of statistics. This paper marks the first introduction of an SG, extending to daily scales, to complement CESM2-LENS2. Both CESM2-LENS2 and SSP370 RF are readily accessible from the \href{https://www.cesm.ucar.edu/community-projects/lens2/data-sets}{NCAR} and \href{https://zenodo.org/record/3515339}{Zenodo} websites, respectively. Our analysis focuses on global surface temperatures for the period 2015–2100, encompassing all $192\times288=55,296$ grid points in space. Existing literature \citep{jeong2018reducing,jeong2019stochastic,Huang'sEmulator} has demonstrated that fewer than ten ensembles are sufficient to construct an efficient SG for a climate variable. Therefore, we select training simulations from a subset of $20$ members micro-initialized from the year 1251, with each member created through random perturbations of the atmospheric potential temperature field. These members are assumed to be independent from each other. We aggregate the ensembles at annual, monthly, and daily scales, despite their original availability at a three-hourly resolution. Consequently, each annual, monthly, and daily aggregated temperature simulation comprises approximately $4.75$ million ($192\times 288\times 86$), $57.07$ million, and $1.73$ billion data points, respectively.

\begin{figure}
\centering
\subfigure[$\Bar{y}_{A,9}(L_i,l_j)$]{
\label{fig:subfig:Annualmean2023}
\includegraphics[scale=0.57]{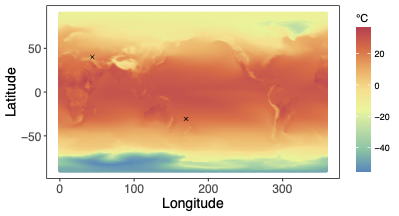}}
\subfigure[$y_{A,9}^{\text{sd}}(L_i,l_j)$]{
\label{fig:subfig:Annualsd2023}
\includegraphics[scale=0.57]{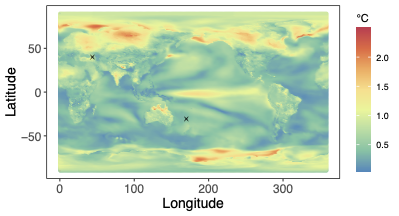}}\\
\subfigure[Annual temperature time series]{
\label{fig:subfig:Annualtemporal}
\includegraphics[scale=0.57]{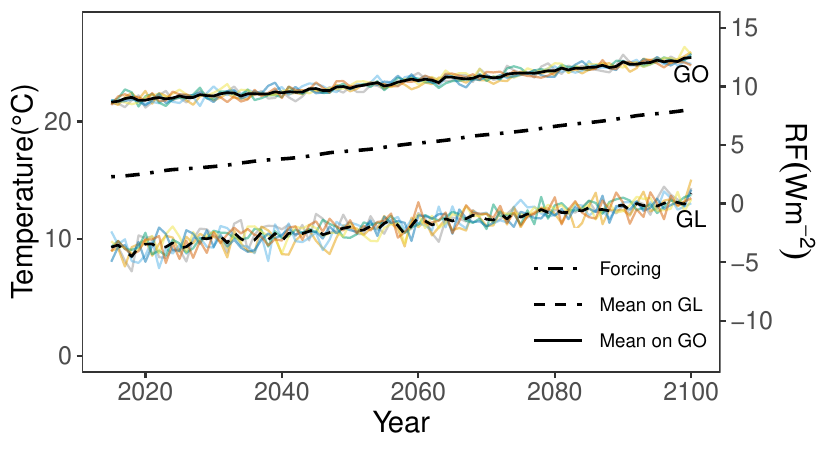}}
\subfigure[Monthly temperature time series]{
\label{fig:subfig:Mionthlytemporal}
\includegraphics[scale=0.57]{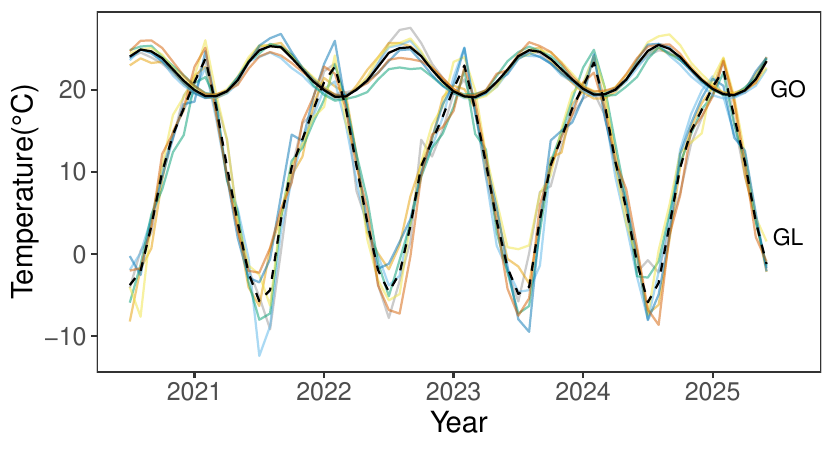}}\\
\subfigure[Annual]{
\label{fig:subfig:histannual}
\includegraphics[scale=0.56]{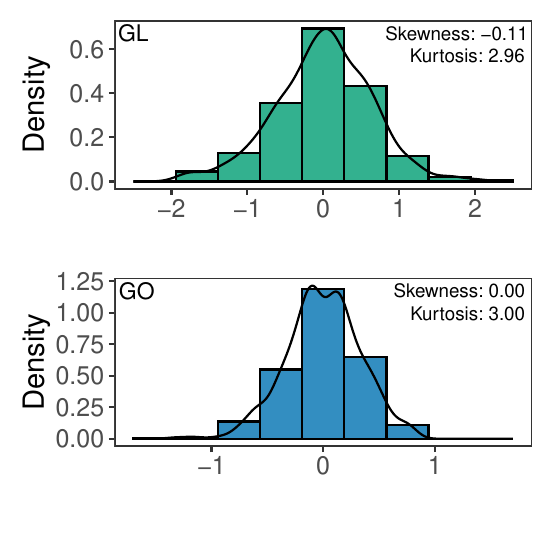}}
\subfigure[Monthly]{
\label{fig:subfig:histMonthly}
\includegraphics[scale=0.56]{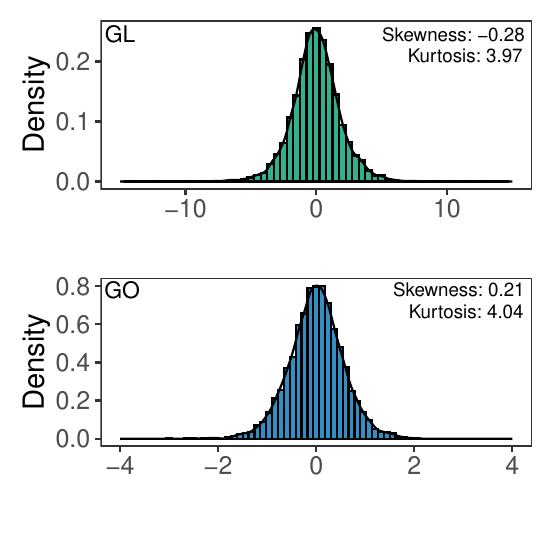}}
\subfigure[Daily]{
\label{fig:subfig:histDaily}
\includegraphics[scale=0.56]{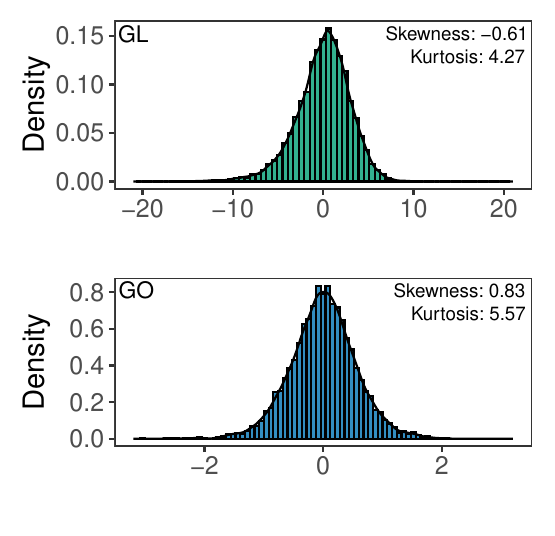}}\\
\caption{Illustration of simulations with different scales. (a) and (b) are maps of the ensemble mean and standard deviation of the annually aggregated temperature in the year 2023. Two black ``$\times$" marks indicate grid points located at coordinates $(40.05,43.75)$ and $(-30.63,170.00)$, designating points on land (GL) and ocean (GO), respectively. (c) shows annual temperature time series at GL and GO, where each ensemble is represented by a distinct color. (d) shows the monthly temperature time series at GL and GO, where the black curves are ensemble means. (e)--(g) are histograms of annual, monthly and daily temperature simulations at GL and GO, which have been detrended by ensemble means. Only data of years 2020, 2040, 2060, 2080, and 2100 are used in daily data. }
\label{Fig:Demo}
\vspace{-5pt}
\end{figure} 

Let $y_t^{(r)}(L_i,l_j)$ denote the temperature in Celsius at latitude $L_i$, longitude $l_j$, time point $t$ after year 2014, and ensemble $r$, where $i=1,\ldots,I$ ($I=192$), $j=1,\ldots,J$ ($J=288$), $t=1,\ldots,T$, and $r=1,\ldots,R$. The value of $T$ varies depending on the temporal resolution chosen, with possible values of $86$, $1032$, or $31,390$. Similarly, let $x_t$ represent the RF at time $t$ for all ensembles, as they are generated with the same external forcing. We use members 11 to 17 (out of 20) to illustrate certain data characteristics. The justification for the number of ensembles will be further discussed in Fig.~\ref{fig:subfig:IfitwithRs}, using an index for goodness-of-fit. Let $y_{A,t}^{(r)}(L_i,l_j)$ be the annually aggregated temperature at grid point $(L_i,l_j)$, year $t+$2014 and ensemble $r$. Consequently, the ensemble mean and standard deviation at $(L_i,l_j)$ and $t+$2014 are denoted as $\Bar{y}_{A,t}(L_i,l_j)=R^{-1}\sum_{r=1}^R y_{A,t}^{(r)}(L_i,l_j)$ and $y_{A,t}^{\text{sd}}(L_i,l_j)=\{R^{-1}\sum_{r=1}^R (y_{A,t}^{(r)}(L_i,l_j)-\Bar{y}_{A,t}(L_i,l_j))^2\}^{1/2}$, respectively. Figs.~\ref{fig:subfig:Annualmean2023} and \ref{fig:subfig:Annualsd2023} depict maps of $\Bar{y}_{A,9}(L_i,l_j)$ and $y_{A,9}^{\text{sd}}(L_i,l_j)$. 
% Notably, two black ``$\times$" marks indicate grid points located at coordinates $(40.05,43.75)$ and $(-30.63,170.00)$, designating points on land (GL) and ocean (GO), respectively. 
The maps align with the intuitive understanding that temperature generally decreases with increasing latitude. The surface temperatures in the Himalayas and Andes ranges are lower than in surrounding regions. Numerical instabilities occur near the two pole regions. Additionally, the surface temperature at the tropical Pacific Ocean exhibits larger variation across ensembles due to the climate phenomenon called El Ni\~{n}o–Southern Oscillation.

Let $\{y_{A,t}^{(r)}(L_i,l_j)\}_{t=1}^T$ represent the annual temperature trajectory of ensemble $r$ at $(L_i,l_j)$. Fig.~\ref{fig:subfig:Annualtemporal} illustrates these time series for both GL (land) and GO (ocean) locations, i.e., $\{y_{A,t}^{(r)}(\text{GL})\}_{t=1}^{T}$ and $\{y_{A,t}^{(r)}(\text{GO})\}_{t=1}^T$, alongside their respective ensemble means $\{\Bar{y}_{A,t}(\text{GL})\}_{t=1}^T$ and $\{\Bar{y}_{A,t}(\text{GO})\}_{t=1}^T$. Take $\{y_{A,t}^{(r)}(\text{GL})\}_{t=1}^T$ as an example. All ensembles exhibit a shared increasing trend, influenced by the RF also depicted in Fig.~\ref{fig:subfig:Annualtemporal}. Moreover, each ensemble exhibits its unique shape, signifying the presence of uncertainty among ensembles. These motivate us to develop an SG incorporating both deterministic and stochastic factors. Similarly, Fig.~\ref{fig:subfig:Mionthlytemporal} provides monthly temperature time series of years 2021--2025 at GL and GO, along with their respective ensemble means. They experience alternating peaks and valleys because they are situated in the northern and southern hemispheres, respectively. In comparison to annual data, Fig.~\ref{fig:subfig:Mionthlytemporal} exhibit a clear seasonality, which is a common feature in monthly and daily data. The seasonality of surface temperature varies across grid points and remains consistent among all ensembles at a fixed grid point. Notably, from Figs.~\ref{fig:subfig:Annualtemporal} and \ref{fig:subfig:Mionthlytemporal}, both the uncertainty and the variation are more pronounced at GL. 

Figs.~\ref{fig:subfig:histannual}--\ref{fig:subfig:histDaily} demonstrate histograms, skewness and kurtosis of the detrended annual, monthly and daily temperature simulations at GL and GO, respectively. For example, the top panel of Fig.~\ref{fig:subfig:histannual} displays the histogram of $\{y_{A,t}^{(r)}(\text{GL})-R^{-1}\sum_{r=1}^R y_{A,t}^{(r)}(\text{GL})\}_{r=1,\ldots,R; t=1,\ldots,T}$ and its skewness and kurtosis. The close-to-zero skewness and the close-to-three kurtosis enable a Gaussian assumption when we model $R$ time series $\{y_{A,t}^{(r)}(\text{GL})-R^{-1}\sum_{r=1}^R y_{A,t}^{(r)}(\text{GL})\}_{t=1}^T$, with an auto-regression. In contrast, the histograms in Figs.~\ref{fig:subfig:histMonthly} and \ref{fig:subfig:histDaily} tend to be skewed and heavy-tailed. Fig.~S1 presents the skewness and kurtosis for time series across all global grid points. 
% Further assessment through a Jarque-Bera test \citep{Jarque1987ATF} indicates that only $23.3\%$ of grid points with annually aggregated simulations reject the Gaussianity. 
The degree of skewness and the presence of heavy tails become more pronounced with increasing temporal resolution. The non-Gaussianity is particularly severe in the band region below latitude $-60^\circ$, termed as Band, and the North Pole region, where the surface temperature simulations exhibit numerical instabilities. The reason will be discussed in Section~\ref{sec:subsec:daily}. 
% The Jarque-Bera test reveals that about $95\%$ grid points reject the Gaussianity for simulations with higher temporal resolutions. 
These observations underscore the necessity for additional transformations and parameters to Gaussianize the monthly and daily simulations, as the first two moments alone are insufficient to characterize data with greater temporal complexity.

\section{Stochastic Generator Methodology}
\label{sec:methodology}
In this section, we detail the procedures of constructing an SG and generating emulations for the temperature simulations described in Section~\ref{sec:datadescription}. Leveraging the SHT, our proposed SG offers efficient and adaptable capabilities for dealing with global climate simulations that exhibit: 1) a substantial size in terms of the number of observations in space and time; 2) non-stationary spatial dependencies among latitudes and land/ocean regions; and 3) non-Gaussian temporal trajectories. 

The data characteristics outlined in Section~\ref{sec:datadescription} underscore the deterministic chaotic nature inherent in climate models \citep{lorenz1963,branstator2010,Stefano_principle}, which motivates us to decouple the data into deterministic and stochastic components as follows:
\vspace{-20pt}
\begin{equation}
    y_t^{(r)}(L_i,l_j)=m_t(L_i,l_j)+\sigma(L_i,l_j)Z_t^{(r)}(L_i,l_j).
    \label{eq:decouple}
    \vspace{-20pt}
\end{equation}
Here, $m_t$ and $\sigma$ are deterministic functions responsible for the mean trend and standard error, respectively, and are shared across all ensembles. $Z_t^{(r)}(L_i,l_j)$ is the stochastic component at grid point $(L_i,l_j)$, time point $t$, and ensemble $r$. Given the expansive parameter space, simultaneously inferring both the deterministic and stochastic components would be computationally infeasible. Therefore, we adopt a two-stage approach, commonly employed in existing work \citep{Stefano_principle,jeong2019stochastic,Huang'sEmulator}. First, we evaluate $m_t(L_i,l_j)$ and $\sigma(L_i,l_j)$ at each grid, making the independence assumption regarding $\{Z_t^{(r)}(L_i,l_j)\}_{i=1,\ldots,I; j=1,\ldots,J; t=1,\ldots,T; r=1,\ldots,R}$. Subsequently, we analyze the dependence structure of $Z_t^{(r)}(L_i,l_j)$ by detrending and rescaling $y_t^{(r)}(L_i,l_j)$ with the estimate of $m_t(L_i,l_j)$ and $\sigma(L_i,l_j)$, respectively.

\subsection{Deterministic component of SG}
\label{sec:subsec:deterministic}
In the first stage of constructing the SG, we focus on determining the mean trend $m_t$. It is crucial that the mean trend is both simple enough to minimize the storage requirements for parameters and informative enough to capture the physical relationships between simulations and essential covariates. Let $y_t^{(r)}(L_i,l_j)$ represent the annually aggregated temperature data. Previous research  \citep{castruccio2014statistical,Huang'sEmulator} has shown its dependence on the RF trajectory and adopted an infinite distributed lag model:
\vspace{-20pt}
\begin{equation*}
m_t(L_i,l_j)=\beta_0(L_i,l_j)+\beta_1(L_i,l_j)x_t+\beta_2(L_i,l_j)\{1-\rho(L_i,l_j)\}\sum_{s=1}^{\infty}\rho(L_i,l_j)^{s-1}x_{t-s},
\vspace{-20pt}
\end{equation*}
where $\beta_0(L_i,l_j)$ is the intercept, $\beta_1(L_i,l_j)$ and $\beta_2(L_i,l_j)$ are slopes for the current and past RFs, respectively. Lag weights $\beta_2(L_i,l_j)\{1-\rho(L_i,l_j)\}\rho(L_i,l_j)^{s-1}$ decrease the impact of past RFs exponentially by $\rho(L_i,l_j)\in[0,1]$. Moreover, if $y_t^{(r)}(l_i,l_j)$ represents the monthly aggregated temperature data, additional harmonic terms $\{\cos(2\pi tk/12),\sin(2\pi tk/12)\}_{k=1}^{K_M}$ should be included to fit the interannual cycle as shown in Fig.~\ref{fig:subfig:Mionthlytemporal}. That is, 
\vspace{-20pt}
\begin{equation*}
\begin{aligned}
m_t(L_i,l_j)=&\beta_0(L_i,l_j)+\beta_1(L_i,l_j)x_{\lceil t/12 \rceil}+\beta_2(L_i,l_j)\{1-\rho(L_i,l_j)\}\sum_{s=1}^{\infty}\rho(L_i,l_j)^{s-1}x_{\lceil t/12 \rceil-s}\\
&+\sum_{k=1}^{K_M}\left\{a_k(L_i,l_j)\cos\left(\frac{2\pi tk}{12}\right)+b_k(L_i,l_j)\sin\left(\frac{2\pi tk}{12}\right)\right\},
\end{aligned}
% \label{eq:Monthly_mean}
\vspace{-10pt}
\end{equation*}
which would also be applied to the daily aggregated temperature by replacing $K_M$ and $t/12$ with $K_D$ and $t/365$, respectively. The larger the value of $K_M$ (or $K_D$), the higher the frequency the harmonic terms can present. For the deterministic component, $\{\beta_0(L_i,l_j),\beta_1(L_i,l_j),\beta_2(L_i,l_j)$, $\rho(L_i,l_j),\sigma(L_i,l_j),a_1(L_i,l_j),\ldots,a_K(L_i,l_j),b_1(L_i,l_j),\ldots,b_K(L_i,l_j)\}_{i=1,\ldots,I; j=1,\ldots,J}$ are all $(5+2K)IJ$ parameters to be evaluated and stored, where $K$ can take values of $0$, $K_M$, and $K_D$ to represent the annual, monthly and daily temperature, respectively. With the independence assumption, we can efficiently estimate the parameters of the deterministic component in parallel for each grid point. The detailed inferential process is given in Section~S3.1 of the Supplementary Materials.

\subsection{Stochastic component of SG}
Now, we sequentially model the spatial and temporal dependence of the stochastic component $Z_t^{(r)}(L_i,l_j)$, which is a general principle for analyzing space-time data \citep{stein2007spatial,Stefano_principle} to bypass the prohibitive computation.
\subsubsection{Modeling the spatial dependence}
% \subsubsection{SHT: permitting a low-rank approximation for gridded simulations over globe}  
The inference of the spatial dependence should consider several crucial factors: 1) The simulations are extensive; 2) The simulations encompass the entire globe, making the use of chordal distance inappropriate \citep{Jeong2017Spherical}; and 3) The simulations exhibit non-stationarity among different latitudes and between land and ocean regions. Given these considerations, we propose using the SHT \citep{jones1963stochastic,stein2007spatial,STEIN20083}, which involves expanding the stochastic component $Z_t^{(r)}(L_i,l_j)$ with spherical harmonics:
\vspace{-10pt}
\begin{equation}
    Z_t^{(r)}(L_i,l_j)=\sum_{q=0}^{Q-1}\sum_{m=-q}^q (s_t^{(r)})_q^m H_q^m(L_i,l_j)+\varepsilon_t^{(r)}(L_i,l_j).
    \label{eq:SHT}
\vspace{-10pt}
\end{equation}
In \eqref{eq:SHT}, $\{H_q^m\}_{q=0,1,2,\ldots; m=-q,\ldots,q}$ is a complete set of orthonormal basis functions defined in $\mathcal{L}^2(\mathbb{S}^2)$ and indexed by integer degree $q$ and order $m$, where  $\mathcal{L}^2(\mathbb{S}^2)$ is the Hilbert space consisting of squared-integrable functions on the sphere $\mathbb{S}^2$. The spherical harmonic $H_q^m$ is a complex-valued function with a closed form $H_q^m(L_i,l_j)=\sqrt{\frac{2q+1}{4\pi}\frac{(q-m)!}{(q+m)!}}P_q^m(\cos\theta_i)\exp(\iota m \psi_j)$,
% \begin{equation*}
%     H_q^m(L_i,l_j)=\sqrt{\frac{2q+1}{4\pi}\frac{(q-m)!}{(q+m)!}}P_q^m(\cos\theta_i)\exp(im \psi_j),
% \end{equation*}
where $\iota^2=-1$, $(\theta_i,\psi_j)=(-\pi L_i/180+\pi/2,\pi l_j/180)$ is a reparameterization of $(L_i,l_j)$ and $P_q^m$ is the associated Legendre polynomial of degree $q$ and order $m$, such that $H_q^{-m}=(-1)^m\overline{H_q^m}$. Corresponding to the $H_q^m$,  $(s_t^{(r)})_q^m$ is the complex-valued spherical harmonic coefficient satisfying $(s_t^{(r)})_q^{-m}=(-1)^m\overline{(s_t^{(r)})_q^{m}}$. The norm of $(s_t^{(r)})_q^{m}$, i.e., $\|(s_t^{(r)})_q^{m}\|_2=\sqrt{\Re^2\{(s_t^{(r)})_q^{m}\}+\Im^2\{(s_t^{(r)})_q^{m}\}}$, indicates the energy gathered at degree $q$ and order $m$. More details about spherical harmonics and their application on spectrum and differentiability analysis are given in Section~S3.2.1 of the Supplementary Materials. $Q$ is an integer ranging from $0$ to $Q_{\max}$, with $Q_{\max}=\min\{I-1,(J+1)/2\}$ derived in Section~S3.2.2 and determined by the spatial resolution of the data. The denser the grid points are, the higher the frequency they capture, and the larger the value of $Q_{\max}$ could be. For CESM2-LENS2, $Q_{\max}=144$. The term $\varepsilon_t^{(r)}(L_i,l_j)$ accounts for the remaining information and is assumed to be independent and follow $\mathcal{N}(0,v^2(L_i,l_j))$. In a physical sense, \eqref{eq:SHT} translates $\{Z_t^{(r)}(L_i,l_j)\}_{i=1,\ldots,I; j=1,\ldots,J}$ from the spatial domain to $\{(s_t^{(r)})_q^m\}_{q=0,\ldots,Q-1; l=-q,\ldots,q}$ in the spectral domain. In a statistical sense, with $Q\ll\sqrt{IJ}$, \eqref{eq:SHT} provides a low-rank approximation for the stochastic process $Z_t$, where $(s_t)_q^{m}$ serves as a random coefficient and $Q$ trades off the quality of approximation against the computational complexity. 

\begin{figure}[!t]
\centering
\subfigure[$Z_9^{(1)}(L_i,l_j)$ in spatial domain]{
\label{fig:subfig:Spatial}
\includegraphics[scale=0.4]{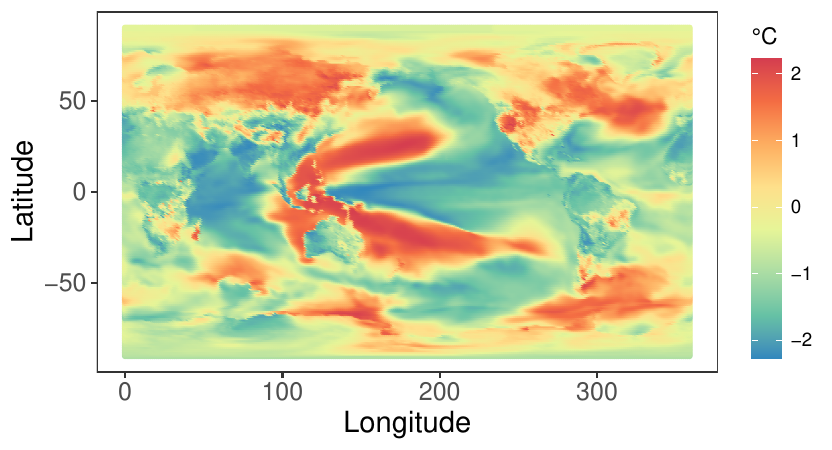}}
\subfigure[$\log_{10}\|(s_9^{(1)})_q^m\|_2$ in spectral domain]{
\label{fig:subfig:Spectral}
\includegraphics[scale=0.4]{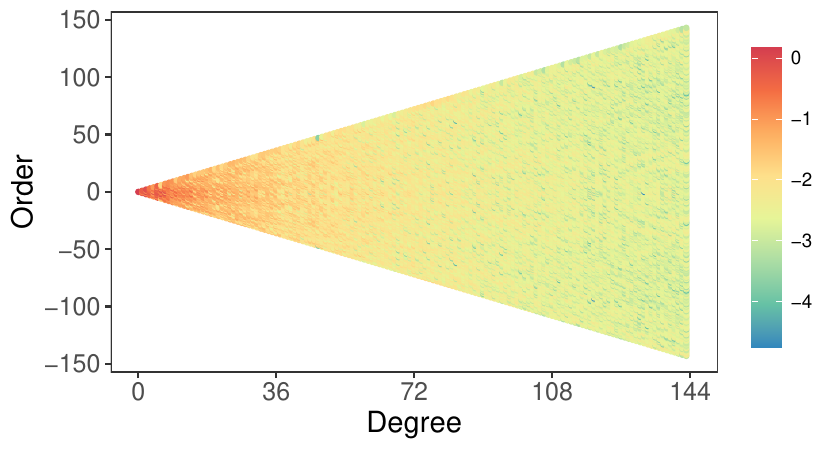}}
\subfigure[$|\varepsilon_9^{(1)}(L_i,l_j)|$ with $Q=36$]{
\label{fig:subfig:SHTinv48}
\includegraphics[scale=0.4]{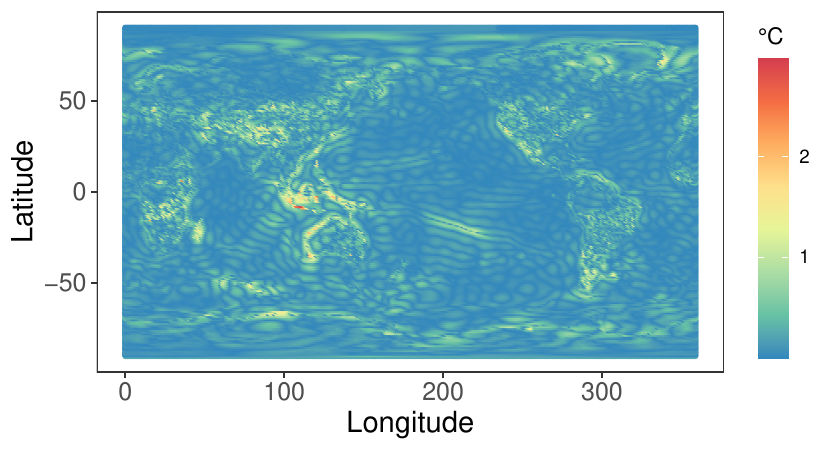}}\\
\vspace{-5pt}
\subfigure[$|\varepsilon_9^{(1)}(L_i,l_j)|$ with $Q=72$]{
\label{fig:subfig:SHTinv96}
\includegraphics[scale=0.4]{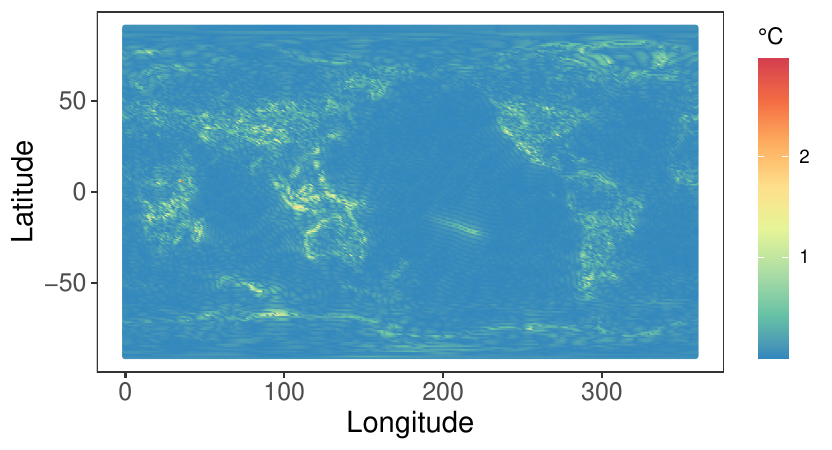}}
\subfigure[$|\varepsilon_9^{(1)}(L_i,l_j)|$ with $Q=116$]{
\label{fig:subfig:SHTinv116}
\includegraphics[scale=0.4]{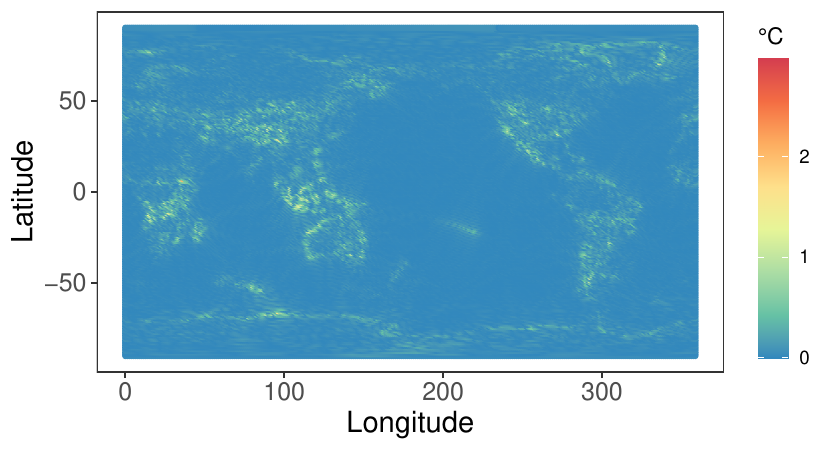}}
\subfigure[$|\varepsilon_9^{(1)}(L_i,l_j)|$ by LatticeKrig]{
\label{fig:subfig:LatticeKrig}
\includegraphics[scale=0.4]{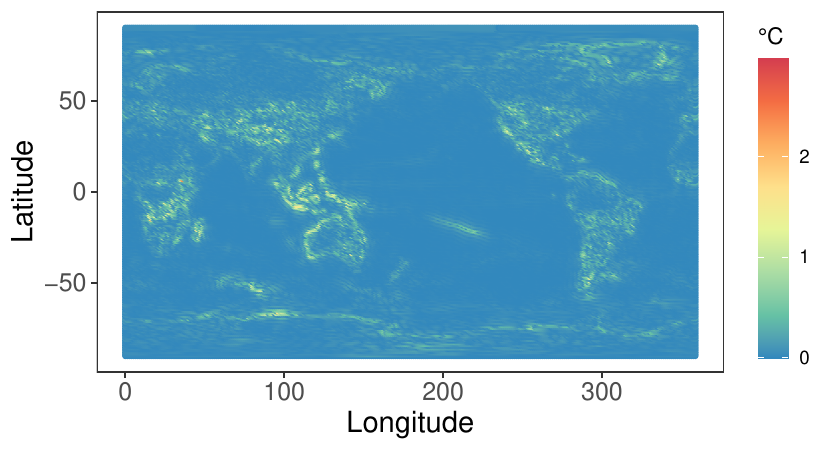}}
\vspace{-10pt}
\caption{Demonstration of SHT. (a) shows a set of stochastic component $\{Z_{9}^{(1)}(L_i,l_j)\}_{i=1,\ldots,I; j=1,\ldots,J}$ in the spatial domain. (b) indicates energy allocation of (a) in spectral domain. (c)--(e) depict the absolute values of $\{\varepsilon_{9}^{(1)}(L_i,l_j)\}_{i=1,\ldots,I; j=1,\ldots,J}$ in \eqref{eq:SHT} with $Q=36$, $72$, and $116$, respectively. (f) depicts the absolute values of $\{\varepsilon_{9}^{(1)}(L_i,l_j)\}_{i=1,\ldots,I; j=1,\ldots,J}$ obtained by using LatticeKrig with $13,658$ bases of $6$ levels, where $13,658>116^2=13,456$.}
\label{Fig:Year2023demo}
\vspace{-10pt}
\end{figure}

We provide Fig.~\ref{Fig:Year2023demo} to facilitate the understanding of SHT. Fig.~\ref{fig:subfig:Spatial} shows a set of stochastic components in the spatial domain, i.e., $\{Z_{9}^{(1)}(L_i,l_j)\}_{i=1,\ldots,I; j=1,\ldots,J}$, which is obtained by detrending and rescaling the annually aggregated temperature at year 2023 and ensemble one with ensemble mean and standard error in Figs.~\ref{fig:subfig:Annualmean2023} and \ref{fig:subfig:Annualsd2023}, respectively. After applying the SHT, Fig.~\ref{fig:subfig:Spectral} presents the pattern of $\{\log_{10}\|(s_{9}^{(1)})_q^m\|_2\}_{q=0,1,\ldots,Q_{max}; m=-q,\ldots,q}$, which indicates the energy allocation in the spectral domain. The energy is more concentrated at lower degrees of lower frequencies, gradually decreasing as the degree increases. The decay rate reflects the differentiability property of the data, as detailed in Section~S3.2.1 of the Supplementary Materials. Figs.~\ref{fig:subfig:SHTinv48}--\ref{fig:subfig:SHTinv116} show the influence of $Q$ on the loss of using SHT for a single set of stochastic components. Section~S3.2.3 provides a more comprehensive illustration of the general approximation performance. In essence, a larger value of $Q$ retains more information and provides a better approximation. Additionally, the spatial structure of surface temperature varies significantly between land and ocean.

We compare SHT with two popular bases used in low-rank approximations, empirical orthogonal functions (EOFs) and those in LatticeKrig \citep{LatticeKrig}, as detailed in Section~S3.2.4. EOFs, as data-driven basis functions, lack a closed form and must be stored. Their evaluation involves computing and eigen-decomposing an $IJ\times IJ$ matrix, leading to constraints on computational time and resources. Figs.~\ref{fig:subfig:SHTinv116} and \ref{fig:subfig:LatticeKrig} illustrate comparable approximation performance of SHT and LatticeKrig on $Z_9^{(1)}(L_i,l_j)$ with similar numbers of bases. However, LatticeKrig entails more time for parameter tuning selection and coefficient calculation, along with  greater storage requirements. For a more detailed and comprehensive comparison, readers are referred to Section~S3.2.4. We emphasize the advantages of employing the spherical harmonics to represent substantial global climate simulations: 1) Natural basis functions. Spherical harmonics are eigenfunctions of the Laplace-Beltrami operator, making them well-suited for analyzing spherical data of various variables, as well as their spectrum; 2) Easily available multi-resolution basis functions. Spherical harmonics have a closed form, which saves the procedures for getting bases, conserves the memory for storing them, and facilitates the derivation of special forms of covariance under certain assumptions. Their resolutions are automatically multiple, eliminating the need for researchers to manually allocate knots or tuning parameters. Instead, one can simply select a degree $Q$ to control the level of detail; 3) Efficient coefficient calculation. As shown in Section~S3.2.2, both the SHT and its inverse can be efficiently calculated with a computational time $O(Q^3)$. Moreover, this computational time can be further reduced to $O(Q^2)$ by parallel computing. On a MacBook with a $3.2$ GHz Apple M1 Pro processor with ten cores, performing the SHT with $Q=Q_{\max}$ in Fig.~\ref{Fig:Year2023demo} takes about $3$ seconds, and the inverse SHT with $Q=36$, $72$, and $144$ take no more than $0.06$, $0.3$, and $2$ seconds, respectively. 
% Unlike existing methods \citep{LK2015,MRA2017,autoFRK} that use basis functions mathematically developed without a clear physical meaning, the SHT utilizes spherical harmonics, which are the eigen-functions of the Laplace-Beltrami operator. This makes the SHT particularly suitable for analyzing spherical data, such as global climate simulations. The SHT transforms data from the spatial domain to the spectral domain, providing a different perspective for data analysis. This spectral view can reveal hidden patterns and characteristics of the data that may not be as apparent in the spatial domain; 2) Automatic multiresolution. Spherical harmonics are naturally organized in order of their degrees. This inherent multiresolution property allows them to capture information at various scales without the need for manually allocating knots or tuning parameters. Researchers need only select the maximum degree $Q$, to control the level of detail; 

Next, we model the spatial dependence structure, which is assumed to remain consistent over time. The covariance between $Z_t^{(r)}(L_i,l_j)$ and $Z_t^{(r)}(L_{i'},l_{j'})$ is 
\vspace{-20pt}
\begin{equation}
    c\{(L_i,l_j),(L_{i'},l_{j'})\}=\mathbf{H}(L_i,l_j)^\top\mathbf{K}\overline{\mathbf{H}(L_i,l_j)}+\delta_{i=i'}\delta_{j=j'}v^2(L_i,l_j),
\label{eq:covariance}
\vspace{-20pt}
\end{equation}
where $\mathbf{K}=\mathbf{K}_t^{(r)}=\mathrm{E}\{\mathbf{s}_t^{(r)}(\mathbf{s}_t^{(r)})^\mathrm{H}\}$ is a $Q^2\times Q^2$ covariance matrix of $\{(s_t^{(r)})_q^m\}_{q=0,1,\ldots,Q; m=-q,\ldots,q}$, $\mathrm{H}$ is the operator of conjugate transpose, $\mathbf{H}(L_i,l_j)$ and $\mathbf{s}_t^{(r)}$ are vectors consisting of basis functions and spherical harmonic coefficients, indexed by $(q^2+q+m+1)$. Specifically, the $(q^2+q+m+1)$th elements of $\mathbf{H}(L_i,l_j)$ and $\mathbf{s}_t^{(r)}$ are $H_q^m(L_i,l_j)$ and $(s_t^{(r)})_q^m$, respectively. 

Further examining the structure of $\mathbf{K}$, it is evident that the stochastic component in Fig.~\ref{fig:subfig:Spatial} exhibits strong heterogeneity across different latitudes, as observed in previous studies of global climate data \citep{stein2007spatial,jeong2018reducing,Huang'sEmulator}. Consequently, the commonly used isotropic assumption for Euclidean data is no longer appropriate. Instead, we adopt the assumption of axial symmetry, which posits that only data on the same latitude are stationary \citep{jones1963stochastic}. This leads to the covariance $c\{(L_i,l_j),(L_{i'},l_{j'})\}$ becoming a function of $L_i$ ($\theta_i$), $L_{i'}$ ($\theta_{i'}$), and $|l_j-l_{j'}|$ ($|\psi_j-\psi_{j'}|$, the central angle between points $(L_i,l_j)$ and $(L_{i'},l_{j'})$ expressed in chordal distance). Using a simple derivation outlined in \citet{jones1963stochastic}, we have
\vspace{-10pt}
\begin{equation}
    c(L_i,L_{i'},|l_j-l_{j'}|)=\sum_{q=0}^{Q-1}\sum_{q'=0}^{Q-1}\sum_{m=-q}^qk_{qq'm}H_q^m(\theta_i,0)H_{q'}^m(\theta_{i'},0)\exp(\iota m|\psi_j-\psi_{j'}|)
    \label{eq:covaxial}
    \vspace{-10pt}
\end{equation}
with $\mathrm{E}\{(s_t^{(r)})_q^m(s_t^{(r)})_{q'}^{m'}\}=\delta_{m=m'}k_{qq'm}$. It means that axial symmetry assumes dependence only on coefficients of the same order. Thus, $\mathbf{K}$ becomes a sparse matrix, consisting of $Q^2+(Q-1)^2+\cdots+1^2=Q^3/3+Q^2/2+Q/6$ non-zero elements.
% and $Q^3/3+2Q/3$ real elements to be stored. 

However, from Figs.~\ref{Fig:Demo} and~\ref{Fig:Year2023demo}, we still observe heterogeneity between land and ocean data on the same latitude. The stochastic components on land and ocean exhibit different spatial structures, requiring different numbers of basis functions to capture them adequately. Therefore, we propose replacing~\eqref{eq:SHT} by 
\vspace{-10pt}
\begin{equation}
    Z_t^{(r)}(L_i,l_j)=\delta_{(L_i,l_j)\in\mathcal{S}_l}\sum_{q=0}^{Q_l-1}\sum_{m=-q}^q(s_t^{(r)})_q^mH_q^m(L_i,l_j)+\delta_{(L_i,l_j)\in\mathcal{S}_{o}}\sum_{q=0}^{Q_o-1}\sum_{m=-q}^q(s_t^{(r)})_q^mH_q^m(L_i,l_j)+\varepsilon_t^{(r)}(L_i,l_j),
    \label{eq:SHTlo}
    \vspace{-10pt}
\end{equation}
where $\mathcal{S}_o$ ($\mathcal{S}_l$) represents the set of grid points over the ocean (land) and $Q_o$ ($Q_l$) denotes the value of $Q$ for ocean (land). We ignore the continuity of temperature changes between land and ocean, since \eqref{eq:SHTlo} is formulated on grid points and our target is not to predict at new points. Accordingly, the covariance $c(L_i,L_{i},|l_j-l_{j'}|)$ is given by plugging the $Q$ values of $(L_i,l_j)$ and $(L_{i'},l_{j'})$ into \eqref{eq:covaxial}. We present the process of choosing appropriate values for $Q_o$ and $Q_l$ in Section~\ref{sec:casestudy}. 

\subsubsection{Modeling the temporal dependence}
\label{sec:subsec:temporal}
Next, we investigate the temporal dependence within the vector time series $\{\mathbf{s}_t^{(r)}\}_{t=1}^T$ in the spectral domain. We will employ an auto-regressive model, but with two additional operations: converting the coefficients to real-valued ones and Gaussianizing them using the TGH transformation. Specifically, for the complex-valued vector $\mathbf{s}_t^{(r)}$ with the special structure of $(s_t^{(r)})_q^{-m}=(-1)^m\overline{(s_t^{(r)})_q^{m}}$, we first linearly transform it into a real-valued vector:
\vspace{-15pt}
\begin{equation*}
   \tilde{\mathbf{s}}_t^{(r)}=\mathbf{A}^{-1} \mathbf{s}_t^{(r)}=[(s_t^{(r)})_0^0,\Im\{(s_t^{(r)})_1^1\},(s_t^{(r)})_1^0,\Re\{(s_t^{(r)})_1^1\},\ldots,\Im\{(s_t^{(r)})_{Q-1}^{Q-1}\},\ldots,\Re\{(s_t^{(r)})_{Q-1}^{Q-1}\}]^\top,
   \vspace{-20pt}
\end{equation*}
where both $\mathbf{A}$ and its inverse are $Q^2\times Q^2$ sparse matrices with known structures. For example, in the case of $Q=2$, $\mathbf{A}$ and $\mathbf{A}^{-1}$ can be represented as 
\vspace{-10pt}
\begin{equation*}
    \left(\begin{array}{cccc}
        1 & 0 & 0 & 0 \\ [-2.5ex]
        0 & \iota & 0 & -1 \\ [-2.5ex]
        0 & 0 & 1 & 0 \\ [-2.5ex]
        0 & \iota & 0 & 1
    \end{array}\right)  
    \quad  \text{and} \quad
    \left(\begin{array}{cccc}
        1 & 0 & 0 & 0 \\ [-2.5ex]
        0 & -\iota/2 & 0 & -\iota/2 \\ [-2.5ex]
        0 & 0 & 1 & 0 \\ [-2.5ex]
        0 & -1/2 & 0 & 1/2 
    \end{array}\right),
    \vspace{-10pt}
\end{equation*}
respectively. For simplicity, we denote the $(q^2+q+m+1)$th element of $\tilde{\mathbf{s}}_t^{(r)}$ as $(\Tilde{s}_t^{(r)})_q^m$. Consequently, we have $Z_t^{(r)}(L_i,l_j)=\mathbf{H}(L_i,l_j)^\top\mathbf{A}\Tilde{\mathbf{s}}_t^{(r)}+\varepsilon_t^{(r)}(L_i,l_j)$, where $\mathbf{A}^\top\mathbf{H}(L_i,l_j)$ and $\tilde{\mathbf{s}}_t^{(r)}$ are vectors of bases and coefficients, respectively. Under the assumption of axial symmetry, the covariance of $\tilde{\mathbf{s}}_t^{(r)}$, denoted as $\tilde{\mathbf{K}}_0=\mathrm{E}(\tilde{\mathbf{s}}_t^{(r)}\tilde{\mathbf{s}}_{t}^{(r)\top})$, is also a sparse matrix. Specifically, $\mathrm{E}\{(\Tilde{s}_t^{(r)})_q^m (\Tilde{s}_{t}^{(r)})_{q'}^{m'}\}=\mathrm{E}\{(\Tilde{s}_t^{(r)})_q^{-m} (\Tilde{s}_t^{(r)})_{q'}^{-m'}\}=\Tilde{k}_{qq'm}\delta_{m=m'}\delta_{m\ge0}$. The derivation is given in Section~S3.3 of Supplementary Materials. The matrix $\tilde{\mathbf{K}}_0$ consists of $2Q^3/3+Q/3$ non-zero elements and only $Q^3/3+Q^2/2+Q/3$ ones need to be stored.  If $Z_t^{(r)}(L_i,l_j)$ represents the stochastic component of annually aggregated temperature, it is reasonable to assume the Gaussianity of $\{Z_t^{(r)}(L_i,l_j)\}_{t=1}^T$, and therefore, the Gaussianity of $\{\Tilde{\mathbf{s}}_t^{(r)}\}_{t=1}^T$. This allows us to model it with a vector auto-regressive model of order $P$ (VAR($P$)): $\tilde{\mathbf{s}}_t^{(r)}=\sum_{p=1}^P\bm\Phi_p\tilde{\mathbf{s}}_{t-p}^{(r)}+\bm\xi_t^{(r)}$, 
% \vspace{-10pt}
% \begin{equation}
%     \tilde{\mathbf{s}}_t^{(r)}=\sum_{p=1}^P\bm\Phi_p\tilde{\mathbf{s}}_{t-p}^{(r)}+\bm\xi_t^{(r)}, 
%     \label{eq:varp}
%     \vspace{-10pt}
% \end{equation}
where $\bm\xi_t^{(r)}\overset{i.i.d.}{\sim}\mathcal{N}_{Q^2}(\mathbf{0},\mathbf{U})$ and $\mathbf{U}=\tilde{\mathbf{K}}_0-\sum_{p=1}^P\bm\Phi_p\tilde{\mathbf{K}}_0\bm\Phi_p^\top-\sum_{p=1}^P\sum_{p'\neq p}^P\bm\Phi_p \tilde{\mathbf{K}}_{|p'-p|}\bm\Phi_{p'}^\top$ with $\tilde{\mathbf{K}}_{|p'-p|}=\mathrm{E}(\tilde{\mathbf{s}}_{t-p}^{(r)}\tilde{\mathbf{s}}_{t-p'}^{(r)\top})$.

However, as in Sections~\ref{sec:datadescription} and S2, when $Z_t^{(r)}(L_i,l_j)$ is the stochastic component of the monthly or daily aggregated temperature, the Gaussianity assumption for time series $\{Z_t^{(r)}(L_i,l_j)\}_{t=1}^T$ at most grid points $(L_i,l_j)$ may not hold. It implies that $\{(\tilde s_t^{(r)})_q^m\}_{t=1}^T$ at some $(q,m)$ pairs are not Gaussian and motivates us to employ a modified TGH transformation. Denote by $\mathcal{S}_{gh}$ the set of $(q,m)$ such that $\{(\tilde s_t^{(r)})_q^m\}_{t=1}^T$ rejects the Gaussianity. We model the coefficients with a TGH auto-regressive model:  $\check{\mathbf{s}}_t^{(r)}=\sum_{p=1}^P\bm\Phi_p\check{\mathbf{s}}_{t-p}^{(r)}+\bm\xi_t^{(r)}$, where the $(q^2+q+m+1)$th element of $\check{\mathbf{s}}_t^{(r)}$ is
\vspace{-10pt}
\begin{equation}
    (\check{s}_t^{(r)})_q^m=\left\{
    \begin{array}{lr}
       \lambda_q^m\tau_{g_q^m,h_q^m}^{-1}\{(\Tilde{s}_t^{(r)})_q^m/\omega_q^m\} , & \text{if}\quad(q,m)\in\mathcal{S}_{gh}, \\
        (\Tilde{s}_t^{(r)})_q^m, & \text{if}\quad(q,m)\notin\mathcal{S}_{gh},
    \end{array}
    \right.
    \vspace{-10pt}
    \label{eq:tukeygh}
\end{equation}
with $\tau_{g_q^m,h_q^m}(s)=(g_q^m)^{-1}\{\exp(g_q^m s)-1\}\exp(h_q^m s^2/2)$. Compared to the regular TGH, the one in \eqref{eq:tukeygh} removes a location parameter and introduces a scale parameter $\lambda_q^m$. The former is due to the zero-mean $\{(\Tilde{s}_t^{(r)})_q^m\}_{t=1,\ldots,T; r=1,\ldots,R}$. The later ensures the standard deviation of $\{(\check{s}_t^{(r)})_q^m\}_{t=1,\ldots,T; r=1,\ldots,R}$ to be equal to that of $\{(\tilde{s}_t^{(r)})_q^m\}_{t=1,\ldots,T; r=1,\ldots,R}$. Four additional parameters describing characteristics beyond the first two moments are to be stored. Based on \eqref{eq:tukeygh}, the covariance matrix of $\bm\xi_t^{(r)}$ is $\mathbf{U}=\check{\mathbf{K}}_0-\sum_{p=1}^P\bm\Phi_p\check{\mathbf{K}}_0\bm\Phi_p^\top-\sum_{p=1}^P\sum_{p'\neq p}^P\bm\Phi_p\check{\mathbf{K}}_{|p-p'|}\bm\Phi_{p'}^\top$, where both $\check{\mathbf{K}}_0=\mathrm{E}(\check{\mathbf{s}}_t^{(r)}\check{\mathbf{s}}_t^{(r)\top})$ with $\mathrm{E}\{(\check{s}_t^{(r)})_q^m (\check{s}_t^{(r)})_{q'}^{m'}\}=\mathrm{E}\{(\check{s}_t^{(r)})_q^{-m} (\check{s}_t^{(r)})_{q'}^{-m'}\}=\check{k}_{qq'm}\delta_{m=m'}\delta_{m\ge0}$ and $\check{\mathbf{K}}_{|p-p'|}=\mathrm{E}(\check{\mathbf{s}}_{t-p}^{(r)}\check{\mathbf{s}}_{t-p'}^{(r)\top})$ have sparse structures under the assumption of axial symmetry.  

Furthermore, we assume that $\bm\Phi_p$ is a diagonal matrix with the $(q^2+q+m+1)$th diagonal element denoted as $(\phi_p)_q^m$, which neglects the cross dependence between $\{(\Tilde{s}_t^{(r)})_q^m\}_{q=0,\ldots,Q-1; m=-q,\ldots,q}$. This assumption enables the independent and parallel evaluation of $\{(\phi_p)_q^m\}_{p=1}^P$ across $q$ and $m$ and ensures $\mathbf{U}$ to inherit the same sparse structure from $\check{\mathbf{K}}_0$ and $\check{\mathbf{K}}_{|p-p'|}$. We briefly validated it in Section~S3.4 by testing the significance of the first temporal lag of the cross-correlation. The proportion of $p$-values less than 0.05 is nearly zero, indicating negligible cross-dependence. The details about the parameter estimation are given in Section~S3.5.

\subsection{Emulation with SG}
\label{sec:subsec:emulationswithSG}
The procedures of constructing an SG for the monthly and daily simulations from CESM2-LENS2 is summarized in Algorithm~S1 of Section~S3.6. We assume that the tuning parameters $K$, $Q_l$, $Q_o$, and $P$ are known. More details about their selection will be provided in Section~\ref{sec:casestudy}. In Stage~1, we calculate and store a total of $(5+2K)IJ$ parameters in the deterministic component. In Step 1) of Stage 2, although we use different $Q$ values for land and ocean data to evaluate $v(L_i,l_j)$, we still need coefficients up to $Q'=\max(Q_l,Q_o)$. The number of stored parameters in Stage 2 is $IJ+(4+P)Q'^2+(Q'^3/3+Q'^2/2+Q'/3)$. Therefore, the total number of parameters is of the order $O(IJ+Q'^3)$. For the annually aggregated temperature, there is no need to perform a TGH transformation in Step 3) of Stage 2. Consequently, $\mathcal{S}_{gh}$ is an empty set, and only $IJ+PQ'^2+(Q'^3/3+Q'^2/2+Q'/3)$ parameters are needed for the stochastic component. In summary, we develop an SG by training on simulations of size $IJRT$ and store it for generating unlimited emulations using at most $(6+2K)IJ+(4+P)Q'^2+(Q'^3/3+Q'^2/2+Q'/3)$ memorized parameters.

Here, we contrast the storage requirements of our SG with that of another efficient SG proposed by \citet{Huang'sEmulator} (referred to as HCBG-SG) for surface temperature simulations from LENS1 \citep{kay2015community}. HCBG-SG was primarily developed within the spatial domain and required memorization of $(2K+P_H+8)IJ+6I+2$ parameters. This consisted of $(5+2K)IJ$ parameters for the deterministic component, $P_H IJ$ parameters for evaluating the temporal dependence with $P_H$ representing the selected order for the autoregressive model, $3IJ$ parameters for TGH transformation, and $6I+2$ parameters for assessing the non-stationary spatial dependence. The storage advantage of our SG stems from its modeling in the spectral domain rather than the spatial domain, with $Q'^2\ll IJ$. Taking the case of $P=P_H=1$ and $Q'=70$ as an example, which maximizes storage requirements of our SG for all temporal resolutions, the daily HCBG-SG necessitates additional $25,747$ parameters. However, for annual simulations, which do not need a TGH, HCBG-SG needs $120,541$ fewer parameters. A further comparison of their performance is presented in Section~\ref{sec:casestudy}.  
% Hence, our SG offers a significant advantage in terms of parameter storage efficiency.

In Algorithm~\ref{alg:EmulationSG}, we outline the procedures for generating monthly and daily temperature emulations for CESM2-LENS2. Note that a Cholesky decomposition should be performed on $\mathbf{U}$ when generating $\check{\mathbf{s}}_t^{(r')}$, which may be time-consuming if $Q'$ is not small. However, the sparse structure of $\mathbf{U}$ avoids the possible computational limitation. For the annually aggregated temperature, we can simply remove Step (b) in Stage 2 and replace $\Tilde{\mathbf{s}}_{SG,t}^{(r')}$ in Step (c) with $\check{\mathbf{s}}_{SG,t}^{(r')}$.

\begin{algorithm}[!t]
\caption{Emulation with SG}
\label{alg:EmulationSG}
\begin{algorithmic}[3] 
\renewcommand{\algorithmicrequire}{ \textbf{Input:}} 
\REQUIRE $\rho(L_i,l_j)$, $\beta_0(L_i,l_j)$, $\beta_1(L_i,l_j)$, $\beta_2(L_i,l_j)$, $a_1(L_i,l_j)$,\ldots, $a_K(L_i,l_j)$, $b_1(L_i,l_j)$,\ldots, $b_K(L_i,l_j)$, $\sigma(L_i,l_j)$, $v(L_i,l_j)$, $\lambda_q^m$, $\omega_q^m$, $g_q^m$, $h_q^m$, $(\phi_1)_q^m,\ldots,(\phi_P)_q^m$, $i=1,\ldots,I$, $j=1,\ldots,J$, $q=0,\ldots,Q_o-1$, $m=-q,\ldots,q$, $\mathbf{U}$, $Q_l$, $Q_o$, $K$, $P$, $\mathbf{A}\in\mathbb{R}^{Q_o^2\times Q_o^2}$, $R'$.
\vspace{5pt}
\renewcommand{\algorithmicrequire}{ \textbf{Stage 1: Preliminary}} 
\REQUIRE \textbf{ }\\   
    \vspace{5pt}
    1) For each grid point $(L_i,l_j)$, calculate $\{m_t(L_i,l_j)\}_{t=1}^T$ with given parameters of the deterministic component, $i=1,\ldots,I$ and $j=1,\ldots,J$.\\
    \vspace{5pt}
    % \textcolor{orange}{Remove: ``2) Calculate $\check{\mathbf{K}}-\sum_{p=1}^P\bm\Phi_p\check{\mathbf{K}}\bm\Phi_p^\top-\sum_{p=1}^P\sum_{p'\neq p}\bm\Phi_p\mathrm{E}(\check{\mathbf{s}}_{t-p}^{(r)}\check{\mathbf{s}}_{t-p'}^{(r)\top})\bm\Phi_{p'}^\top$." as we give $\mathbf{U}$ directly.}\\
    % \vspace{5pt}
\renewcommand{\algorithmicrequire}{ \textbf{Stage 2: Emulation}} 
\REQUIRE \textbf{ }\\     
\vspace{5pt}
\hspace{-15pt}For $r'=1,\ldots,R'$:\\
 \vspace{5pt}
    1) generate $\check{\mathbf{s}}_{SG,-(p-1)}^{(r')}\sim\mathcal{N}_{Q_o^2}(\mathbf{0},\mathbf{U})$, $p=1,\ldots,P$,\\
    \vspace{5pt}
    2) for $t=1,\ldots,T$: \\
    \vspace{5pt}
    \hspace{15pt} (a) generate $\bm\xi_{SG,t}^{(r')}\sim\mathcal{N}_{Q_o^2}(\mathbf{0},\mathbf{U})$, and calculate  $\check{\mathbf{s}}_{SG,t}^{(r')}=\sum_{p=1}^P\Phi_p\check{\mathbf{s}}_{SG,t-p}^{(r')}+\bm\xi_{SG,t}^{(r')}$,\\
    \vspace{5pt}
    \hspace{15pt} (b) calculate $\tilde{\mathbf{s}}_{SG,t}^{(r')}$, where $(\tilde{s}_{SG,t}^{(r')})_q^m=\omega_q^m\tau_{g_q^m,h_q^m}\{(\check{s}_{SG,t}^{(r')})_q^m/\lambda_q^m\}$,\\
    \vspace{5pt}
    \hspace{15pt} (c) calculate $\mathbf{s}_{SG,t}^{(r')}=\mathbf{A}\tilde{\mathbf{s}}_{SG,t}^{(r')}$, and perform the inverse SHT on $\mathbf{s}_{SG,t}^{(r')}$ with \eqref{eq:SHTlo} to obtain $\{\mathrm{invSHT}(\mathbf{s}_{SG,t}^{(r')})(L_i,l_j)\}_{i=1,\ldots,I; j=1,\ldots,J}$, \\
    \vspace{5pt}
    \hspace{15pt} (d) generate $\varepsilon_{SG,t}^{(r')}(L_i,l_j)\sim\mathcal{N}(0,\hat v(L_i,l_j)^2)$, $i=1,\ldots,I;j=1,\ldots,J$;\\
    \vspace{5pt}
    \hspace{15pt} (e) generate emulations 
    \vspace{-5pt}
    \begin{equation*}
        y_{SG,t}^{(r')}(L_i,l_j)=m_t(L_i,l_j)+\sigma(L_i,l_j)\{\mathrm{invSHT}(\mathbf{s}_{SG,t}^{(r')})(L_i,l_j)+\varepsilon_{SG,t}^{(r')}(L_i,l_j)\}.
        \vspace{-5pt}
    \end{equation*}
\renewcommand{\algorithmicensure}{ \textbf{Output:}} 
\ENSURE \textbf{$\{y_{SG,t}^{(r')}(L_i,l_j)\}_{i=1,\ldots,I;j=1,\ldots,J;r'=1,\ldots,R'}$.}\\ 
\end{algorithmic}
\end{algorithm}

% \vspace{-10pt}

\section{Case Study: CESM2-LENS2}
\label{sec:casestudy}
In this section, we develop emulators and generate emulations for surface temperature simulations from CESM2-LENS2 using the proposed algorithms. We delve into the the implementation of our SG (referred to as SHT-SG in this section), covering the choice of tuning parameters and the modeling of the dependence structure, and the assessment of their performance. Additionally, we include the results of HCBG-SG for comparison. Discussion about the monthly SG is presented in Section~S4.2 of the Supplementary Materials to save space.

The assessment of an SG relies on the specific purpose for which the emulator is intended \citep{castruccio2014statistical}. Our main goal is to generate emulations that closely resemble the given simulations. To achieve this, our SG should efficiently decouple and capture the deterministic and stochastic components of the data. First, the estimated mean trend  $\hat m_t$ should mimic and extract the variation of the ensemble mean in the simulations. After adding the stochastic component, the variability of emulations should be similar to the internal variability observed in simulations. Therefore, we will discuss two key factors: goodness-of-fit and variability, quantified with 
\vspace{-10pt}
\footnotesize
\begin{equation*}
    \mathrm{I}_{\rm fit}(L_i,l_j)=\frac{\sum_{r=1}^R\sum_{t=1}^T\{y_t^{(r)}(L_i,l_j)-\hat m_t(L_i,l_j)\}^2}{R/(R-1)\sum_{r=1}^R\sum_{t=1}^T\{y_t^{(r)}(L_i,l_j)-\Bar{y}_t(L_i,l_j)\}^2} 
    % \quad \text{and} \quad \mathrm{I}_{\rm uq}(L_i,l_j)=\frac{\mathrm{CRA}\{y_{SG,t}^{(1)}(L_i,l_j),\ldots,y_{SG,t}^{(R)}(L_i,l_j)\}}{\mathrm{CRA}\{y_t^{(1)}(L_i,l_j),\ldots, y_t^{(R)}(L_i,l_j)\}},
\end{equation*} 
\normalsize
and 
\vspace{-15pt}
\footnotesize
\begin{equation*}
    \mathrm{I}_{\rm uq}(L_i,l_j)=\frac{\mathrm{CRA}\{y_{SG,t}^{(1)}(L_i,l_j),\ldots,y_{SG,t}^{(R)}(L_i,l_j)\}}{\mathrm{CRA}\{y_t^{(1)}(L_i,l_j),\ldots, y_t^{(R)}(L_i,l_j)\}},
\end{equation*}
\normalsize
respectively. Given a grid point $(L_i,l_j)$, the index $\mathrm{I}_{\rm fit}(L_i,l_j)$ compares $\hat m_t(L_i,l_j)$ with $2K+4$ parameters with the ensemble mean $\bar y_{t}(L_i,l_j)=R^{-1}\sum_{r=1}^R y_t^{(r)}(L_i,l_j)$ with $T$ parameters among all ensembles and time points. The closer the $\mathrm{I}_{\rm fit}$ value to $1$, the better the ability of the SG to capture the variability of the ensemble mean. Index $\mathrm{I}_{\rm uq}$, a measure proposed in \citet{Huang'sEmulator}, leverages the concept of functional data depths \citep{lopez2009concept}. In particular, we treat $\{y_{SG,t}^{(r')}(L_i,l_j)\}_{r'=1}^{R}$ as $R$ functions of $t$ at $(L_i,l_j)$ and measure the interquartile range \citep{sun2011functional} of these functions using the central region area (CRA). Then, $\mathrm{I}_{\rm uq}(L_i,l_j)$ is the uncertainty qualification of $\{y_{SG,t}^{(r')}(L_i,l_j)\}_{r'=1}^{R}$ against that of $\{y_t^{(r)}(L_i,l_j)\}_{r=1}^R$. A value of $\mathrm{I}_{\rm uq}$ close to $1$ suggests that the variability of the generated emulations closely reflects the internal variability of the training simulations. 

Moreover, we will visually and numerically compare various statistical characteristics of the emulations to those of simulations. This ensures that our emulations faithfully capture the essential features of the temperature data from CESM2-LENS2. For example, at each grid point, we compare the empirical distribution of $\{y_{SG,t}^{(r')}(L_i,l_j)\}_{t=1,\ldots,T; r'=1,\ldots,R}$ to that of $\{y_{t}^{(r)}(L_i,l_j)\}_{t=1,\ldots,T; r=1,\ldots,R}$ by the Wasserstein distance \citep{santambrogio2015optimal}, which is denoted as $\mathrm{WD}_{S}(L_i,l_j)$. Similarly, we calculate the Wasserstein distance between the empirical distribution of $\{y_{SG,t}^{(r')}(L_i,l_j)\}_{i=1,\ldots,I; j=1,\ldots,J; r'=1,\ldots,R}$ and $\{y_{t}^{(r)}(L_i,l_j)\}_{i=1,\ldots,I; j=1,\ldots,J; r=1,\ldots,R}$ at each time point $t$, which is denoted as $\mathrm{WD}_T(t)$.

\subsection{Annually aggregated temperature}
\label{sec:subsec:annual}
\begin{figure}[!b]
\centering
\subfigure[$\mathrm{I}_{\rm fit}$ versus $R$]{
\label{fig:subfig:IfitwithRs}
\includegraphics[scale=0.5]{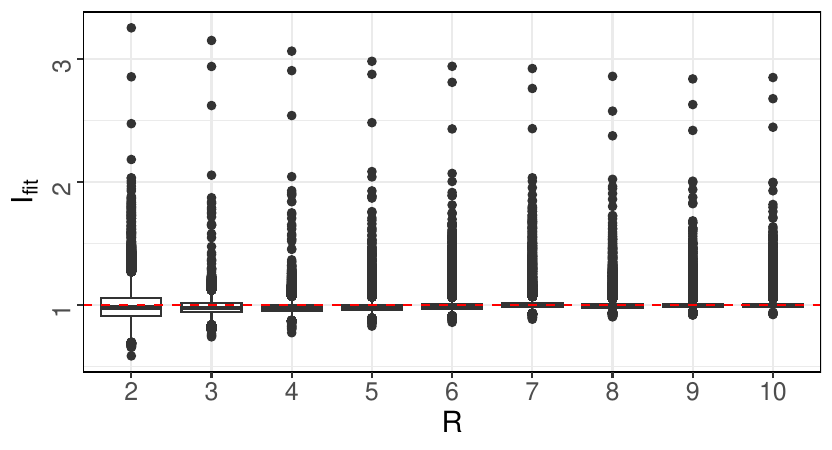}}
\subfigure[$\hat\sigma(L_i,l_j)$]{
\label{fig:subfig:sigma_demo}
\includegraphics[scale=0.5]{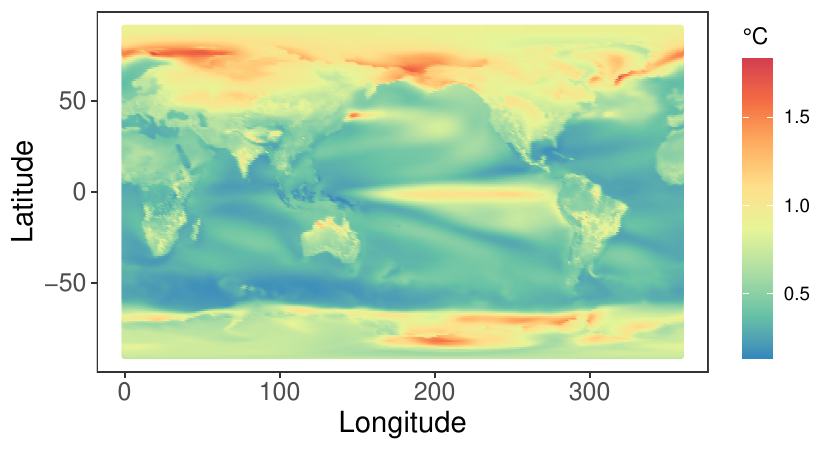}}\\
\vspace{-5pt}
\subfigure[Choice of $Q$]{
\label{fig:subfig:BIC_annual}
\includegraphics[scale=0.5]{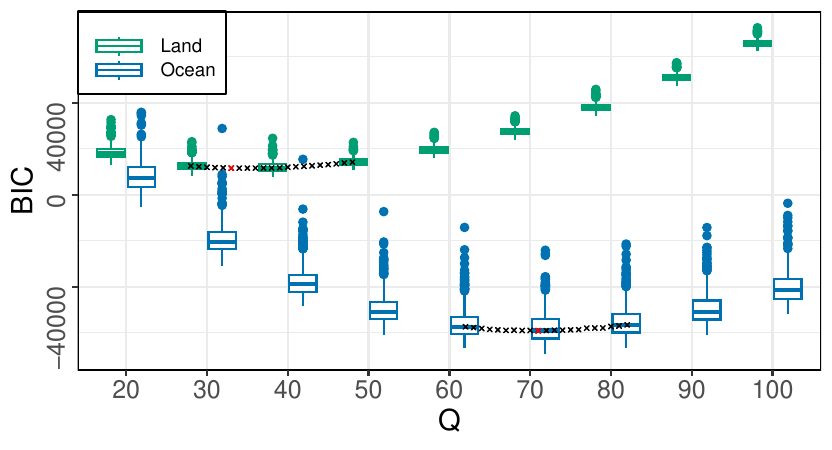}}
\subfigure[$\hat v(L_i,l_j)$]{
\label{fig:subfig:v2hat_annual}
\includegraphics[scale=0.5]{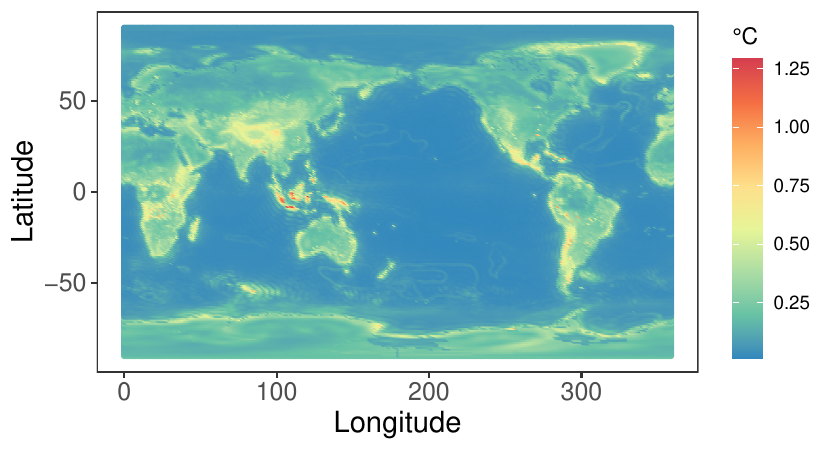}}\\
\vspace{-5pt}
\caption{Inference process for the annual SG. (a) shows boxplots of $\{\mathrm{I}_{\rm fit}(L_i,l_j)\}_{i=1,\ldots,I; j=1,\ldots,J}$ for different values of $R$. (b) is the map of the estimated standard deviation $\{\hat\sigma(L_i,l_j)\}_{i=1,\ldots,I; j=1,\ldots,J}$. (c) shows boxplots of $\{\mathrm{BIC}_{*,t}^{(r)}(Q)\}_{r=1,\ldots,R; t=1,\ldots,T}$ for different values of $Q$, where $*$ takes ``l" and ``o". Points $\times$ show medians of BIC values. The red points indicate the minimum values. (d) is the map of $\{\hat v(L_i,l_j)\}_{i=1,\ldots,I; j=1,\ldots,J}$.}
\label{Fig:annual_tuning}
\vspace{-10pt}
\end{figure}

First, we evaluate the deterministic component $m_t$ and $\sigma$ of the annually aggregated temperature simulations $\{y_t^{(r)}(L_i,l_j)\}_{i=1,\ldots,I; j=1,\ldots,J; t=1,\ldots,T; r=1,\ldots,R}$, with a total of $T=86$ time points.
% , which we denoted as  $\hat m_t$ and $\hat\sigma$. Ideally, we hope that $\hat m_t$ with $4IJ$ parameters could perform comparably to the ensemble mean $\bar y_{t}=R^{-1}\sum_{r=1}^R y_t^{(r)}$ with $TIJ$ parameters. As in \citet{castruccio2014statistical,castruccio2019reproducing}, we use 
% \vspace{-10pt}
% \begin{equation*}
%     \mathrm{I}_{\rm fit}(L_i,l_j)=\frac{\sum_{r=1}^R\sum_{t=1}^T\{y_t^{(r)}(L_i,l_j)-\hat m_t(L_i,l_j)\}^2}{R/(R-1)\sum_{r=1}^R\sum_{t=1}^T\{y_t^{(r)}(L_i,l_j)-\Bar{y}_t(L_i,l_j)\}^2}
%     \vspace{-10pt}
% \end{equation*}
% to assess the goodness-of-fit of the SG. The closer the $\mathrm{I}_{\rm fit}$ value to $1$, the better the ability of the SG to capture the variability of the ensemble mean. 
Fig.~\ref{fig:subfig:IfitwithRs} presents boxplots of  $\{\mathrm{I}_{\rm fit}(L_i,l_j)\}_{i=1,\ldots,I; j=1,\ldots,J}$ for different values of $R$. These values tend to get closer to $1$ as $R$ increases, and they stabilize after $R\geq7$, which means that using seven training simulations is sufficient to stabilize the inference. Therefore, we select $R=7$ for the subsequent analysis. The corresponding $\{\mathrm{I}_{\rm fit}(L_i,l_j)\}_{i=1,\ldots,I; j=1,\ldots,J}$ values are shown in Fig.~S5(a), with a median value of $1.001$. Temperature can be fitted well at most grid points. Furthermore, Fig.~\ref{fig:subfig:sigma_demo} displays the map of $\hat\sigma(L_i,l_j)$, illustrating the non-stationarity of the data. It is evident that larger variations of surface temperatures are observed over land.
% Note that $\hat\sigma(L_i,l_j)$ at the polar regions are even higher, since the very close physical distance between neighboring locations brings numerical instabilities.

\begin{figure}[!t]
\centering
\subfigure[$(\hat\phi_1)_q^m$]{
\label{fig:subfig:Phihat_annual}
\includegraphics[scale=0.45]{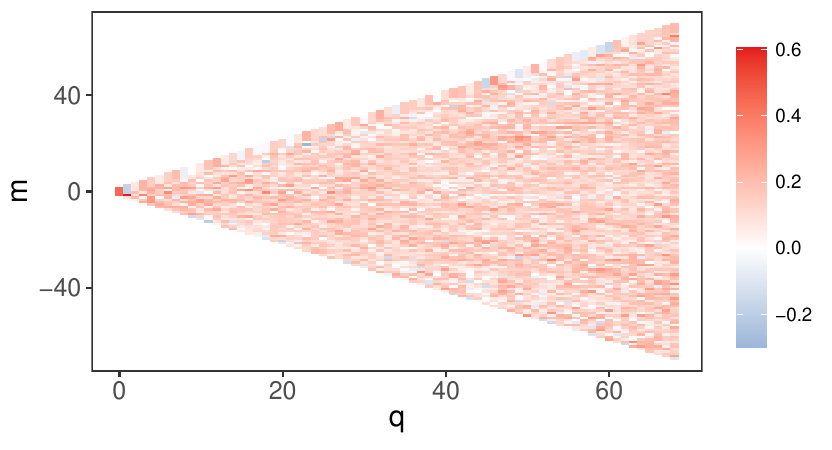}}
\subfigure[$\tilde{\mathbf{K}}_{\mathrm{emp}}$]{
\label{fig:subfig:TKsample}
\includegraphics[scale=0.45]{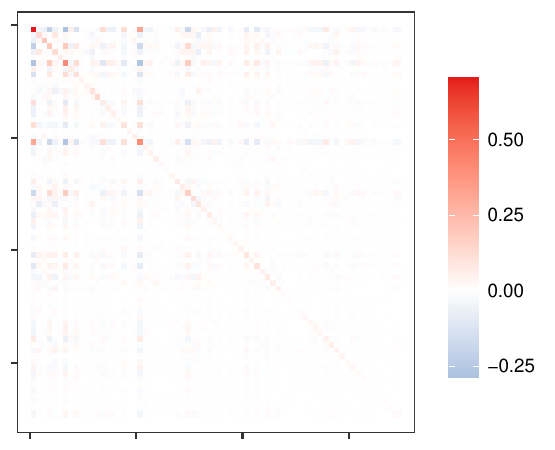}}
\subfigure[$\tilde{\mathbf{K}}_{\mathrm{axl}}$]{
\label{fig:subfig:TKaxial}
\includegraphics[scale=0.45]{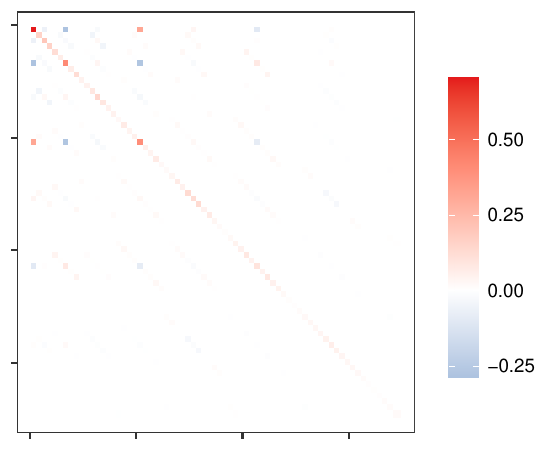}}\\
\vspace{-5pt}
\subfigure[Auto-covariance at $L_i=-11.8^\circ$]{
\label{fig:subfig:Autocov_118}
\includegraphics[scale=0.5]{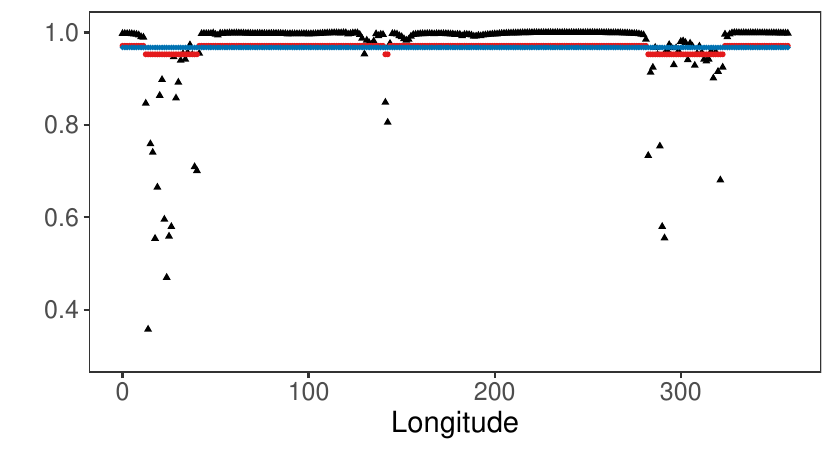}}
\hspace{10pt}
\subfigure[Auto-covariance at $L_i=36.3^\circ$]{
\label{fig:subfig:Autocov_363}
\includegraphics[scale=0.5]{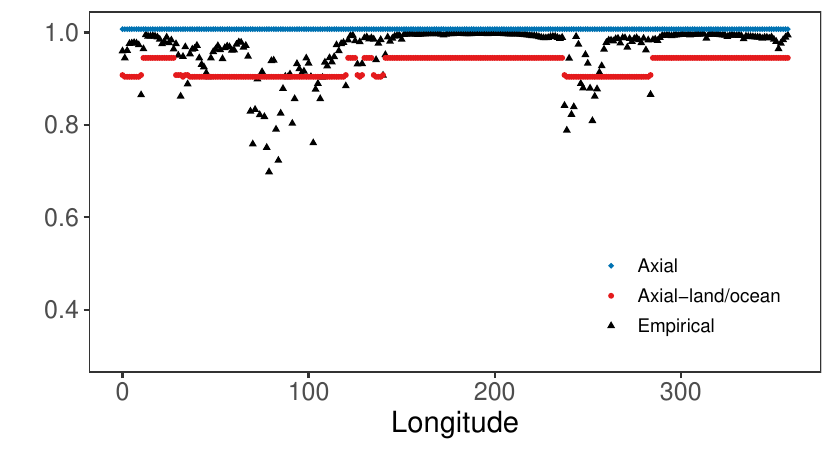}}
\vspace{-5pt}
\caption{Temporal and spatial dependence structure. (a) is the map of the estimated $\{(\phi_1)_q^m\}_{q=0,\ldots,Q_o-1; m=-q,\ldots,q}$ for the annual simulations. (b) is the first $69$ rows and columns of $\tilde{\mathbf{K}}_0$ empirically evaluated by $(RT)^{-1}\sum_{r=1}^R\sum_{t=1}^T\tilde{\mathbf{s}}_t^{(r)}\tilde{\mathbf{s}}_t^{(r)\top}$ and denoted as $\tilde{\mathbf{K}}_{\mathrm{emp}}$. (c) is the first $69$ rows and columns of $\tilde{\mathbf{K}}_0$ evaluated under the axial symmetric assumption and denoted as $\tilde{\mathbf{K}}_{\mathrm{axl}}$. (d) and (e) show empirical and fitted auto-covariance at latitudes $-11.8^\circ$ and $36.3^\circ$, respectively.}
\label{Fig:annual_dependence}
\vspace{-10pt}
\end{figure}

Then, we perform the SHT on the stochastic component, necessitating the selection of appropriate values for $Q_l$ and $Q_o$. Based on the Gaussian assumption of $\varepsilon_t^{(r)}(L_i,l_j)$, we propose a Bayesian information criterion (BIC) for $\{Z_t^{(r)}(L_i,l_j)\}_{i=1,\ldots,I; j=1,\ldots,J}$ at time $t$ of ensemble $r$:
\vspace{-10pt}
\begin{equation*}
    \mathrm{BIC}_{*,t}^{(r)}(Q)=\log(n_*)Q^2+n_*\log(2\pi)+\delta_{(L_i,l_j)\in\mathcal{S}_*}\sum_{i=1}^{I}\sum_{j=1}^J\log\{\hat v^2(L_i,l_j)\}+\delta_{(L_i,l_j)\in\mathcal{S}_*}\sum_{i=1}^I\sum_{j=1}^J\left\{\frac{\varepsilon_t^{(r)}(L_i,l_j)}{\hat v(L_i,l_j)}\right\}^2,
    \vspace{-10pt}
\end{equation*}
where $\hat v^2(L_i,l_j)=(RT)^{-1}\sum_{r=1}^R\sum_{t=1}^T\{\varepsilon_t^{(r)}(L_i,l_j)\}^2$. When $*$ takes ``o", the above BIC with $n_o=\sum_{i=1}^I\sum_{j=1}^J\delta_{(L_i,l_j)}\in\mathcal{S}_o$ helps select an appropriate $Q_o$ for $Z_t^{(r)}(L_i,l_j)$ over the ocean. Similarly, when $*$ is ``l", it leads to the determination of $Q_l$. Fig.~\ref{fig:subfig:BIC_annual} illustrates boxplots of $\{\mathrm{BIC}_{*,t}^{(r)}\}_{r=1,\ldots,R; t=1,\ldots,T}$ against different values of $Q$, revealing a substantial discrepancy in preferred $Q$ values between land and ocean. The stochastic components on land exhibit a preference for a smaller $Q$ value compared to those on the ocean. This might seem counterintuitive, given that Figs.~\ref{fig:subfig:SHTinv48} and \ref{fig:subfig:SHTinv96} suggest that the stochastic component on land requires more harmonic terms for a better approximation. However, it is essential to note that the introduction of $Q$ aims not only for accurate approximation but also to strike a balance between accuracy and the number of parameters, leading to improved model selection. Examining the boxplots for land, the improvement gained by including additional terms does not fully compensate for the associated drawbacks, such as increased storage requirements and accumulated errors from evaluation. The subsequent Figs.~\ref{fig:subfig:Autocov_118} and \ref{fig:subfig:Autocov_363} further validate this model selection result. Finally, the $\times$ points in Fig.~\ref{fig:subfig:BIC_annual} suggest selecting $Q_l=35$ and $Q_o=69$ to minimize the medians of BIC for land and ocean, respectively.

We further analyze the values of $\hat v^2(L_i,l_j)$, which serves as an assessment of the low-rank approximation on $(L_i,l_j)$. There are several regions on land that exhibit significantly larger $\hat v^2(L_i,l_j)$: 1) High-altitude regions such as the Himalayas and the Andes ranges with altitudes exceeding $3500$ meters above sea level; and  2) Regions characterized by diverse topography, such as the Indonesian archipelago, encompassing islands, mountains, and coastal areas. These geographic factors contribute to a more complex spatial and temporal structure in surface temperature in these regions. To capture more details, one may consider including additional covariates in the deterministic component and choosing specific values of $Q$ for these regions in the stochastic component. These considerations are avenues for future exploration.
% \textcolor{red}{We further analyze the values of $\hat v^2(L_i,l_j)$ on land. In Fig.~\ref{fig:subfig:v2hat_annual}, certain regions exhibit significantly larger $\hat v^2(L_i,l_j)$. For example, areas at the Himalayas and Andes ranges with altitudes exceeding $3500$ meters above sea level. Another distinct area with large $\hat v^2(L_i,l_j)$ is the Indonesian archipelago, characterized by diverse geography, including islands, mountains, and coastal areas. Various regions may undergo distinct temperature patterns owing to their geographical features, resulting in more complex temperature changes. Note that $\hat v^2(L_i,l_j)$ serves as an assessment of the low-rank approximation on $(L_i,l_j)$. To achieve a more accurate approximation, it may be worthwhile to consider specific values of $Q$ for high-altitude regions. The impact of altitude and the complex geographical structure should be further examined through the analysis of emulations.}    

Next, we assess the temporal dependence by fitting $\{(\tilde s_t^{(r)})_q^m\}_{t=1,\ldots,T; r=1,\ldots,R}$ with the VAR($P$) model. For each $(q,m)$, we suggest to choose $P$ by minimizing $\mathrm{BIC}_q^m(P)=P\log\{(T-P)R\}+R(T-P)\{\log(2\pi)+1\}+R(T-P)\log((\hat u_q^m)^2)$,
% \begin{equation}
%     \mathrm{BIC}_q^m(P)=P\log\{(T-P)R\}+R(T-P)\log(2\pi)+R(T-P)\log((\hat u_q^m)^2),
%     \label{eq:bicp}
% \end{equation}
where $(\hat u_q^m)^2=R^{-1}(T-P)^{-1}\sum_{r=1}^R\sum_{t=P+1}^T\{(\tilde s_t^{(r)})_q^m-\sum_{p=1}^P(\hat\phi_p)_q^m(\tilde s_{t-p}^{(r)})_q^m\}^2$. For briefness, we use the conditional log-likelihood (given $(\tilde s_1^{(r)})_q^m,\ldots,(\tilde s_P^{(r)})_q^m$) rather than an exact log-likelihood in the above BIC. The proportions of $P=1,\ldots,5$ are $99.3\%$, $0.2\%$, $0.2\%$, $0.1\%$, and $0.2\%$, respectively. Therefore, we opt for  $P=1$ for the annual case. Fig.~\ref{fig:subfig:Phihat_annual} further demonstrates the estimates of $\{(\phi_1)_q^m\}_{q=0,\ldots,Q_o-1; m=-q,\ldots,q}$.

Finally, we evaluate the spatial dependence by examining the matrix $\tilde{\mathbf{K}}_0$. Figs.~\ref{fig:subfig:TKsample} and \ref{fig:subfig:TKaxial} show top-left corners of two evaluations of $\tilde{\mathbf{K}}_0$. $\Tilde{\mathbf{K}}_{\mathrm{emp}}$ is the sample covariance matrix of $\tilde{\mathbf{s}}_t^{(r)}$, forming a dense $Q_o^2\times Q_o^2$ matrix. In contrast, $\Tilde{\mathbf{K}}_{\rm axl}$ is a sparse matrix. The sparsity of $\Tilde{\mathbf{K}}_{\rm axl}$ would be inherited by $\mathbf{U}$, facilitating storage and enabling fast Cholesky decomposition using efficient algorithms \citep{furrer2010spam}. Figs.~\ref{fig:subfig:Autocov_118} and \ref{fig:subfig:Autocov_363} display empirical and fitted covariances for $Z(L_i,l_j)$ and $Z(L_i,l_{j+1})$, specifically the auto-covariance $\{\mathrm{cov}\{Z(L_i,l_j),Z(L_i,l_{j+1})\}\}_{j=1}^{J-1}$ at $L_i=-11.8^\circ$ and $36.3^\circ$. These plots highlight the distinction between the proposed separate SHT for land and ocean \eqref{eq:SHTlo} (Axial-land/ocean) and the original SHT \eqref{eq:SHT} (Axial). The empirical auto-covariance over the ocean appears almost flat, supporting the assumption of axial symmetry over the ocean. In contrast, the auto-covariance over land significantly differs from that over the ocean. This difference is not captured by the fitted covariance of Axial, which selects $Q=77$ via BIC. Axial-land/ocean fits the covariance separately for land and ocean, resulting in segmented lines in Figs.~\ref{fig:subfig:Autocov_118} and \ref{fig:subfig:Autocov_363}. However, land temperature exhibits a complex and non-stationary dependence structure, with covariance weakening as one moves farther inland. This intricacy warrants further investigation in future research.

\begin{figure}[!b]
\centering
\subfigure[$\mathrm{I}_{\rm uq}(L_i,l_j)$ of SHT-SG]{
\label{fig:subfig:Iuq_annual}
\includegraphics[scale=0.5]{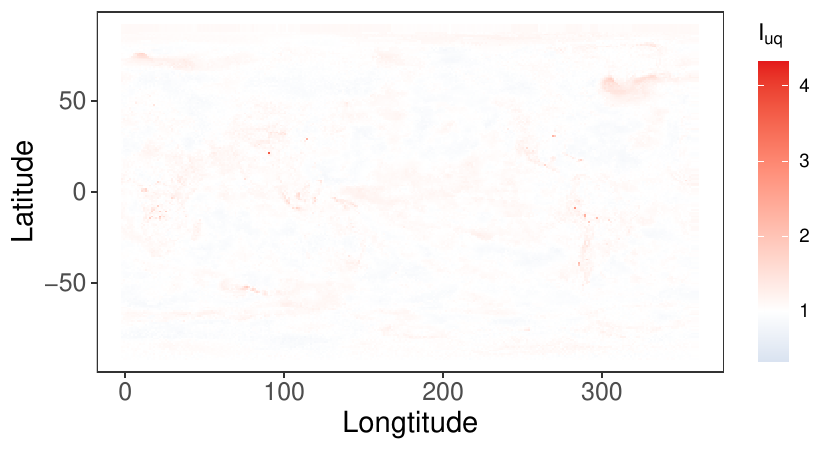}}
\subfigure[$\mathrm{I}_{\rm uq}(L_i,l_j)$ of HCBG-SG]{
\label{fig:subfig:Iuq_annual_Huang}
\includegraphics[scale=0.5]{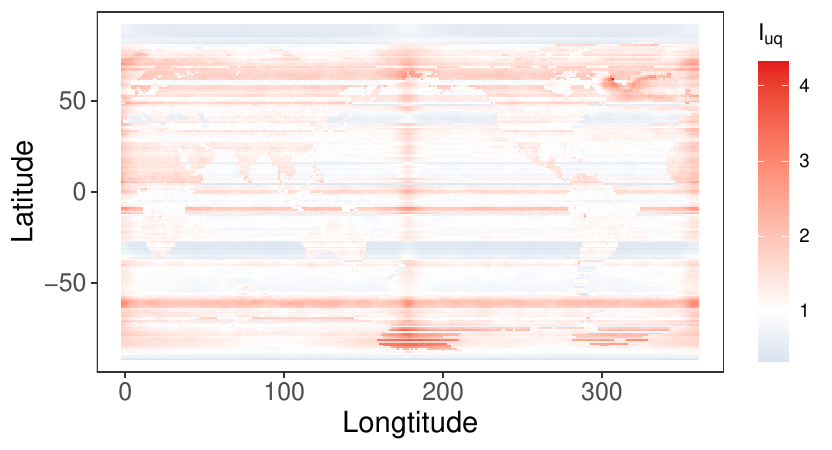}}\\
\subfigure[$\mathrm{WD}_{S}(L_i,l_j)$ of SHT-SG]{
\label{fig:subfig:WDs}
\includegraphics[scale=0.5]{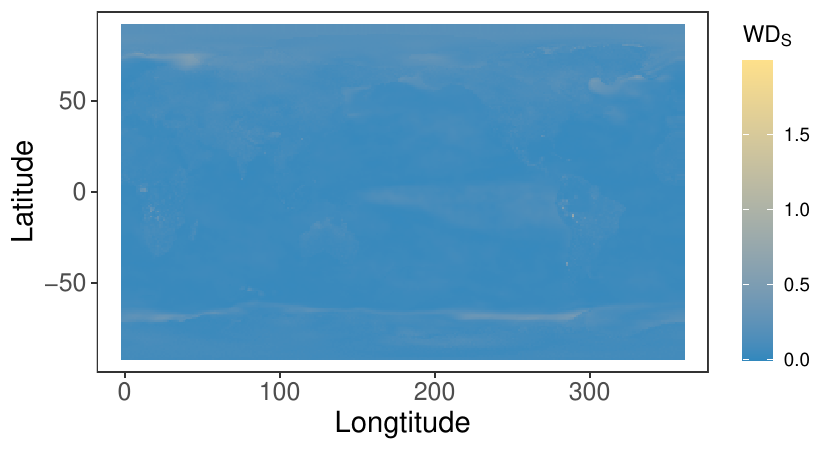}}
\subfigure[$\mathrm{WD}_{S}(L_i,l_j)$ of HCBG-SG]{
\label{fig:subfig:WDs_Huang}
\includegraphics[scale=0.5]{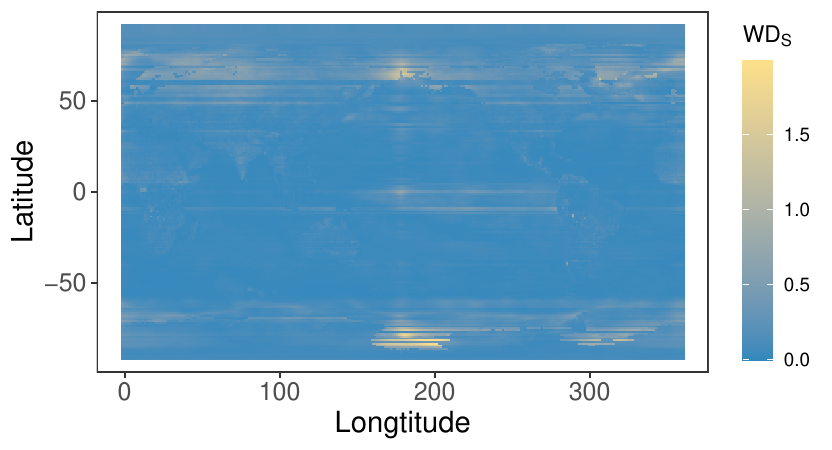}}
\vspace{-5pt}
\caption{(a) and (b) are maps of $\{\mathrm{I}_{\rm uq}(L_i,l_j)\}_{i=1,\ldots,I; j=1,\ldots,J}$ for the annual temperature emulations. (c) and (d) are maps of $\{\mathrm{WD}_{S}(L_i,l_j)\}_{i=1,\ldots,I; j=1,\ldots,J}$ for the annual temperature emulations. }
\label{Fig:fitanduq}
\vspace{-10pt}
\end{figure}

Inputting the above-estimated parameters, we generate $R'=R=7$ emulations with Algorithm~\ref{alg:EmulationSG}. Each emulation takes approximately $30$ seconds to produce without parallel computation. Additionally, we include emulations generated by HCBG-SG for comparison, with detailed implementation procedures shown in the Supplementary Materials. Figs.~\ref{fig:subfig:Iuq_annual} and \ref{fig:subfig:Iuq_annual_Huang} compare $\mathrm{I}_{\rm uq}$ values for the proposed SHT-SG and HCBG-SG, with median values of $1.013$ and $1.085$, respectively. Our emulations can successfully capture the variability of simulations at most grid locations, including polar regions. A high-$\mathrm{I}_{\rm uq}$ region near the location $(50.00,300.00)$ is caused by the poor evaluation of the mean trend, which is discussed in Section~S4.1.1 of the Supplementary Materials. Fig.~\ref{fig:subfig:Iuq_annual_Huang} shows a poor spatial continuity for HCBG-SG, which may be caused by the independent evaluation of parameters at each latitude and a relatively weak model of dependence across latitudes. A similar pattern can be seen in Fig.~\ref{fig:subfig:WDs_Huang}, which depicts the map of $\mathrm{WD}_{S}$ for HCBG-SG, with a median value of $0.076$. The $\mathrm{WD}_{S}$ values for SHT-SG in Fig.~\ref{fig:subfig:WDs}, which are closer to zero with a median value of $0.062$, provide further evidence of the similarity between our emulations and the simulations. From Fig.~\ref{Fig:fitanduq}, geographic factors have no obvious impact on both two SGs regarding $\mathrm{I}_{\rm uq}$ and $\mathrm{WD}_S$ values. The $\mathrm{WD}_T$ values for SHT-SG and HCBG-SG are comparable, with medians $0.126$ and $0.108$, respectively. In the Supplementary Materials, we further compare other basic statistical characteristics of our emulations to those of the simulations by replicating the procedures outlined in Fig.~\ref{Fig:Demo}.

\subsection{Daily aggregated temperature}
\label{sec:subsec:daily}
Now, we develop an SG for the daily aggregated temperature simulations with $T=31,390$. Analyzing such a vast amount of data presents a computational challenge. Therefore, we adopt a strategy similar to \citet{Huang'sEmulator} to perform inference only on data for the years 2020, 2040, 2060, 2080, and 2100. The procedure of inference closely follows that of the monthly case. Therefore, we put the details of development to Section~S4.3 of the Supplementary Materials and focus on assessing the performance of the emulations and a further exploration of data characteristics.

\begin{figure}[!b]
\centering
\subfigure[$\mathrm{I}_{\rm uq}(L_i,l_j)$ for SHT-SG(with TGH)]{
\label{fig:subfig:Iuq_Tukey_daily}
\includegraphics[scale=0.47]{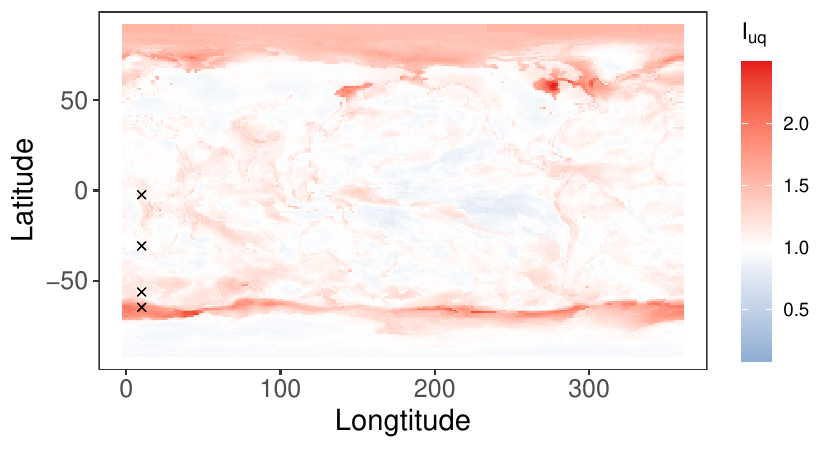}}
\hspace{5pt}
\subfigure[$\mathrm{I}_{\rm uq}(L_i,l_j)$ for HCBG-SG]{
\label{fig:subfig:Iuq_Huang_daily}
\includegraphics[scale=0.47]{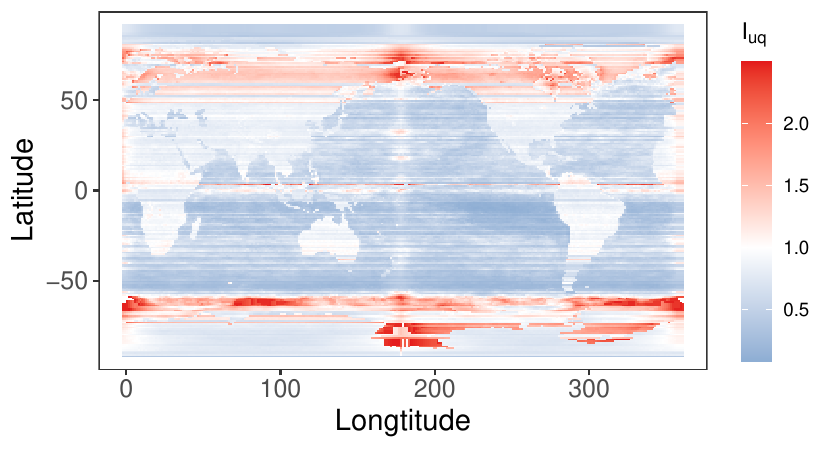}}\\
\subfigure[$\mathrm{WD}_{S}(L_i,l_j)$ for SHT-SG(with TGH)]{
\label{fig:subfig:WDtime_Tukey_daily}
\includegraphics[scale=0.47]{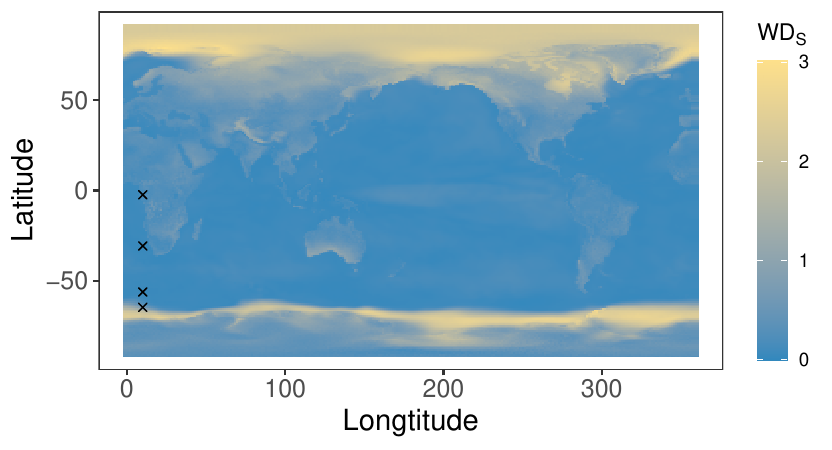}}
\hspace{5pt}
\subfigure[$\mathrm{WD}_{S}(L_i,l_j)$ for HCBG-SG]{
\label{fig:subfig:WD_Huang_daily}
\includegraphics[scale=0.47]{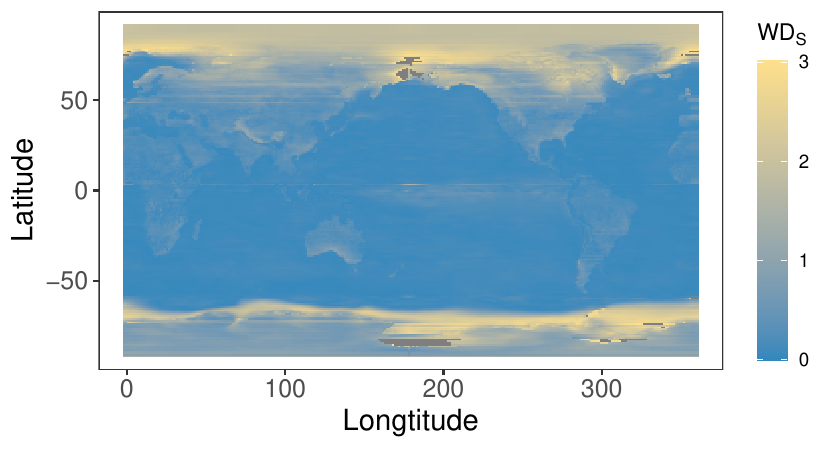}}\\
\subfigure[$\mathrm{I}_{\rm uq}(L_i,l_j)$ for SHT-SG]{
\label{fig:subfig:Box_Iuq_daily}
\includegraphics[scale=0.47]{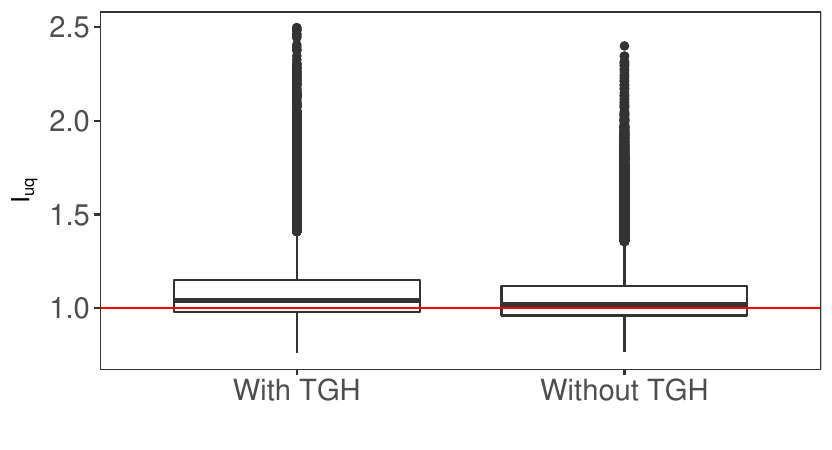}}
\hspace{5pt}
\subfigure[$\mathrm{WD}_{S}(L_i,l_j)$ for SHT-SG]{
\label{fig:subfig:Box_WD_daily}
\includegraphics[scale=0.47]{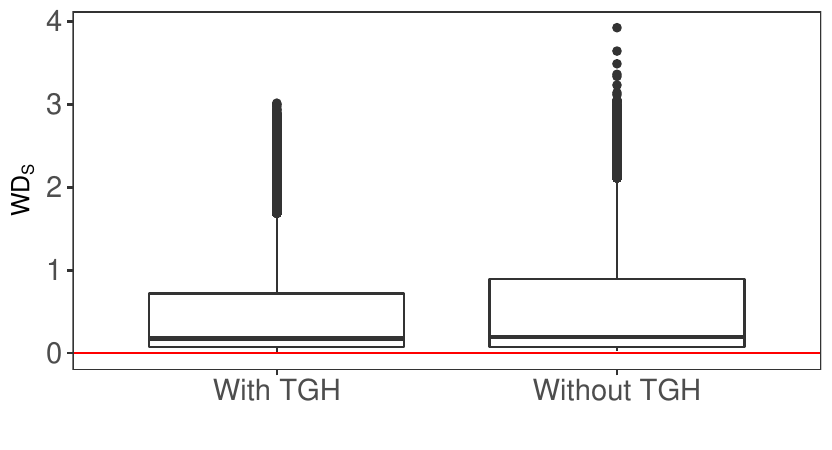}}\\
\vspace{-5pt}
\caption{Performance assessment of $R'=7$ daily emulations generated by various methods. (a) and (b) are maps of $\{\mathrm{I}_{\rm uq}(L_i,l_j)\}_{i=1,\ldots,I; j=1,\ldots,J}$ for the SHT-SG(with TGH) and HCBG-SG, respectively. (c) and (d) are maps of $\{\mathrm{WD}_{S}(L_i,l_j)\}_{i=1,\ldots,I; j=1,\ldots,J}$ for SHT-SG(with TGH) and HCBG-SG, respectively. Points $\times$ represent four testing grid points TGL=$(-2.36,10.00)$, TGO=$(-30.63,10.00)$, TGN=$(-56.07,10.00)$, and TGB=$(-64.55,10.00)$. (e) and (f) are boxplots of $\{\mathrm{I}_{\rm uq}(L_i,l_j)\}_{i=1,\ldots,I; j=1,\ldots,J}$ and $\{\mathrm{WD}_{S}(L_i,l_j)\}_{i=1,\ldots,I; j=1,\ldots,J}$ for comparing the SHT-SG(with TGH) and SHT-SG(without TGH).}
\label{Fig:daily_perform}
\vspace{-10pt}
\end{figure}

For the daily data, we choose $K_D=4$, $Q_l=36$, $Q_o=68$, and $P=1$. The choice of $K_D>K_M=3$ reveals the presence of more complex fluctuations in daily time series, requiring harmonic terms with higher frequencies in the model. The more structured and higher $(\hat\phi_1)_q^m$ in  Fig.~S14(f) indicates a stronger dependence among time points. The performance of our SG, both with and without the use of TGH (referred to as SHT-SG(with TGH) and SHT-SG(without TGH), respectively), along with HCBG-SG, is illustrated in Fig.~\ref{Fig:daily_perform}. For better demonstration, indices larger than maximum values in Figs.~\ref{fig:subfig:Iuq_Tukey_daily} and \ref{fig:subfig:WDtime_Tukey_daily} are excluded from Figs.~\ref{fig:subfig:Iuq_Huang_daily} and \ref{fig:subfig:WD_Huang_daily}, respectively. Comparing emulations generated by HCBG-SG with those of SHT-SG(with TGH), the latter illustrates a closer resemblance to simulations, greater robustness in numerically unstable regions, and better spatial continuity. From Fig.~\ref{fig:subfig:Box_WD_daily}, the incorporation of TGH in our SG results in an improvement in $\mathrm{WD}_S$. The empirical distribution of emulations from SHT-SG(with TGH) is closer to that of simulations, aligning with the correct assumption. This advantage may not be apparent in Fig.~\ref{fig:subfig:Box_Iuq_daily}, as $\mathrm{I}_{\rm uq}$ is used mainly for assessing the variabilities of emulations.

As in the monthly case, the daily emulations exhibit challenges in accurately representing certain regions, notably the North Pole and the Band regions. These regions display larger $\mathrm{I}_{\rm uq}$ values, indicating higher uncertainty. Additionally, they may deviate from the behavior observed in the simulations, as reflected by the larger $\mathrm{WD}_S$ values. Moreover, $\mathrm{WD}_S$ values are generally higher over land compared to ocean areas, which may be attributed to the complex spatial and temporal structures and larger variation of the temperature on land.

\begin{figure}[!b]
\centering
% \subfigure[$\mathrm{I}_{\rm uq}(L_i,l_j)$]{
% \label{fig:subfig:Iuq_Tukey_daily}
% \includegraphics[scale=0.5]{Fig_LENS2/Iuq_Tukey_daily.pdf}}
% \subfigure[$\mathrm{WD}_{S}(L_i,l_j)$]{
% \label{fig:subfig:WDtime_Tukey_daily}
% \includegraphics[scale=0.5]{Fig_LENS2/WDtime_Tukey_daily.pdf}}\\
\subfigure[TGL]{
\label{fig:subfig:Demo_GL}
\includegraphics[scale=0.45]{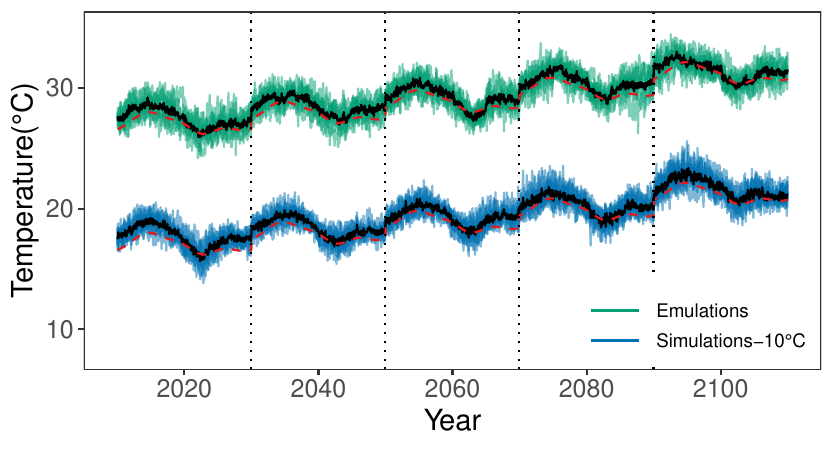}}
\subfigure[TGO]{
\label{fig:subfig:Demo_GO}
\includegraphics[scale=0.45]{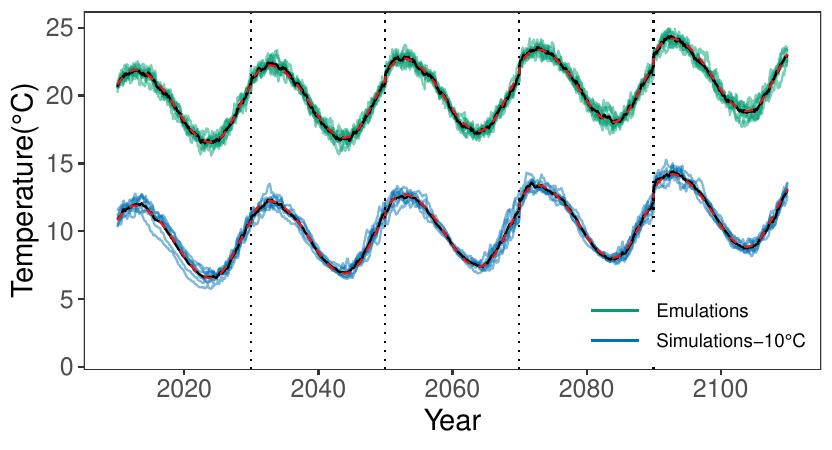}}\\
\vspace{-5pt}
\subfigure[TGN]{
\label{fig:subfig:Demo_GN}
\includegraphics[scale=0.45]{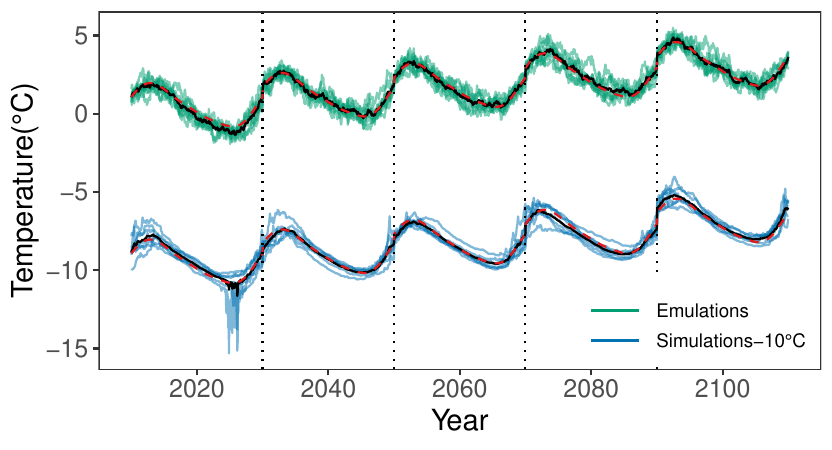}}
\subfigure[TGB]{
\label{fig:subfig:Demo_GB}
\includegraphics[scale=0.45]{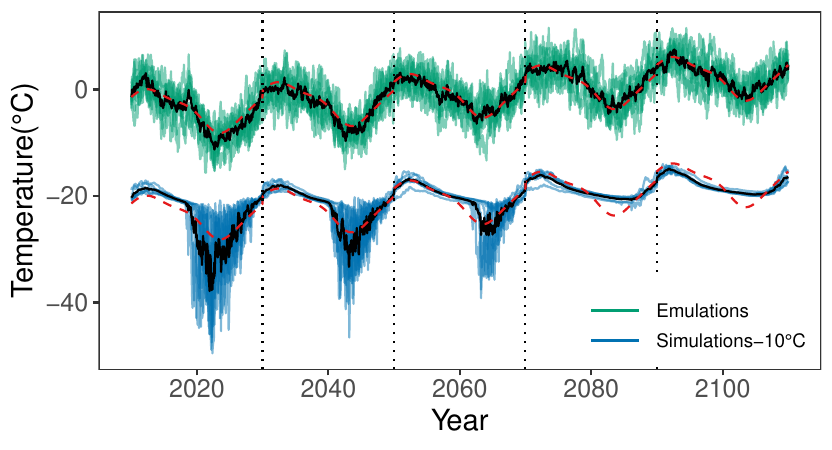}}\\
\vspace{-5pt}
\caption{Comparison of daily emulations and simulations at four testing grid points. The black solid curves are ensemble means for simulations and emulations. The red dashed curves are $\hat m_t$s.}
\label{Fig:abnormal}
\vspace{-10pt}
\end{figure}

We further explore these by intuitively displaying and comparing simulations and emulations at four testing grid points selected from the southern hemisphere with the same longitude in Fig.~\ref{Fig:abnormal}. TGL is a land point near the equator. In Fig.~\ref{fig:subfig:Demo_GL}, temperature simulations on TGL exhibit larger variability but can be well replicated by the emulations. TGO is an ocean point in the middle latitudes. From Fig.~\ref{fig:subfig:Demo_GO}, temperature simulations on TGO have smaller variability overall, with relatively larger variability at peaks and valleys of each year. These patterns can also be observed at emulations on TGO. TGN is also an ocean point in the middle latitudes, but very close to the Band region. In Fig.~\ref{fig:subfig:Demo_GN} (and Fig.~S14(a)), some temperature simulations exhibit sudden drops at a few days in year 2020, which leads to an overestimate of $\hat\sigma(\rm TGN)$, and hence a relatively higher variability in the emulations. Despite this, with $\mathrm{I}_{\rm uq}(\rm TGN)=1.200$ and $\mathrm{WD}_{S}(\rm TGN)=0.078$, the abnormal drops do not significantly affect these indices. Both $\mathrm{I}_{\rm uq}$ and $\mathrm{WD}_{S}$ exhibit a certain tolerance to isolated ``outliers". TGB is an ocean point within the Band region. Compared with the scenarios in Fig.~\ref{fig:subfig:Demo_GN}, temperature simulations at TGB display much more rapid and significant fluctuations at lots of time slots in years 2020, 2040, and 2060. In contrast, in years 2080 and 2100, the temperature simulations behave as other grid points on ocean with small variability. As shown in Fig.~S14(b), simulations at the North Pole region exhibit similar patterns, i.e., surface temperature experiences sharp fluctuations on numerous days. According to the explanation kindly provided by NCAR, these temperature fluctuations are caused by the way of data generation.  Specifically, the surface temperature in simulations is the spatial average of the surface temperature of whatever medium is in the grid box, whether it is water, land, or sea ice. Grid points at Band and North Pole regions undergo a dynamic transition, shifting back and forth between being partial sea ice and being open ocean. Consequently, sometimes we get the temperature of sea ice, and sometimes we get the temperature of the top layer of the ocean (or a mixture of the two). Combining this with Fig.~S14, the numerical instabilities exhibit four features: they do 1) not occur in each ensemble; 2) not happen during a fixed period; 3) not vary within a similar extent; and 4) not exhibit a consistent pattern across different grid points. Therefore, although the high $\mathrm{I}_{\rm uq}$ and $\mathrm{WD}_S$ values at the North Pole and Band regions directly stem from the inadequate assumption of a fixed $\hat \sigma$ over time, developing an SG that can accurately model the temperature fluctuations in these two regions is a challenge.

% \textcolor{orange}{(Even if they could, they would not mimic a real ESM for the surface temperature.)}
\vspace{-10pt}
\section{Conclusion and Future Work}
\label{sec:discussion}
In this paper, we presented an efficient SG for global temperature simulations from the newly published CESM2-LENS2. With at most $O(IJ+Q_o^3)$ parameters to be stored, an unlimited number of emulations of size $O(IJT)$, even at a daily scale, can be generated to help with the investigation of climate internal variability. Such a saving of computational time and resources comes from the use of SHT, which expands data with $Q_o^2\ll IJ$ spherical harmonics and represents a practical low-rank approximation on the sphere. By customizing $Q$ values for grid points on land and ocean and leveraging the axial symmetry, the proposed SG can properly capture the complex non-stationary dependencies among different latitudes and land/ocean regions. To account for the non-Gaussian nature of the high-resolution time series, we introduced a modified TGH transformation into the SG. In our case study, we developed SGs based on $R=7$ annually, monthly and daily aggregated simulations, respectively. We evaluated the efficiency and accuracy of the proposed SG by comparing the generated emulations to the original simulations and emulations generated by \citet{Huang'sEmulator} using various indexes and visual inspections. 

While the proposed SGs have demonstrated commendable performance, there remains room for further enhancements: 1) Including altitude factors in both the deterministic and stochastic components. The influence of altitude has been evident in the inference process and the performance of emulations. Surface temperatures are sensitive to altitude \citep{castruccio2016compressing}; 2) Improving the efficiency of SG on land, particularly for monthly and daily simulations. The larger variability observed in these cases suggests the need for more suitable models and additional factors to account for the complicated spatial and temporal structures on land; 3) Replacing the constant standard error with a temporal-varying one, denoted as $\sigma_t$. For simulations with higher temporal resolution, the assumption of a constant $\sigma$ leads to an overestimate of standard error, further amplifying the variability of emulations, especially on land; and 4) Extending the current work to include hourly simulations. Hourly data may exhibit more intriguing characteristics, and exploring their emulation presents an opportunity for further investigation and understanding.

% The daily SG has suboptimal performance in regions such as the North Pole and Band regions. These are primarily attributed to the violation of the assumption of a fixed $\sigma$ for all time points in these regions. To alleviate this limitation, introducing a function $\sigma_t$ that varies temporally could provide a more accurate representation of temperature variations; 3) An SG for simulations with higher spatial and temperature . A natural extension is to apply the proposed algorithms to hourly temperature simulations, which may exhibit even more pronounced skewness, kurtosis, and other intriguing characteristics. 4) More accurate model on land.

Beyond surface temperature, the idea of constructing an SG with SHT can be extended to various climate variables such as wind speed and atmospheric gas concentrations. After the removal of the deterministic component, the stochastic component can be efficiently expanded using spherical harmonics. The model in the spectral domain is contingent on the specific nature of the variable, which may ask for including other influential factors. Moreover, for variables that may exhibit interdependence, such as temperature and precipitation, exploring methods for their joint modeling in the spectral domain is also of interest. Investigating regional climate change, especially on land, is crucial for several reasons. On the one hand, climate change does not affect all regions equally. On the other hand, different regions have varying levels of vulnerability to climate change, e.g., extreme weather events, sea-level rise, or precipitation patterns. Local communities, governments, and businesses may be interested in how specific areas are being impacted so that they can adapt and plan for the future. Therefore, another possible extension is to build an SG for high-resolution climate simulations on a specific region.
\vskip -40pt
\begin{center}
{\bf Supplementary Materials}
\end{center}
\vskip -10pt
Additional information about the CESM2-LENS2 data, inference process, derivation, validation, and results for the case studies are contained in the online supplementary materials. (.pdf file)

The R code and instructions for downloading, processing the data, and reproducing the results in the article are available at the GitHub repository:\\
\href{https://github.com/SpatialTemporalStats/LENS2_Emulator_Reproducibility_Materials}{https://github.com/SpatialTemporalStats/LENS2\underline{ }Emulator\underline{ }Reproducibility\underline{ }Materials}.
\vskip -40pt
\begin{center}
{\bf Disclosure Statement}
\end{center}
\vskip -10pt
The authors report there are no competing interests to declare.

\vspace{-0.5cm}

\baselineskip 25.7 pt

\bibliographystyle{asa}
%\citationstyle{asa}
\bibliography{Ref}

\newpage
\smallskip
\begin{center}
{\large\bf Supplementary Materials for ``Efficient stochastic generators with spherical harmonic transformation for high-resolution global climate simulations from CESM2-LENS2"}
\end{center}

\setcounter{lemma}{0}  
\renewcommand{\thelemma}{S\arabic{lemma}}
\setcounter{section}{0}  
\renewcommand{\thesection}{S\arabic{section}}
\setcounter{algorithm}{0}  
\renewcommand{\thealgorithm}{S\arabic{algorithm}}
\setcounter{equation}{0} 
\renewcommand{\theequation}{S\arabic{equation}}
\setcounter{figure}{0} 
\renewcommand{\thefigure}{S\arabic{figure}}
\setcounter{page}{1} 
\renewcommand{\thepage}{S\arabic{page}}

\section{Introduction}
This document supplements the main manuscript, providing more details about the CESM2-LENS2 data, inference process, derivation, validation, and results in case studies. 

\section{Supplement to Section~2}
\label{sec:subsec:supplement_datadescription}
\begin{figure}[!h]
\centering
\subfigure[Skewness of annual data]{
\label{fig:subfig:Skew_annual}
\includegraphics[scale=0.4]{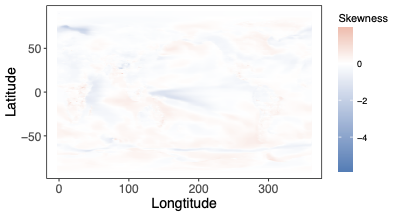}}
\subfigure[Skewness of monthly data]{
\label{fig:subfig:Skew_monthly}
\includegraphics[scale=0.4]{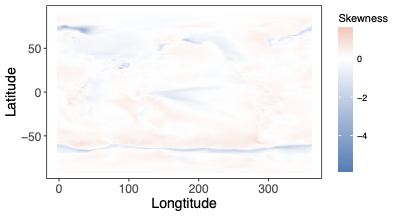}}
\subfigure[Skewness of daily data]{
\label{fig:subfig:Skew_daily}
\includegraphics[scale=0.4]{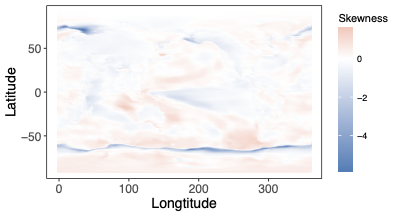}}\\
\subfigure[Kurtosis of annual data]{
\label{fig:subfig:Kurt_annual}
\includegraphics[scale=0.4]{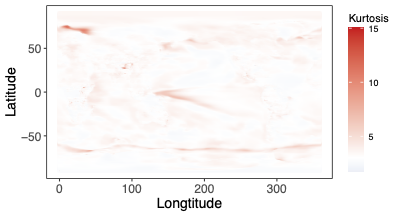}}
\subfigure[Kurtosis of monthly data]{
\label{fig:subfig:Kurt_monthly}
\includegraphics[scale=0.4]{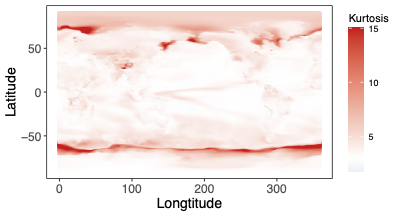}}
\subfigure[Kurtosis of daily data]{
\label{fig:subfig:Kurt_daily}
\includegraphics[scale=0.4]{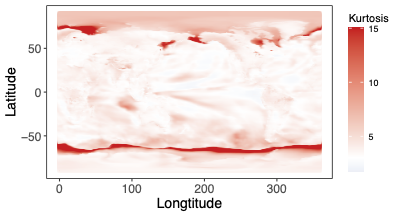}}\\
\caption{Skewness and kurtosis of annually, monthly, and daily aggregated detrended temperature simulations over the globe.}
\label{Fig:SkewandKurt}
\end{figure}

We further explore the characteristics of simulations with different temporal resolutions. Fig.~\ref{Fig:SkewandKurt} illustrates their skewness and kurtosis after detrending by ensemble means. For daily data, only temperature at years 2020, 2040, 2060, 2080, and 2100 are used. Kurtosis of monthly and daily data larger than $15$ are displayed as $15$ in Figs.~S\ref{fig:subfig:Kurt_monthly} and~S\ref{fig:subfig:Kurt_daily} for better illustration. With more prominently colored regions transitioning from  Fig.~S\ref{fig:subfig:Skew_annual} (and S\ref{fig:subfig:Kurt_annual}) to S\ref{fig:subfig:Skew_daily} (and S\ref{fig:subfig:Kurt_daily}), the extents of both skewness and heavy tails become more pronounced with the increase in temporal resolution. These non-Gaussian characteristics arise from the intricate temporal dynamics in monthly and daily data. Additionally, the presence of numerical instabilities further exacerbates the non-Gaussianity in the North Pole region and the band region below latitude $-60^\circ$ (referred to as the Band region).

\section{Supplement to Section~3}
\subsection{Evaluation of parameters in the deterministic component}
\label{sec:subsec:supplement_deterministic}
For writing simplicity, we take the annually aggregated temperature as an example to illustrate the inferential process, which follows previous works \citep{castruccio2014statistical,Huang'sEmulator}. First, we fix $\rho(L_i,l_j)$ so that the Model~(1) simplifies into an ordinary linear regression
\begin{equation*}
    y_t^{(r)}(L_i,l_j)=\mathbf{x}_t(L_i,l_j)^\top\bm\beta(L_i,l_j)+\sigma(L_i,l_j)Z_t^{(r)}(L_i,l_j), \quad t=1,\ldots,T,
\end{equation*}
where $\mathbf{x}_t(L_i,l_j)=[1,x_t,\{1-\rho(L_i,l_j)\}\sum_{s=1}^{\infty}\rho(L_i,l_j)^{s-1}x_{t-s}]^\top$, $\bm\beta(L_i,l_j)=[\beta_0(L_i,l_j),\beta_1(L_i,l_j)$, $\beta_2(L_i,l_j)]^\top$, and $Z_t^{(r)}(L_i,l_j)$ is the white noise. By maximizing the log likelihood function $l_{\rho}\{\bm\beta(L_i,l_j)$, $\sigma^2(L_i,l_j)\}$, we get estimates of $\bm\beta(L_i,l_j)$ and $\sigma^2(L_i,l_j)$: $$\hat{\bm\beta}_{\rho}(L_i,l_j)=\{\mathbf{X}(L_i,l_j)^\top\mathbf{X}(L_i,l_j)\}^{-1}\mathbf{X}(L_i,l_j)^\top\mathbf{y}^{(r)}(L_i,l_j),$$ $$\hat\sigma_{\rho}^2(L_i,l_j)=T^{-1}\{\mathbf{y}^{(r)}(L_i,l_j)-\mathbf{X}(L_i,l_j)\hat{\bm\beta}(L_i,l_j)\}^\top\{\mathbf{y}^{(r)}(L_i,l_j)-\mathbf{X}(L_i,l_j)\hat{\bm\beta}(L_i,l_j)\},$$
where $\mathbf{X}(L_i,l_j)=[\mathbf{x}_1(L_i,l_j),\ldots,\mathbf{x}_T(L_i,l_j)]^\top$ and  $\mathbf{y}^{(r)}(L_i,l_j)=[y_1^{(r)}(L_i,l_j),\ldots, y_T^{(r)}(L_i,l_j)]^\top$. Then, we combine the information from all ensembles to estimate $\hat\rho(L_i,l_j)$, which is achieved by maximizing the profile log likelihood:
\begin{equation*}
\begin{aligned}
   &l_{\rho}\{\hat{\bm\beta}_{\rho}(L_i,l_j),\hat\sigma_{\rho}^2(L_i,l_j)\}\propto\\
   &\sum_{r=1}^R-\log[\mathbf{y}^{(r)\top}(L_i,l_j)\{\mathbf{I}_{T}-\mathbf{X}(L_i,l_j)(\mathbf{X}(L_i,l_j)^\top \mathbf{X}(L_i,l_j))^{-1}\mathbf{X}(L_i,l_j)^\top\}\mathbf{y}^{(r)}(L_i,l_j)].
\end{aligned}
\end{equation*}
Finally, we evaluate the deterministic component with the estimated parameters $\hat{\bm\beta}(L_i,l_j)=\hat{\bm\beta}_{\hat\rho}(L_i,l_j)$ and $\hat\sigma^2(L_i,l_j)=\hat\sigma_{\hat\rho}^2(L_i,l_j)$. 

\subsection{Details of spherical harmonics}
As a supplement to Section~3.2.1, this subsection provides additional aspects of spherical harmonics, offering a comprehensive understanding of their application, efficient coefficient calculation, and a comparative analysis with alternative low-rank approximation techniques, such as empirical orthogonal functions (EOFs) and LatticeKrig \citep{LatticeKrig}.

\subsubsection{Application of spherical harmonics}
Spherical harmonic functions (or spherical harmonics for short) constitute a set of orthonormal basis functions defined on the unit sphere $\mathbb{S}^2$, with closed form:
\begin{equation}
    H_q^m(\theta,\psi)=\sqrt{\frac{2q+1}{4\pi}\frac{(q-m)!}{(q+m)!}}P_q^m(\cos\theta)\exp(\iota m\psi),
    \label{eq:Slm}
\end{equation}
comprising a constant, a Legendre polynomial as a function of latitude $\theta$, and a complex exponential associated with longitude $\psi$. They are eigen-functions of the Laplace-Beltrami operator, i.e., the restriction of the Laplace operator on $\mathbb{S}^2$ \citep{atkinson2012spherical}, and can be regarded as the spherical analogs of Fourier series. Leveraging spherical harmonics, any function $f\in\mathcal{L}^2(\mathbb{S}^2)$ can be expanded as follows: 
\begin{equation}
    f(\theta,\psi)=\sum_{q=0}^{\infty}\sum_{m=-q}^q f_q^m H_q^m(\theta,\psi),
    \label{eq:isht}
\end{equation}
where 
\begin{equation}
 f_q^m=\int_{\theta\in[0,\pi]}\int_{\psi\in[0,2\pi]}f(\theta,\psi)\overline{H_q^m(\theta,\psi)}\sin\theta\mathrm{d}\theta\mathrm{d}\psi.
 \label{eq:coeff}
\end{equation}
This expansion finds wide application in solving partial differential equations and exploring spectral properties across various scientific disciplines \citep{pavlis2019spherical}. 

We further illustrate how the spherical harmonic expansion facilitates the investigation of properties within the function $f$. The total power of $f$ over $\mathbb{S}^2$ is given by 
\begin{equation*}
    \int_{\theta\in[0,\pi]}\int_{\psi\in[0,2\pi]}|f(\theta,\psi)|^2\sin\theta\mathrm{d}\theta\mathrm{d}\psi=\sum_{q=0}^{\infty}\mathcal{P}(q),
\end{equation*}
where $\mathcal{P}(q)=\sum_{m=-q}^q \|f_q^m\|_2^2$ represents the total power (or variance) per spherical harmonic degree and is known as degree variance in geodetic literature \citep{pavlis2019spherical}. Fig.~2(b) depicts a power distribution within $f$, demonstrating a decrease with increasing $q$. Furthermore, the decay rate reflects the differentiability property of $f$. Specifically, if $\mathcal{P}(q)$ satisfies $\sum_{q=0}^{\infty}(q+1/2)^{2s}\mathcal{P}(q)<\infty$, then 
$f$ belongs to the Sobolev space $\mathcal{H}^s(\mathbb{S}^2)$ \citep{atkinson2012spherical}. 

\subsubsection{Efficient coefficient calculation}
Now, we explore an efficient method for calculating the coefficients of spherical harmonics. Here we use $f$ to represent $Z_t^{(r)}$. By substituting \eqref{eq:Slm} into \eqref{eq:coeff}, we reformulate the SHT as
\begin{equation}
    f_{q}^m = \int_{\theta \in [0, \pi]} G_m(\theta) H_q^m( \theta, 0 )  \sin\theta  \mathrm{d} \theta 
    \label{coeff2},
\end{equation}
with $H_q^m( \theta, \psi ) = H_q^m( \theta, 0 ) \exp(\iota m\psi)$ and 
\begin{equation}
G_m(\theta)  = \int_{\psi \in [0, 2 \pi)} f( \theta, \psi ) \exp(-\iota m\psi)\mathrm{d} \psi.
\label{Eq:Gm1}
\end{equation}
Leveraging the representation of $f$ in \eqref{eq:isht} and the orthonormality of complex exponentials, $G_m(\theta)$ can be represented as
\begin{equation}
G_m(\theta)  = 2\pi \sum_{q=|m|}^{Q-1} f_q^m H_q^m(\theta, 0),
\label{Eq:Gm2}
\end{equation}
which can be used to represent the data in the form:
\begin{equation}
f(\theta,\psi)  = \sum_{m=-(Q-1)}^{Q-1} G_m(\theta) \exp(-\iota m\psi).
\label{Eq:isht2}
\end{equation}
Here we have assumed that $f$ is band-limited at degree $Q$.
% We present a relationship between the maximum possible band-limit $Q$ and the number of points on the grid in the next section.

With the relationship between spherical harmonics and the Wigner-$d$ function, denoted by $d^{q}_{m,n}(\theta)$ for degree $q$ and orders $|m|,|n|\le q$ \citep{kennedy2013hilbert}:
\begin{equation}
    H_q^m( \theta, 0 ) = \sqrt{\frac{2 q + 1}{4 \pi}} d^{q}_{m,0}(\theta)
    \label{Eq:SlmWigner}
\end{equation}
and the expansion of the Wigner-$d$ function in terms of complex exponentials \citep{risbo1996fourier,kennedy2013hilbert}:
\begin{equation}
    d^{q}_{m,0}(\theta) = \iota^{-m} \sum_{m'=-q}^q d^{q}_{m',0}\left(\frac{\pi}{2}\right) \,\,d^{q}_{m',m}\left(\frac{\pi}{2}\right) \,\,
    \exp(\iota m'\theta),
    \label{Eq:WignerdExpansion}
\end{equation}
we can write \eqref{Eq:Gm2} after interchanging the order of summation as 
\begin{equation}
G_m(\theta)  =  \sum_{m'=-(Q-1)}^{Q-1} K_{m,m'}
    \exp(\iota m'\theta),
    \label{Eq:Gm3}
\end{equation}
where $K_{m,m'}=\iota^{-m}\sum_{q=\max(|m'|,|m|)}^{Q-1}  \sqrt{(2 q + 1)\pi} f_q^m d^{q}_{m',0}(\frac{\pi}{2})d^{q}_{m',m}(\frac{\pi}{2})$. Substituting \eqref{Eq:Gm3} into \eqref{coeff2} yields 
\begin{align}
    f_q^m &=    
    \sum_{m''=-(Q-1)}^{Q-1} \mathrm{Q}_{q,m,m''}  \sum_{m'=-(Q-1)}^{Q-1} K_{m,m'} \,\mathcal{I}(m'+m''),
    \label{coeff3}
\end{align}
where
\begin{align*}
\label{Eq:Qlmm}
\mathrm{Q}_{q,m,m''} = \iota^{-m}  \sqrt{\frac{2 q + 1}{4 \pi}}  d^{q}_{m'',0}\left(\frac{\pi}{2}\right)  \,\,d^{q}_{m'',m}\left(\frac{\pi}{2}\right)
\end{align*}
and we have used \eqref{Eq:SlmWigner} and \eqref{Eq:WignerdExpansion}. $\mathcal{I}(m)$ is an integral of the form
\begin{equation*}
\mathcal{I}(m) = \int_{\theta \in [0, \pi]} \exp(\iota m\theta) \sin\theta  \mathrm{d} \theta = \begin{cases} \delta_{|m|,1}\,\frac{\iota m \pi}{2}, & m\,\,{\rm odd} \\ \frac{2}{1-m^2}, & m\,\,{\rm even}
\end{cases}
\end{equation*}
with $\delta_{|m|,1}$ being the Kronecker delta function.

Instead of evaluating integrals along latitude and longitude in \eqref{coeff2} and \eqref{Eq:Gm1}, our approach involves computing $G_m(\theta)$ for each $\theta_i=\pi/2-\pi L_i/180$, $i=1,\ldots,I$, and $K_{m,m'}$ for all $|m|,|m'|<Q$. Since the representation in \eqref{Eq:isht2} indicates that $f(\cdot,\psi)$, with complex exponentials $\{-\exp(\iota m\psi)\}$ as basis functions, is band-limited at $Q$, $G_m(\theta)$ can be recovered accurately using the Fast Fourier transform (FFT) if $J\ge 2Q-1$. We use \eqref{Eq:Gm3} to recover $K_{m,m'}$ by employing inverse FFT provided $G_m(\theta)$ is given for a sufficient number of co-latitude points over the domain $[0,2\pi)$. We have $G_m(\theta)$ over $I$ equiangular points over the domain $[0,\pi]$ with two samples at the poles~($\theta=0,\pi$). Noting \eqref{Eq:Gm2}, we extend the domain of $G_m(\theta)$ to include the points along co-latitude in $(\pi,2\pi)$ for $\theta \in (0,\pi)$ as $G_m(2\pi-\theta)= 2\pi \sum_{q=|m|}^{Q-1} f_q^m H_q^m(\theta, \pi)= (-1)^m \,G_m(\theta)$, where we have used $(-1)^m  H_q^m(\theta, \pi) =  H_q^m(\theta, 0)$. Here we require $2I-2\ge 2Q-1$. Once we have $K_{m,m'}$ for all $|m|,|m'|<Q$, we use \eqref{coeff3} for the computation of SHT. Note that $G_m(\theta)$ and $K_{m,m'}$ can be exactly computed if
\begin{equation*}
    Q \le \min\left(I-1,\frac{J+1}{2}\right).
    \label{Eq:bandlimit}
    \vspace{-10pt}
\end{equation*}

\subsubsection{Approximation performance of SHT}
\begin{figure}[!h]
\centering
\subfigure[$|\varepsilon_9^{(3)}(L_i,l_j)|$ with $Q=36$]{
\label{fig:subfig:SHTinv36_E3}
\includegraphics[scale=0.5]{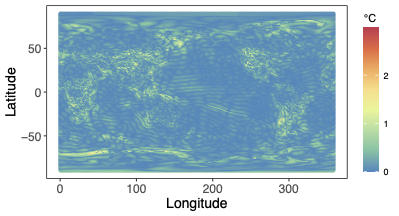}}
\subfigure[$|\varepsilon_{69}^{(1)}(L_i,l_j)|$ with $Q=36$]{
\label{fig:subfig:SHTinv36_T2083}
\includegraphics[scale=0.5]{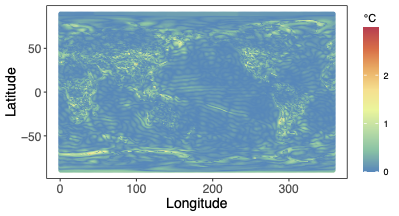}}\\
% \subfigure[$\hat v^2(L_i,l_j)$ with $Q=36$]{
% \label{fig:subfig:v2hat_36}
% \includegraphics[scale=0.4]{Fig_LENS2/v2hat_36.pdf}}
\subfigure[$|\varepsilon_9^{(3)}(L_i,l_j)|$ with $Q=72$]{
\label{fig:subfig:SHTinv72_E3}
\includegraphics[scale=0.5]{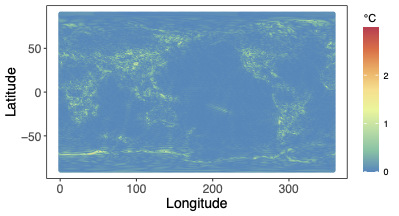}}
\subfigure[$|\varepsilon_{69}^{(1)}(L_i,l_j)|$ with $Q=72$]{
\label{fig:subfig:SHTinv72_T2083}
\includegraphics[scale=0.5]{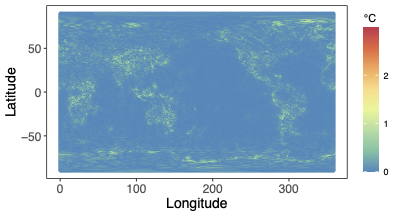}}\\
% \subfigure[$\hat v^2(L_i,l_j)$ with $Q=72$]{
% \label{fig:subfig:v2hat_72}
% \includegraphics[scale=0.4]{Fig_LENS2/v2hat_72.pdf}}
\caption{Approximation performance of SHT with different $Q$ values.}
\label{Fig:Supplement_to_Fig2}
\end{figure}
Figures~2(c) and 2(d) demonstrate the approximation performance of SHT with $Q=36$ and $72$ only for the stochastic component at ensemble $r=1$ and time point $t=9$. Here we supplement results on other ensembles and time points for more general analysis. Specifically, Figs.~S\ref{fig:subfig:SHTinv36_E3} and S\ref{fig:subfig:SHTinv36_T2083} illustrate the absolute values of $\{\varepsilon_9^{(3)}(L_i,l_j)\}_{i=1,\ldots,I; j=1,\ldots,J}$ and $\{\varepsilon_{69}^{(1)}(L_i,l_j)\}_{i=1,\ldots,I; j=1,\ldots,J}$ with $Q=36$, respectively. Similarly, Figs.~S\ref{fig:subfig:SHTinv72_E3} and S\ref{fig:subfig:SHTinv72_T2083} show those with $Q=72$. 
% Furthermore, we use $\hat v^2(L_i,l_j)=\sum_{r=1}^R\sum_{t=1}^T\{\varepsilon_t^{(r)}(L_i,l_j)\}^2$ to see the general performance across all ensembles and time points, which are shown in Figs.~S\ref{fig:subfig:v2hat_36} and S\ref{fig:subfig:v2hat_72}.

\subsubsection{Comparison with other low-rank approximations}
In equation (2), we adopt spherical harmonics as basis functions in the low-rank approximation technique to overcome the computational limitations. When employing low-rank approximation methods, the choice of basis functions provides flexibility. These basis functions can either be orthogonal, such as Fourier functions, orthogonal polynomials, and eigenvectors from a specified covariance \citep{HBS}, or non-orthogonal, as seen in R packages \textit{FRK} \citep{FRK_intro} and \textit{LatticeKrig} \citep{LatticeKrig}. Here we compare spherical harmonics with two popular bases: empirical orthogonal functions (EOFs) and those used in LatticeKrig.

First, we highlight the challenges associated with employing EOFs in the context of our large global simulations and compare EOFs with spherical harmonics. The first challenge is that EOFs themselves are not directly available and need additional storage. To be more specific, EOFs approximate eigen-functions of the covariance function $c(\cdot,\cdot)$, which are unknown and lack a closed form. Utilizing replicates of $Z_t^{(r)}(\cdot)$ enables the computation of a sample covariance matrix $\hat{\mathbf{C}}\in\mathbb{R}^{IJ\times IJ}$. However, performing eigen-decomposition on $\hat{\mathbf{C}}$ suffers from a storage limitation and an impractical computational complexity of $O((IJ)^3)$. While there are faster algorithms \citep{irlba,RSpectra} for calculating the first $\hat Q$ eigen-functions, determining an appropriate $\hat Q$ requires additional procedures. In contrast, spherical harmonics have closed forms, which are readily available. The second challenge is that the coefficients before EOFs for each $Z_t^{(r)}$ have to be evaluated by solving a large linear system. However, the coefficients of spherical harmonics can be efficiently calculated by the algorithm outlined in Section~S3.2.2. The third challenge is the data-driven nature of EOFs, requiring the calculation and storage of a new set of EOFs for each variable (e.g., wind speed or precipitation). In contrast, spherical harmonics maintain applicability across different variables on the sphere.

Next, we illustrate the challenges of applying LatticeKrig to the stochastic component. LatticeKrig uses compactly supported Wendland functions as bases, with their center points allocated in grids of multiple levels. These multi-resolution basis functions are adept at capturing different scales of variation present in the data. The coefficients of these bases are assumed as a Gaussian Markov random field, resulting in a sparse precision matrix and an efficient evaluation procedure. For spherical data, the R package \textit{LatticeKrig} offers geometric options, and in this case, we choose \textit{LKGeometry=``LKSphere"}.

We exemplify the implementation details of LatticeKrig using $\{Z_9^{(1)}(L_i,l_j)\}_{i=1,\ldots,I;j=1,\ldots,J}$ in Fig.2(a). Several tuning parameters are essential for constructing basis functions and the model. The \textit{startingLevel} (i.e.,the first level of lattice) and \textit{nlevel} (i.e., number of levels) determine the resolutions and the number of basis functions. The \textit{a.wght}, representing the weight given to the central lattice point in the spatial autoregressive model, is set at $5.482$. The \textit{nu} is a smoothness parameter controlling relative variances for different multi-resolution levels (i.e., the parameter \textit{alpha}). Here, we set \textit{nu}=1, which optimizes the fitting results. The \textit{lambda} is a parameter in model describing the variance relationship among different parts in the stochastic component. It is evaluated from the given data by maximizing a log likelihood function.

\begin{figure}[!h]
\centering
\subfigure[LatticeKrig with $3420$ bases]{
\label{fig:subfig:Lattice_Demo_2}
\includegraphics[scale=0.5]{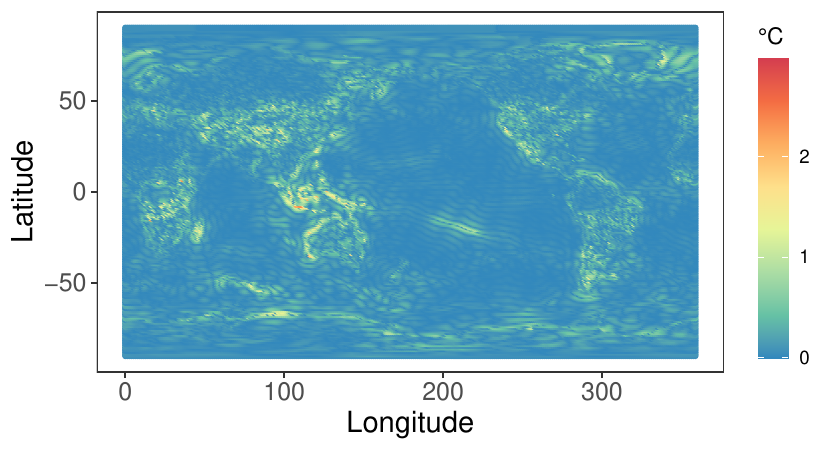}}
\subfigure[LatticeKrig with $13658$ bases]{
\label{fig:subfig:Lattice_Demo_3}
\includegraphics[scale=0.5]{Fig_LENS2/Lattice_Demo_3.pdf}}\\
\subfigure[SHT with $3364$ bases]{
\label{fig:subfig:SHTinv58}
\includegraphics[scale=0.5]{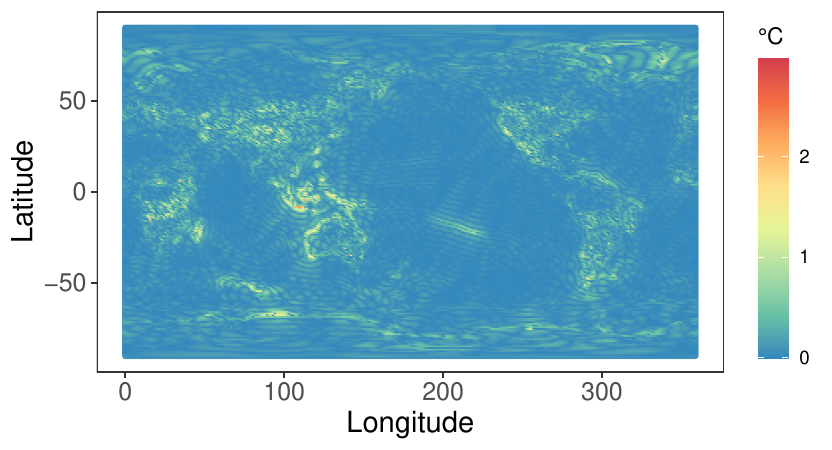}}
\subfigure[SHT with $13456$ bases]{
\label{fig:subfig:SHTinv1161}
\includegraphics[scale=0.5]{Fig_LENS2/SHTinv116.pdf}}\\
\subfigure[Comparison in space]{
\label{fig:subfig:LatticeKrig_Space}
\includegraphics[scale=0.5]{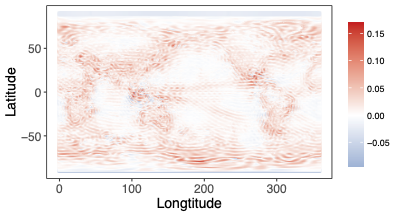}}
\subfigure[Comparison in time]{
\label{fig:subfig:LatticeKrig_Time}
\includegraphics[scale=0.5]{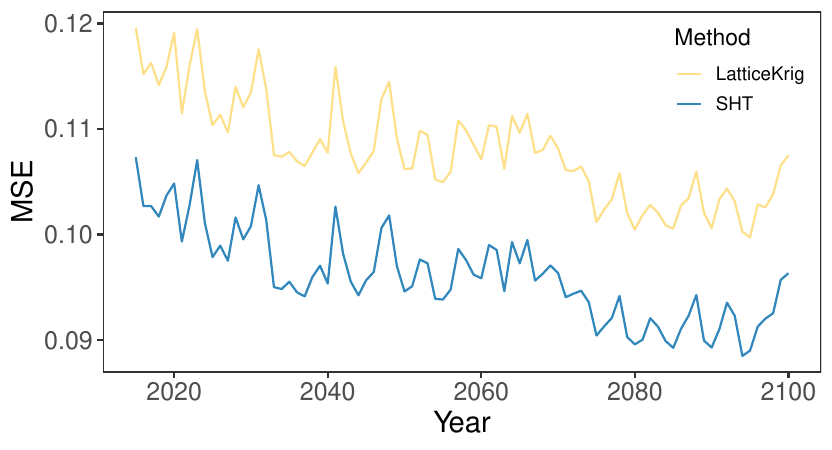}}
\caption{Comparison between LatticeKrig and SHT. (a) and (b) show the absolute values of $\{\varepsilon_9^{(1)}(L_i,l_j)\}_{i=1,\ldots,I; j=1,\ldots,J}$ obtained by LatticeKrig with \textit{startingLevel}=1, \textit{nlevel}=5 and 6, respectively. (c) and (d) show the absolute values of $\{\varepsilon_9^{(1)}(L_i,l_j)\}_{i=1,\ldots,I; j=1,\ldots,J}$ obtained by SHT with $Q=58$ and $116$, respectively. (e) depicts the difference between  $\{\sum_{r=1}^R\sum_{t=1}^T|\varepsilon_t^{(r)}(L_i,l_j)|/(RT)\}_{i=1,\ldots,I; j=1,\ldots,J}$ of LatticeKrig and that of SHT; (f) shows time series $\{\sum_{r=1}^R\sum_{i=1}^I\sum_{j=1}^J|\varepsilon_t^{(r)}(L_i,l_j)|/(RIJ)\}_{t=1,\ldots,T}$ for LatticeKrig and SHT.}
\label{Fig:LatticeKrig_SHT}
\end{figure}

Figures~S\ref{fig:subfig:Lattice_Demo_2}--S\ref{fig:subfig:SHTinv1161} compare the performance of LatticeKrig and SHT on a single set of stochastic component, with the basis numbers of SHT specified to be as close as possible but less than those of LatticeKrig. In this specific example, SHT achieves superior approximations with fewer bases and less time. LatticeKrig requires about $19$ and $186$ minutes to obtain the approximations in Figs.~S\ref{fig:subfig:Lattice_Demo_2} and S\ref{fig:subfig:Lattice_Demo_3}, respectively, with the majority of computational cost coming from the evaluation of \textit{lambda}. In contrast, the implementation of SHT and inverse SHT takes no more than 6 seconds.

We proceed to demonstrate the performance of LatticeKrig on stochastic components of all simulations, assumed to be independent. Consequently, tuning parameters are held constant across all $t$ and $r$. We set \textit{lambda} to be $0.000124893$, an average over \textit{lambda}s of randomly selected stochastic components. With known tuning parameters, the approximation in Fig.~S\ref{fig:subfig:Lattice_Demo_2} takes approximately $7$ seconds, comparable to SHT. Figs.~S\ref{fig:subfig:LatticeKrig_Space} and \ref{fig:subfig:LatticeKrig_Time} compare the performance of LatticeKrig to that of SHT on all stochastic components, where SHT slightly outperforms in most cases with fewer bases and less computational time. It is essential to note that our objective is not to assert the superiority of SHT in accuracy. Instead, we aim to emphasize that using SHT results in substantial storage savings. Specifically, for our simulations, LatticeKrig requires at least five levels, i.e., $3420$ bases, to achieve comparable results and computational time to SHT. Despite LatticeKrig offering a sparse precision matrix, it still involves storing $2,582,885$ non-zero elements. However, under the assumption of axial symmetry, the subsequent section reveals that $\tilde{\mathbf{K}}$ has no more than $57,155$ non-zero elements to be memorized.

\subsection{Structure derivation for $\tilde{\mathbf{K}}$ under the axial symmetry}
\label{sec:subsec:supplement_derivative}
In this subsection, we derive the structure of $\tilde{\mathbf{K}}$ under the assumption of axial symmetry. First, we expand $Z(\theta,\psi)$ with spherical harmonics as

\begin{equation*}
\begin{aligned}
Z(\theta,\psi)&=\sum_{q=0}^{Q-1}\sum_{m=-q}^qH_q^m(\theta,\psi)s_q^m\\
&=\sum_{q=0}^{Q-1}[\sum_{m=1}^q \{H_q^m(\theta,\psi)s_q^m+H_q^{-m}(\theta,\psi)s_q^{-m}\}+H_q^0(\theta,\psi)s_q^0]\\
&=\sum_{q=0}^{Q-1}\{\sum_{m=1}^q F(q,m)+H_q^0(\theta,\psi) s_q^0\}.
\end{aligned}
\end{equation*}
With the closed-form  
\begin{equation*}
    H_q^m(\theta,\psi)=\sqrt{\frac{2q+1}{4\pi}\frac{(q-m)!}{(q+m)!}}P_q^m(\cos\theta)\exp(\iota m \psi),
\end{equation*}
we have $H_q^m(\theta,\psi)=H_q^m(\theta,0)\exp(\iota m\psi)$. Moreover, $H_q^{-m}(\theta,0)=(-1)^mH_q^m(\theta,0)$ by $P_q^{-m}=(-1)^m\frac{(q-m)!}{(q+m)!}P_q^m$. Therefore,
\begin{equation*}
\begin{aligned}
    F(q,m)&=H_q^m(\theta,0)\exp(\iota m\psi)\{\Re(s_q^m)+\iota\Im(s_q^m)\}+H_q^{-m}(\theta,0)\exp(-\iota m\psi)(-1)^m\{\Re(s_q^{m})-\iota\Im(s_q^{m})\}\\
    &=H_q^m(\theta,0)\{\cos(m\psi)+\iota\sin(m\psi)\}(\tilde s_q^m+\iota\tilde s_q^{-m})\\
    &\quad +(-1)^m H_q^{-m}(\theta,0)\{\cos(m\psi)-\iota\sin(m\psi)\}(\tilde s_q^m-\iota\tilde s_q^{-m})\\
    &=H_q^m(\theta,0)\cos(m\psi)(\tilde s_q^m+\iota \tilde s_q^{-m})+(-1)^mH_q^{-m}(\theta,0)\iota\sin(m\psi)(\tilde s_q^m+\iota\tilde s_q^{-m})\\
    &\quad +H_q^m(\theta,0)\cos(m\psi)(\tilde s_q^m-\iota \tilde s_q^{-m})-(-1)^mH_q^{-m}(\theta,0)\iota\sin(m\psi)(\tilde s_q^m-\iota\tilde s_q^{-m})\\
    &=2H_q^m(\theta,0)\cos(m\psi)\tilde s_q^m-2(-1)^{m}H_q^{-m}(\theta,0)\sin(m\psi)\tilde s_q^{-m}.
\end{aligned}
\end{equation*}

Now, we calculate the covariance between $Z(\theta,\psi)$ and $Z(\theta',\psi')$:
\begin{equation*}
\begin{aligned}
     \mathrm{E}\{Z(\theta,\psi)Z(\theta',\psi')\}&=\sum_{q=1}^{Q-1}\sum_{q'=1}^{Q-1}[\sum_{m=1}^{q}\sum_{m'=1}^{q'}\mathrm{E}\{F(q,m)F'(q',m')\}+\mathrm{E}\{H_q^0(\theta,\psi)s_q^0H_{q'}^0(\theta',\psi')s_{q'}^0\}\\
      &\quad+\sum_{m'=1}^{q'}\mathrm{E}\{F'(q',m')H_q^0(\theta,\psi)s_q^0\}+\sum_{m=1}^{q}\mathrm{E}\{F(q,m)H_{q'}^0(\theta',\psi')s_{q'}^0\}].\\
\end{aligned}
\end{equation*}
For the first term in the bracket, we have
\begin{equation*}
\begin{aligned}
     \frac{1}{4}\mathrm{E}\{F(q,m)F'(q',m')\}&=\mathrm{E}[\{H_q^m(\theta,0)\cos(m\psi)\tilde s_q^m-(-1)^{m}H_q^{-m}(\theta,0)\sin(m\psi)\tilde s_q^{-m}\}\\
     &\quad\{H_{q'}^{m'}(\theta',0)\cos(m'\psi')\tilde s_{q'}^{m'}-(-1)^{m'}H_{q'}^{-m'}(\theta',0)\sin(m'\psi')\tilde s_{q'}^{-m'}\}]\\
     &=H_q^m(\theta,0)H_{q'}^{m'}(\theta',0)\cos(m\psi)\cos(m'\psi')\mathrm{E}(\tilde{s}_q^m\Tilde{s}_{q'}^{m'})\\
     &\quad-(-1)^{m'}H_q^m(\theta,0)H_{q'}^{-m'}(\theta',0)\cos(m\psi)\sin(m'\psi')\mathrm{E}(\tilde{s}_q^m\Tilde{s}_{q'}^{-m'})\\
     &\quad-(-1)^{m}H_q^{-m}(\theta,0)H_{q'}^{m'}(\theta',0)\sin(m\psi)\cos(m'\psi')\mathrm{E}(\tilde{s}_q^{-m}\Tilde{s}_{q'}^{m'})\\
     &\quad+(-1)^{m+m'}H_q^{-m}(\theta,0)H_{q'}^{-m'}(\theta',0)\sin(m\psi)\sin(m'\psi')\mathrm{E}(\tilde{s}_q^{-m}\Tilde{s}_{q'}^{-m'}).
\end{aligned}
\end{equation*}
To let the above equality only depend on $\theta$, $\theta'$, and $|\psi-\psi'|$, we need $\mathrm{E}(\tilde{s}_q^m\Tilde{s}_{q'}^{m'})=\mathrm{E}(\tilde{s}_q^{-m}\Tilde{s}_{q'}^{-m'})=k_{qq'm}\delta_{m=m'}\delta_{m>0}$. For the second term in bracket, we have
\begin{equation*}
    \mathrm{E}\{H_q^0(\theta,\psi)s_q^0H_{q'}^0(\theta',\psi')s_{q'}^0\}=H_q^0(\theta,0)H_{q'}^0(\theta',0)\mathrm{E}(s_q^0s_{q'}^0),
\end{equation*}
which only depend on $\theta$ and $\theta'$. For the third term in the bracket, we have
\begin{equation*}
\begin{aligned}
    \mathrm{E}\{F'(q',m')H_q^0(\theta,\psi)s_q^0\}&=2\mathrm{E}[\{H_{q'}^{m'}(\theta',0)\cos(m'\psi')\tilde s_{q'}^{m'}-(-1)^{m'}H_{q'}^{-m'}(\theta',0)\sin(m'\psi')\tilde s_{q'}^{-m'}\}H_q^0(\theta,\psi)s_q^0]\\
    &=2 H_{q'}^{m'}(\theta',0)H_q^0(\theta,\psi)\cos(m'\psi')\mathrm{E}(\tilde s_{q'}^{m'}s_q^0)\\
    &\quad-2(-1)^{m'}H_{q'}^{-m'}H_q^0(\theta,\psi)(\theta',0)\sin(m'\psi')\mathrm{E}(\tilde s_{q'}^{-m'}s_q^0).
\end{aligned}
\end{equation*}
To let the above equality only depend on $\theta$, $\theta'$, and $|\psi-\psi'|$, we need $\mathrm{E}(\tilde s_{q'}^{m'}s_q^0)=\mathrm{E}(\tilde s_{q'}^{-m'}s_q^0)=0$, where $m$ and $m'>1$. In general, to satisfy the assumption of axial symmetry, we should let $\mathrm{E}(\Tilde{s}_q^m\Tilde{s}_{q'}^{m'})=\mathrm{E}(\Tilde{s}_q^{-m}\Tilde{s}_{q'}^{-m'})=k_{qq'm}\delta_{m=m'}\delta{m\ge 0}$.

We calculate the number of non-zero elements in $\Tilde{\mathbf{K}}$. When $m=0$, $\forall q,q'=0,1,\ldots,Q-1$, the $(q^2+q+1,q'^2+q'+1)$th element of $\Tilde{\mathbf{K}}$ is non-zero. Therefore, $Q^2$ non-zero elements for $m=0$. When $m\geq 1$, $\forall q,q'=m,\ldots,Q-1$, the $(q^2+q+m+1,q'^2+q'+m+1)$th and $(q^2+q-m+1,q'^2+q'-m+1)$th elements of $\Tilde{\mathbf{K}}$ are non-zero. Therefore, $2(Q-m)^2$ non-zero-elements for $m\geq 1$. The total number of non-zero elements in $\Tilde{\mathbf{K}}$ is $Q^2+2(Q-1)^2+2(Q-2)^2+\ldots 2=2Q^3/3+Q/3$.

However, not all non-zero elements need to be stored. Some element values are repeated. For $m=0$, only $Q(Q+1)/2$ elements of $\Tilde{\mathbf{K}}$ need to be stored, which are indexed by  $(q^2+q+1,q'^2+q'+1)$ with $q=0,\ldots,Q-1$ and $q'=q,\ldots,Q-1$. For $m\geq 1$,  $(Q-m)(Q-m+1)/2$ elements indexed by $(q^2+q+m+1,q'^2+q'+m+1)$ or $(q^2+q-m+1,q'^2+q'-m+1)$ need to be stored, where $q=m,\ldots,Q-1$ and $q'=q,\ldots,Q-1$. Therefore, the total number of elements to be memorized is $Q(Q+1)/2+(Q-1)(Q-1+1)/2+\ldots,1\times2/2=Q^3/6+Q^2/2+Q/3$.

\subsection{Validation of diagonal matrix $\Phi_p$}
\label{sec:subsec:supplement_validate}
\begin{figure}[ht]
\centering
\subfigure[Annual]{
\label{fig:subfig:CrossP_annual}
\includegraphics[scale=0.6]{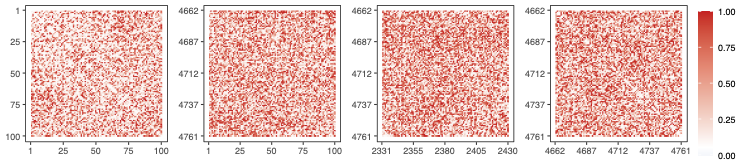}}\\
\subfigure[Monthly]{
\label{fig:subfig:CrossP_monthly}
\includegraphics[scale=0.6]{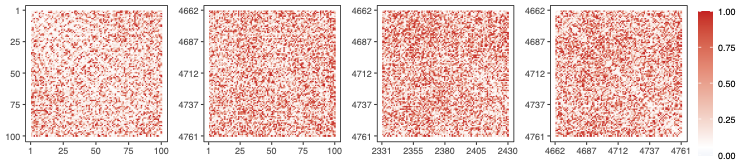}}\\
% \subfigure[Daily]{
% \label{fig:subfig:CrossP_daily}
% \includegraphics[scale=0.4]{Fig_LENS2/CrossP_daily.pdf}}
\caption{$P$-values of the first temporal lag of the cross-correlation for annual and monthly data.}
\label{Fig:crosscovariance}
\end{figure}

We briefly validate the assumption of diagonal matrices $\{\Phi_p\}_{p=1}^P$, where $P=1$ is obtained by the inference in our case study. That is, we assume that the cross-correlation between elements of $\check{\mathbf{s}}_t$ ($\tilde{\mathbf{s}}_t$ for the annual data) can be neglected. The detailed procedure of validation is as follows. First, we fit any two time series $(\check{s}_t)_q^m$ and $(\check{s}_t)_{q'}^{m'}$ with their respective AR($1$) models, and obtain their residuals. Then, we calculate the $p$-values for the first temporal lag of the cross-correlation between these two residuals. We put the result in the $(q^2+q+m+1,{q'}^2+q'+m'+1)$th element of matrix $\mathbf{PV}\in\mathbb{R}^{Q^2\times Q^2}$. Fig.~\ref{Fig:crosscovariance} shows parts of $\mathbf{PV}$ matrices for annual and monthly data. For most pairs of $(\check{s}_t)_q^m$ and $(\check{s}_t)_{q'}^{m'}$, especially those with higher degrees, there is not enough evidence to conclude that a significant auto-correlation exists and the residuals are approximately independent. The proportions of $p$-values larger than $0.05$ are almost $1$ for all scales.

\subsection{Evaluation of parameters in the modified TGH transformation}
\label{sec:subsec:supplement_Tukey}
For a time series $\{\tilde{s}_t\}_{t=1}^T$ satisfying a TGH auto-regressive model, we can estimate the parameters $\bm\mu_{1}=(\omega,g,h)^\top$ and $\bm\mu_{2}=(\phi_1,\ldots,\phi_P)^\top$ by maximizing the following log-likelihood function \citep{Yan2019Non-Gaussian}:
\begin{equation*}
\begin{aligned}
\mathcal{L}_T(\bm\mu_{1},\bm\mu_{2};\{\tilde{s}_t\}_{t=1}^T)&=f(\ddot{s}_{1,\bm\mu_{1}})+f(\ddot{s}_{2,\bm\mu_{1}} | \ddot{s}_{1,\bm\mu_{1}})+\cdots+f(\ddot{s}_{T,\bm\mu_{1}} | \ddot{s}_{T-1,\bm\mu_{1}},\ldots,\ddot{s}_{T-P,\bm\mu_{1}})\\
&-T\log(\omega)-\frac{h}{2}\sum_{t=1}^T\ddot{s}_{t,\bm\mu_{1}}^2-\sum_{t=1}^T\log[\exp(g\ddot{s}_{t,\bm\mu_{1}})+\frac{h}{g}\{\exp(g\ddot{s}_{t,\bm\mu_{1}})-1\}\ddot{s}_{t,\bm\mu_{1}}],
\end{aligned}
\end{equation*}
where $f(\cdot|\cdot)$ is the conditional log-likelihood for the Gaussian AR($P$) process $\check{s}$ and $\ddot{s}_{t,\bm\mu_1}=\tau_{g,h}^{-1}\{\tilde{s}_{t}/\omega\}$. 

For $(q,m)\in\mathcal{S}_{g,h}$, the estimate of $(\bm\mu_1)_q^m$ and $(\bm\mu_2)_q^m$ is the maximizer of $\sum_{r=1}^R\mathcal{L}_T(\cdot,\cdot;\{(\tilde{s}_t^{(r)})_q^m\}_{t=1}^T)$, which contains information of all $R$ members. Then, $\lambda_q^m$ can be evaluated by $\mathrm{sd}\{(\tilde s_t^{(r)})_q^m\}/\mathrm{sd}\{(\ddot{s}_t^{(r)})_q^m\}$, where $\mathrm{sd}$ is the calculation of standard deviation over $t$ and $r$, and $(\check{s}_t^{(r)})_q^m=\lambda_q^m(\ddot{s}_t^{(r)})_q^m$. The value of $P$ can be chosen by using  $\check{\mathrm{BIC}}_q^m(P)=-2\sum_{r=1}^R\mathcal{L}_T(\bm\mu_1,\bm\mu_2;\{(\Tilde{s}_t^{(r)})_q^m\}_{t=1}^T)+(P+3)\log\{R(T-P)\}$. These computations are expensive due to the non-analytic form of $\tau_{g,h}^{-1}$. Therefore, we follow \citet{XU201578} to find a maximum approximated likelihood estimator (MALE) by approximating $\tau_{g,h}^{-1}$ with a piecewise linear function. It significantly reduces the computational burden and takes only $O(T)$ computational time. For $(q,m)$ in the complementary set of $\mathcal{S}_{gh}$, we set $(\check{s}_t^{(r)})_q^m=(\tilde{s}_t^{(r)})_q^m$, $\bm\mu_1=\mathbf{0}$, $\lambda_q^m=1$, and evaluate $\bm\mu_2$ by VAR($P$).

\subsection{Algorithm for developing SG}
\label{sec:subsec:supplement_algorithms}
This subsection provides a summary for procedures of developing SGs in Algorithm~\ref{alg:ConstructionSG}, which supplements to Section~3.3.
\begin{algorithm}
\caption{Construct an SG for CESM2-LENS2}
\label{alg:ConstructionSG}
\begin{algorithmic}[3] 
\renewcommand{\algorithmicrequire}{ \textbf{Input:}} 
\REQUIRE \textbf{$\{y_t^{(r)}(L_i,l_j)\}_{t=1,\ldots,T;i=1,\ldots,I;j=1,\ldots,J;r=1,\ldots,R}$, $K$, $Q_l$, $Q_o$, $P$, $\mathbf{A}^{-1}\in\mathbb{R}^{Q_o^2\times Q_o^2}$ }
\vspace{5pt}
\renewcommand{\algorithmicrequire}{ \textbf{Stage 1: Deterministic component $m_t$ and $\sigma$}} 
\REQUIRE \textbf{ }\\   
    \vspace{5pt}
    \hspace{-15pt}For each grid point $(L_i,l_j)$, $i=1,\ldots,I$ and $j=1,\ldots,J$:\\
    \vspace{5pt}
    1) evaluate parameters $\rho(L_i,l_j)$, $\beta_0(L_i,l_j)$, $\beta_1(L_i,l_j)$, $\beta_2(L_i,l_j)$, $a_1(L_i,l_j)$,\ldots, $a_K(L_i,l_j)$, $b_1(L_i,l_j)$,\ldots, $b_K(L_i,l_j)$ in $m_t(L_i,l_j)$ and $\sigma(L_i,l_j)$ by the method in Section~3.1.\\ 
    \vspace{5pt}
    2) for each time point $t$, calculate $m_t(L_i,l_j)$ by plugging evaluated parameters into its expression, $t=1,\ldots,T$.\\
    \vspace{5pt}
    3) for each ensemble $r$, obtain the stochastic component $\{Z_t^{(r)}(L_i,l_j)\}_{t=1}^T$ by detrending and rescaling $\{y_t^{(r)}(L_i,l_j)\}_{t=1}^T$ with $m_t(L_i,l_j)$ and $\sigma(L_i,l_j)$, respectively, $r=1,\ldots,R$.\\
    \vspace{5pt}
\renewcommand{\algorithmicrequire}{ \textbf{Stage 2: Stochastic component $Z_t^{(r)}(L_i,l_j)$}} 
\REQUIRE \textbf{ }\\  
 \vspace{5pt}
    1) For each ensemble $r$ and time point $t$, perform SHT on $\{Z_t^{(r)}(L_i,l_j)\}_{i=1,\ldots,I; j=1,\ldots,J}$ to obtain coefficients $\{(s_t^{(r)})_q^m\}_{q=0,\ldots,Q_o-1; m=-q,\ldots,q}$ and standard deviation $v(L_i,l_j)$.\\
    \vspace{5pt}
    2) Transform the complex-valued coefficient $(s_t^{(r)})_q^m$ to be the real-valued one $(\tilde{s}_t^{(r)})_q^m$ with $\mathbf{A}^{-1}$.\\
    \vspace{5pt}
    3) Evaluate the temporal dependence structure. For each $(q,m)$, test the Gaussianity of $\{(\tilde{s}_t^{(r)})_q^m\}_{t=1,\ldots,T;r=1,\ldots,R}$ and find the set $\mathcal{S}_{gh}$. Then, for $(q,m)\in\mathcal{S}_{gh}$, find the MALE of parameters $\omega_q^m$, $g_q^m$,  $h_q^m$ and $\lambda_q^m$ in TGH transformation and $(\phi_1)_q^m,\ldots,(\phi_P)_q^m$ in auto-regressive model using method in Section~3.2.2. Meanwhile, obtain $\{(\check{s}_t^{(r)})_q^m\}_{t=1,\ldots,T; r=1,\ldots,R}$.\\
    \vspace{5pt}
    4) Evaluate the spatial dependence structure. Empirically estimate the covariance matrix $\check{\mathbf{K}}_0$ and $\check{\mathbf{K}}_{|p-p'|}$, $p,p'=1,\ldots,P$, $p'\neq p$. For example, estimate $\check{k}_{qq'm}$ with 
    \begin{equation*}
        (2RT)^{-1}\sum_{r=1}^R\sum_{t=1}^T\{(\check{s}_t^{(r)})_q^m(\check{s}_t^{(r)})_{q'}^m+(\check{s}_t^{(r)})_q^{-m}(\check{s}_t^{(r)})_{q'}^{-m}\}.
    \end{equation*}
    \vspace{5pt}
    5) Calculate $\mathbf{U}=\check{\mathbf{K}}_0-\sum_{p=1}^P\bm\Phi_p\check{\mathbf{K}}_0\bm\Phi_p^\top-\sum_{p=1}^P\sum_{p'\neq p}^P\bm\Phi_p\check{\mathbf{K}}_{|p-p'|}\bm\Phi_{p'}^\top$.
    \vspace{5pt}
\renewcommand{\algorithmicensure}{ \textbf{Output:}} 
\ENSURE $\rho(L_i,l_j)$, $\beta_0(L_i,l_j)$, $\beta_1(L_i,l_j)$, $\beta_2(L_i,l_j)$, $a_1(L_i,l_j)$,\ldots, $a_K(L_i,l_j)$, $b_1(L_i,l_j)$,\ldots, $b_K(L_i,l_j)$, $\sigma(L_i,l_j)$, $v(L_i,l_j)$, $\lambda_q^m$, $\omega_q^m$, $g_q^m$, $h_q^m$, $(\phi_1)_q^m,\ldots,(\phi_P)_q^m$, $\mathbf{U}$, $i=1,\ldots,I$, $j=1,\ldots,J$, $q=0,\ldots,Q_o-1$, $m=-q,\ldots,q$. 
\end{algorithmic}
\end{algorithm}

\section{Supplement to Section~4}
\subsection{Annually aggregated temperature}
\label{sec:supplement_annual}
This subsection serves as a supplement to Section~4.1, including more details about our SG based on SHT (referred to as SHT-SG), the one proposed by \citet{Huang'sEmulator} (referred to as HCBG-SG), and their emulations for the annually aggregated temperature.

\subsubsection{Details in $\mathrm{I}_{\rm fit}$ of the proposed SG}
\begin{figure}[!ht]
\centering
\subfigure[$\mathrm{I}_{\rm fit}(L_i,l_j)$]{
\label{fig:subfig:Ifit_annual}
\includegraphics[scale=0.5]{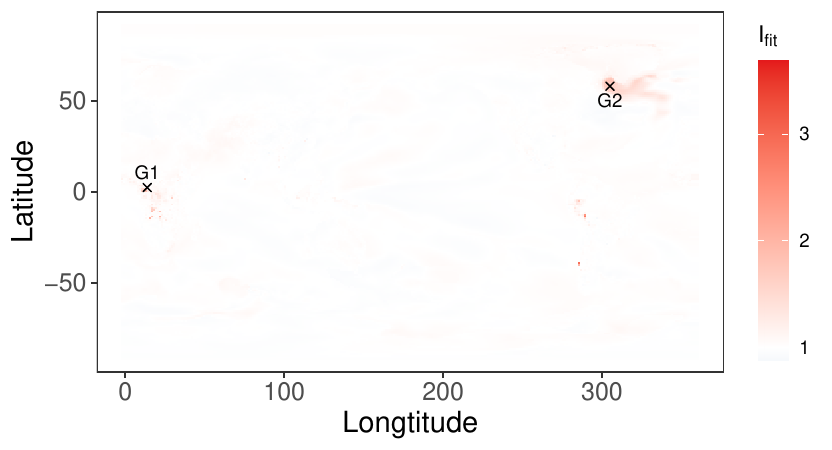}}
\subfigure[Time series]{
\label{fig:subfig:Check_Ifit}
\includegraphics[scale=0.5]{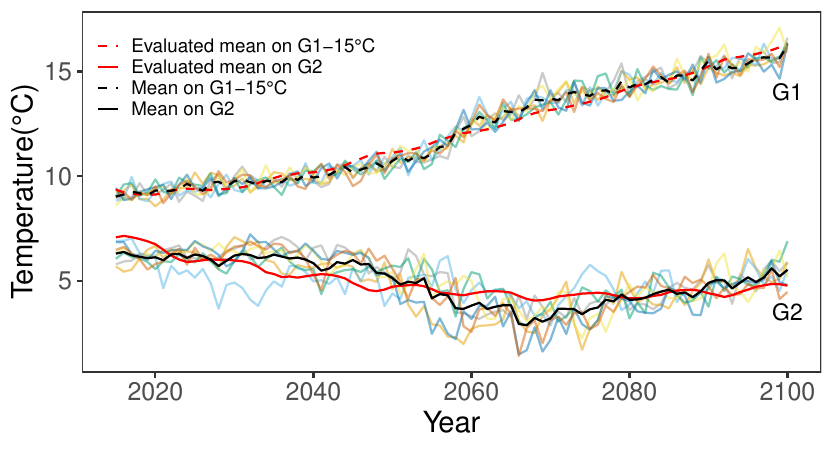}}
\caption{Assessment of goodness-of-fit. (a) is the map of $\{\mathrm{I}_{\rm fit}(L_i,l_j)\}_{i=1,\ldots,I; j=1,\ldots,J}$ for the annual temperature with $R=7$. Two black ``$\times$" represent grid points G1 and G2 with high $\mathrm{I}_{\rm fit}$ values. (b) shows annual temperature time series at G1 and G2, respectively, with their ensemble and evaluated mean.}
\label{Fig:fitanduqS}
\end{figure}

The map of $\{\mathrm{I}_{\rm fit}(L_i,l_j)\}_{i=1,\ldots,I; j=1,\ldots,J}$ in Fig.~S\ref{fig:subfig:Ifit_annual} reveals that for most grids, the evaluated mean $\hat m_t$ closely aligns with the ensemble mean $\Bar{y}_t$. However, in regions around $(56.00,65.00)\times(300.00,330.00)$ (parts of the Labrador Sea and North Atlantic) and $(-13.00,2.00)\times(12.00,20.00)$ (covering parts of the Democratic Republic of the Congo, Republic of the Congo, and Angola), the $\mathrm{I}_{\rm fit}$ values are notably higher. To delve deeper, Fig.~S\ref{fig:subfig:Check_Ifit} provides time series at two grid points G1=$(2.36,13.75)$ and G2=$(57.96,305.00)$ chosen from these areas. In comparison with the time series at GL and GO shown in Fig.~1(c), those in Fig.~S\ref{fig:subfig:Check_Ifit} exhibit more complex trends. For example, the temperature at G1 experiences a rapid increase from about 2055 to 2060, while the temperature at G2 sharply drops after 2040 and starts to rise from 2070. These nuances cannot be fully captured by the proposed $m_t$ and may require additional information. 

\subsubsection{Implementation of HCBG-SG}
We provide a brief illustration of constructing an SG for the annual surface temperature from CESM2-LENS2 using the method outlined in \citet{Huang'sEmulator}. With the exception of the temperature $y_t^{(r)}(L_i,l_j)$ and the mean trend $m_t$, the notations used in this subsection align with those in \citet{Huang'sEmulator} and are independent of those used elsewhere in this paper. This allows readers to compare parameter evaluation results with those presented in \citet{Huang'sEmulator} if they are interested. While the parameters shown in \citet{Huang'sEmulator} are for the monthly surface temperature from CESM-LENS1, they still share some commonalities with those presented here.

\begin{figure}[!ht]
\centering
\subfigure[$\hat\phi(L_i,l_j)$]{
\label{fig:subfig:Phi_Huang_Annual}
\includegraphics[scale=0.5]{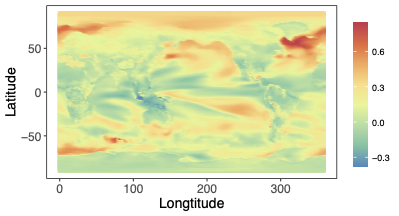}}
\subfigure[$\hat\sigma(L_i,l_j)$]{
\label{fig:subfig:Sig_Huang_Annual}
\includegraphics[scale=0.5]{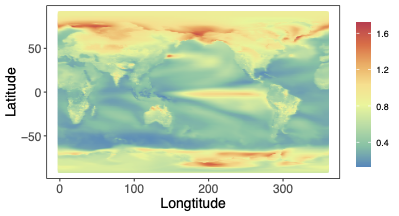}}\\
\subfigure[$\hat{\bm\theta}_i^{\rm land}$ and $\hat{\bm\theta}_i^{\rm ocean}$]{
\label{fig:subfig:Theta_Huang_Annual}
\includegraphics[scale=0.6]{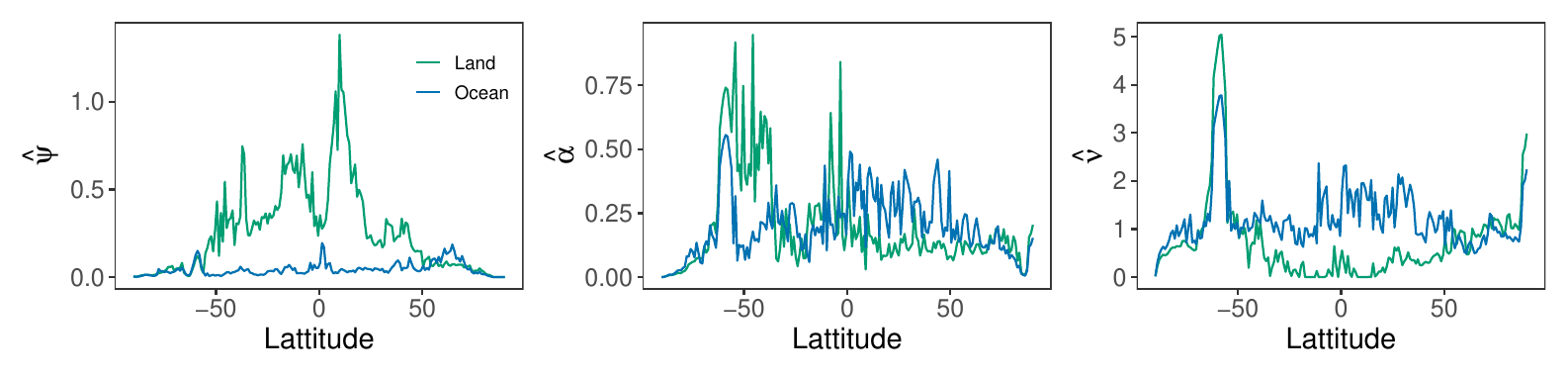}}
\caption{Inference results in the first step for building HCBG-SG. (a) and (b) are maps of the evaluated auto-correlation $\phi(L_i,l_j)$ and standard deviation $\sigma(L_i,l_j)$. (c) shows estimates of $\bm\theta_i^{\rm land}$ and $\bm\theta_i^{\rm ocean}$ for each latitude.}
\label{Fig:Mean_Huang_Annual}
\end{figure}

The approach of \citet{Huang'sEmulator} involves three main steps. First, they evaluate the mean trend and temporal dependence by assuming that $\{y_t^{(r)}(L_i,l_j)\}_{i=1,\ldots,I; j=1,\ldots,J; t=1,\ldots,T; r=1,\ldots,R}$ satisfies
\begin{equation*}
y_t^{(r)}(L_i,l_j)=m_t(L_i,l_j)+\epsilon_t^{(r)}(L_i,l_j),
\end{equation*}
where $\epsilon_t^{(r)}(L_i,l_j)=\phi(L_i,l_j)\epsilon_{t-1}^{(r)}(L_i,l_j)+\sigma(L_i,l_j)\eta_{t}^{(r)}(L_i,l_j)$. They then simplify the computation by ignoring the spatial dependence among $\eta_{t}^{(r)}(L_i,l_j)$ at different grid points and proceed to evaluate parameters in $m_t$, $\phi(L_i,l_j)$, and $\sigma(L_i,l_j)$ at each grid point in parallel. Similar to \citet{Huang'sEmulator}, we present the evaluated $\phi(L_i,l_j)$ and $\sigma(L_i,l_j)$ in Figs.~S\ref{fig:subfig:Phi_Huang_Annual} and S\ref{fig:subfig:Sig_Huang_Annual}, where lower auto-correlation and higher variation over land. There are $6IJ$ parameters in this step. 

Second, they evaluate the spatial dependence of $\eta(L_i,l_j)$ across longitudes for each latitude. Specifically, at latitude $L_i$, they expand $\eta_t^{(r)}(L_i,l_j)$ as 
\begin{equation*}
    \eta_t^{(r)}(L_i,l_j)=J^{-1/2}\sum_{c=0}^{J-1}\{f(c,l_j;\bm\theta_i^{\rm land},\bm\theta_i^{\rm ocean})\}^{1/2}\exp(2\pi\iota cj/J)\tilde{\eta}_t^{(r)}(L_i,c),
\end{equation*}
where $c$ is a wave number, $f(c,l_j;\bm\theta_i^{\rm land},\bm\theta_i^{\rm ocean})=\delta_{(L_i,l_j)\in\mathcal{S}_l}f(c;\bm\theta_i^{\rm land})+\{1-\delta_{(L_i,l_j)\in\mathcal{S}_l}\}f(c;\bm\theta_i^{\rm ocean})$ with $\bm\theta=(\psi,\alpha,\nu)$ and Kronecker delta $\delta_{(L_i,l_j)\in\mathcal{S}_l}$ for identifying grid points on land. It is further assumed that $f(c;\bm\theta_i^{\rm land})=\psi_i^{\rm land}/[\{{(\alpha_i^{\rm land})^2 +4\sin^2(c\pi/J)}\}^{\nu_i^{\rm land}+1/2}]$, and a similar expression is used for $f(c;\bm\theta_i^{\rm ocean})$. The evaluation of $\bm\theta_i^{\rm land}$ and $\bm\theta_i^{\rm ocean}$ is presented in Fig.~S\ref{fig:subfig:Theta_Huang_Annual}. This step involves $6I$ parameters.

Third, they evaluate the dependence across latitudes. Assume that the coherence $$\mathrm{corr}\{\tilde{\eta}_t^{(r)}(L_i,c),\tilde{\eta}_{t'}^{(r)}(L_{i'},c')\}=\delta_{c=c',t=t'}\rho_{L_i,L_{i'}}(c)=[\xi/\{1+4\sin^2(c\pi/J)\}^{\kappa}]^{|i-i'|},$$ where $\xi\in(0,1)$ and $\kappa>0$ control the overall coherence decay rate with larger latitude differences and the faster coherence decay for larger $c$, respectively. The estimated values are $\hat\xi=0.93$ and $\hat\kappa=0.50$. Both values need to be stored.

For monthly and daily aggregated simulations, the residuals $\{\epsilon_t^{(r)}(L_i,l_j)\}_{t=1,\ldots,T; r=1,\ldots,R}$ may deviate from normality. Consequently, a TGH auto-regressive model, incorporating three additional parameters to facilitate the Gaussianization of residuals, is required at each grid point.

\subsubsection{Basic statistical characteristics}
\begin{figure}[!b]
\centering
\subfigure[$|\Bar{y}_{SG,9}(L_i,l_j)-\Bar{y}_{A,9}(L_i,l_j)|$]{
\label{fig:subfig:GenAnnualmean2023}
\includegraphics[scale=0.5]{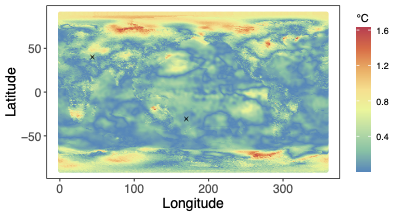}}
\subfigure[$|\Bar{y}_{SG,69}(L_i,l_j)-\Bar{y}_{A,69}(L_i,l_j)|$]{
\label{fig:subfig:GenAnnualmean2083}
\includegraphics[scale=0.5]{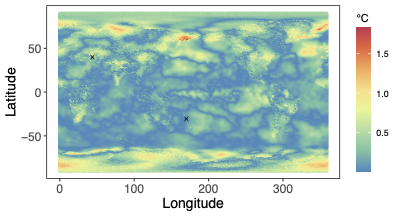}}\\
\subfigure[$|y_{SG,9}^{\rm sd}(L_i,l_j)-y_{A,9}^{\rm sd}(L_i,l_j)|$]{
\label{fig:subfig:GenAnnualsd2023}
\includegraphics[scale=0.5]{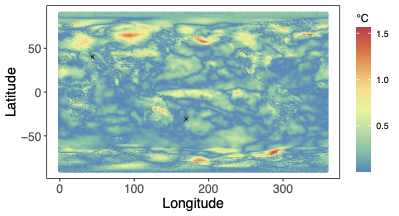}}
\subfigure[$|y_{SG,69}^{\rm sd}(L_i,l_j)-y_{A,69}^{\rm sd}(L_i,l_j)|$]{
\label{fig:subfig:GenAnnualsd2083}
\includegraphics[scale=0.5]{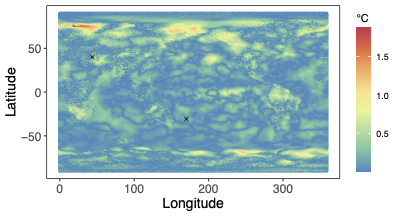}}\\
\subfigure[Time series at GO]{
\label{fig:subfig:GenAnnualTemporalO}
\includegraphics[scale=0.5]{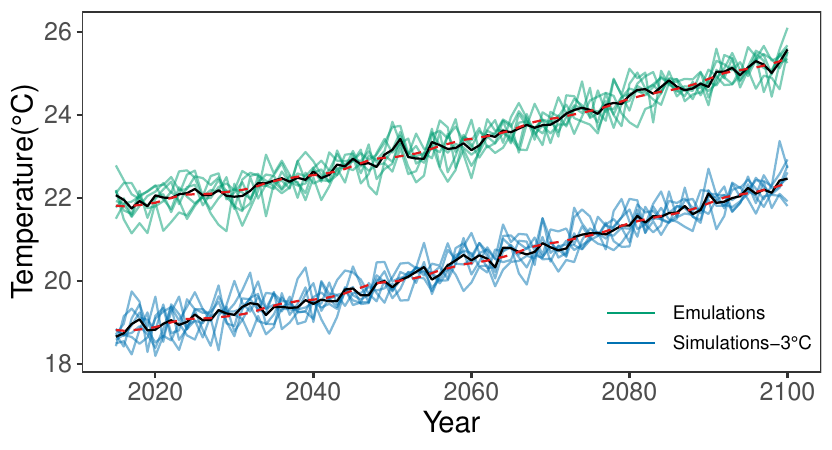}}
\subfigure[Time series at GL]{
\label{fig:subfig:GenAnnualTemporalL}
\includegraphics[scale=0.5]{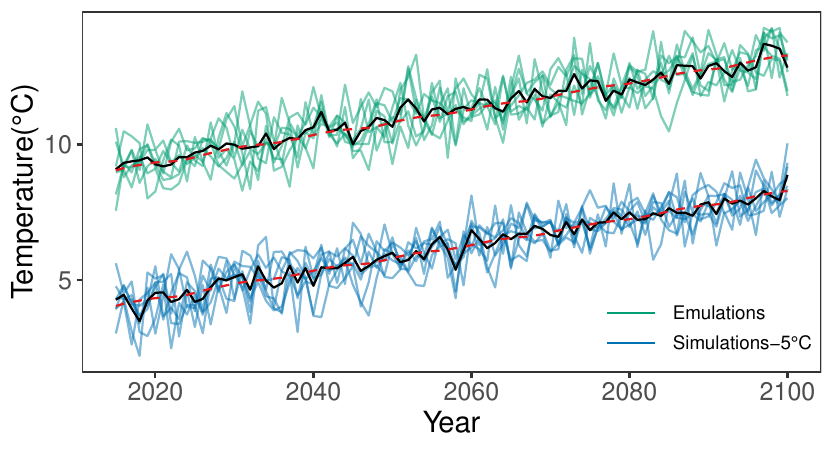}}\\
\caption{Comparison between annual emulations generated by the proposed SG and the training simulations. (a) and (b) show the difference between the ensemble mean of emulations and that of simulations for the years 2023 and 2083, respectively. (b) and (d) show the difference between the ensemble standard deviation of emulations and that of simulations for years 2023 and 2083, respectively. (e) and (f) depict emulations and simulations at grid points GO and GL. The black solid curves are ensemble means for simulations and emulations. The red dashed curves represent $\hat m_t$(GO) and $\hat m_t$(GL).}
\label{Fig:emulation_annual_supplement}
\end{figure}

We replicate the procedures outlined in Figs.~1(a)--1(c) to compare some basic statistical characteristics of our annual emulations to those of simulations. Figs.~S\ref{fig:subfig:GenAnnualmean2023} and S\ref{fig:subfig:GenAnnualsd2023} compare ensemble means $\Bar{y}_{SG,t}(L_i,l_j)=R^{-1}\sum_{r'=1}^R y_{SG,t}^{(r')}(L_i,l_j)$ and standard deviations $y_{SG,t}^{\rm sd}(L_i,l_j)=\{R^{-1}\sum_{r'=1}^R(y_t^{(r')}(L_i,l_j)-\Bar{y}_{SG,t}(L_i,l_j))^2\}^{1/2}$ of emulations with those of simulations at year 2023, i.e., $t=9$. Results at another randomly selected year are shown in Figs.~S\ref{fig:subfig:GenAnnualmean2083} and S\ref{fig:subfig:GenAnnualsd2083} to further aid in examining random variation from underlying issues with our emulations. Both the ensemble means and standard errors of emulations closely align with those of simulations at most grid points, except for some areas near the North and South Pole. Figs.~S\ref{fig:subfig:GenAnnualTemporalO} and S\ref{fig:subfig:GenAnnualTemporalL} provide time series of emulations and (shifted) simulations at two grid points. In both cases, the $\hat m_t$ fits the ensemble mean $\Bar{y}_t$ well, and emulations and simulations exhibit similar variability.

\begin{figure}[!ht]
\centering
\subfigure[$L_i=-11.8^\circ$]{
\label{fig:subfig:Periodogram_118}
\includegraphics[scale=0.5]{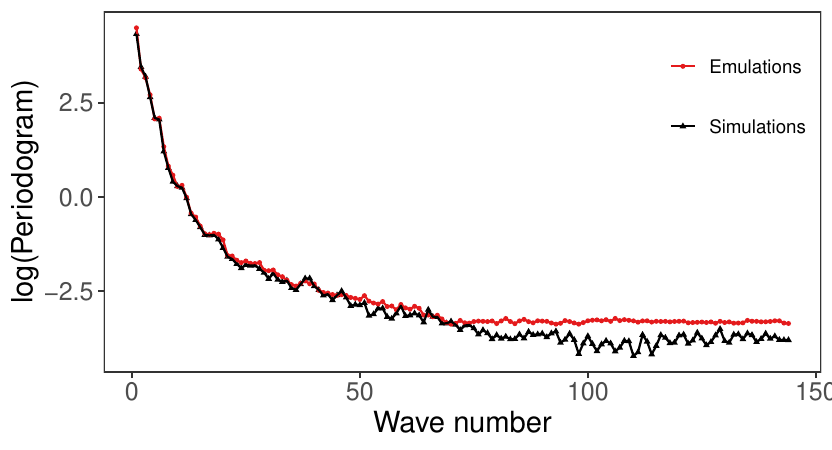}}
\subfigure[$L_i=36.3^\circ$]{
\label{fig:subfig:Periodogram_363}
\includegraphics[scale=0.5]{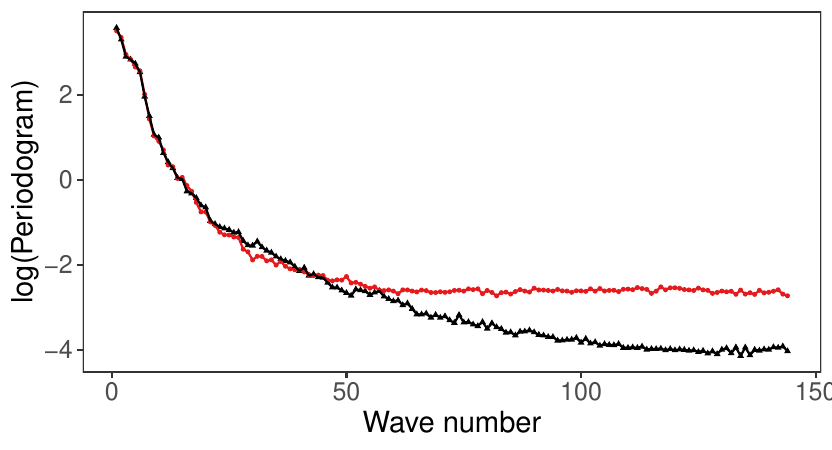}}\\
\caption{Logarithmic periodograms for emulations and simulations at (a) $L_i=-11.8^\circ$ and (b) $L_i=36.3^\circ$.}
\label{Fig:Periodogram_Annual}
\end{figure}

We further examine whether emulations efficiently capture the main spatial dependence structure in simulations by comparing their longitudinal periodograms. Specifically, for a given latitude $L_i$, the longitudinal periodogram at wave number $c$ is calculated as $\sum_{j=0}^{J-1}\exp(-2\pi\iota cj/J)$ $\mathrm{cov}\{Z(L_i,l_1),Z(L_i,l_{j+1})\}$. Fig.~\ref{Fig:Periodogram_Annual} illustrates the empirically evaluated periodograms of emulations and simulations, revealing a close alignment at lower wave numbers that gradually diverges beyond a certain $c$ between $Q_l$ and $Q_o$. This observation aligns with Model~(5), where we retain the first $Q_l$ and $Q_o$ terms for grid points on land and ocean, respectively. These terms serve as low-rank approximations, capturing the large-scale spatial structure of $Z(L_i,l_j)$. The remaining $Q_{\max}^2-Q_l^2$ or $Q_{\max}^2-Q_o^2$ terms are then allocated to be the nugget $\varepsilon(L_i,l_j)$ with variance $v^2(L_i,l_j)$, accounting for fine-scale spatial structure. Therefore, the selection of $Q_l$ and $Q_o$ is crucial in achieving a balance between different scales. Consider other popular low-rank approximations such as LatticeKrig mentioned in Section~S3.2.4. Their basis functions contain unknown parameters that control their smoothness, range, and variance. Given the number of bases, the estimates of these parameters, along with the nugget variance, are typically achieved by maximizing likelihood functions. This process  distinguishes between large- and fine-scale structures. In contrast, spherical harmonics have a known form and require only $Q_l$ and $Q_o$ values. In this work, boxplots of $\{\mathrm{BIC}_{*,t}^{(r)}(Q)\}_{r=1,\ldots,R; t=1,\ldots,T}$ are employed to determine $Q_l$ and $Q_o$, considering the number of parameters in addition to the likelihood functions. These help prevent the low-rank approximation from struggling to describe fine-scale spatial dependence and avoid a potential limitation in low-rank approximations \citep{STEIN20141}.

\subsection{Monthly aggregated temperature}
\label{sec:subsec:supplement_monthly}
\begin{figure}[!b]
\vspace{-10pt}
\centering
\subfigure[$|\hat m_{105}(L_i,l_j)-\bar y_{105}(L_i,l_j)|$]{
\label{fig:subfig:goodnessoffit}
\includegraphics[scale=0.5]{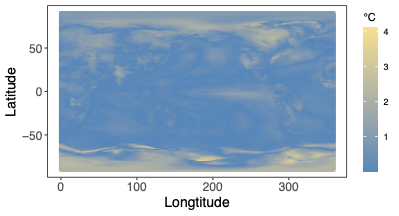}}
\subfigure[$\hat{\sigma}(L_i,l_j)$]{
\label{fig:subfig:Sig_monthly}
\includegraphics[scale=0.5]{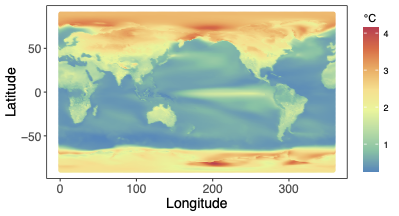}}\\
\caption{Evaluation of the deterministic component of monthly simulations. (a) shows the gap between the estimated mean and ensemble mean of the monthly aggregated temperature simulations for September, 2023. (b) is the map of the estimated standard deviation $\hat\sigma(L_i,l_j)$ for the monthly aggregated temperature.}
\label{Fig:Supplement_monthly}
\vspace{-10pt}
\end{figure}

Let $\{y_t^{(r)}(L_i,l_j)\}_{i=1,\ldots,I; j=1,\ldots,J; t=1,\ldots,T; r=1,\ldots,R}$ represent the monthly aggregated temperature simulations with $T=1032$. In this subsection, we focus on the impact of increasing temporal resolution. We start by modeling the deterministic component, where $K_M$ harmonic terms are added in $m_t(L_i,l_j)$ to account for interannual trends. Based on BIC, as shown in Fig.~\ref{fig:subfig:Monthly_BIC}, the proportions of selecting $K_M=1,\ldots,5$ are $0\%$, $16\%$, $34\%$, $34\%$ and $16\%$, respectively. For the sake of efficient storage and analysis,  we choose $K_M=3$. From Figs.~S\ref{fig:subfig:Skew_monthly} and S\ref{fig:subfig:Kurt_monthly}, the stochastic components at most grid points do not conform to a Gaussian distribution. Consequently, the index $\mathrm{I}_{\rm fit}$ is no longer an appropriate measure for assessing the goodness-of-fit of the SG \citep{castruccio2019reproducing}. Therefore, we randomly choose a time point $105$ and compare the difference between $\hat m_{105}$ and the ensemble mean at $t=105$ in Fig.~S\ref{fig:subfig:goodnessoffit}. The proposed SG can fit most regions well except for two poles and the Band regions because of the numerical instabilities. We also show the map of $\hat \sigma(L_i,l_j)$ in Fig.~S\ref{fig:subfig:Sig_monthly} and compare it with that in Fig.~3(b). They have similar patterns but the variation of monthly temperature is larger.

\begin{figure}[!b]
\centering
\subfigure[BIC]{
\label{fig:subfig:BIC_monthly}
\includegraphics[scale=0.4]{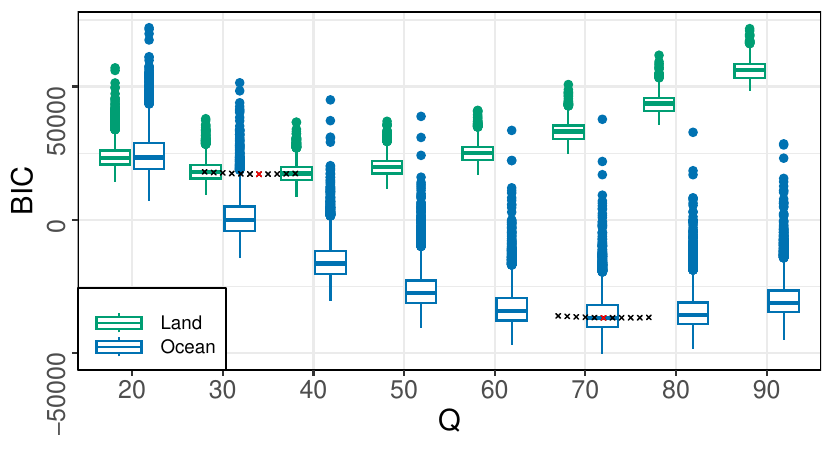}}
\subfigure[$\hat v(L_i,l_j)$]{
\label{fig:subfig:v2hat_monthly}
\includegraphics[scale=0.4]{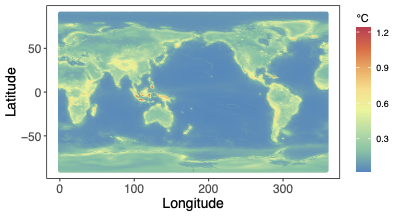}}
% \subfigure[Jarque-Bera test]{
% \label{fig:subfig:Bera_spectral}
% \includegraphics[scale=0.35]{Fig_LENS2/Bera_spectral.pdf}}
\subfigure[$(\hat\phi_1)_q^m$ for monthly data]{
\label{fig:subfig:TPhihat_monthly_Tukey}
\includegraphics[scale=0.4]{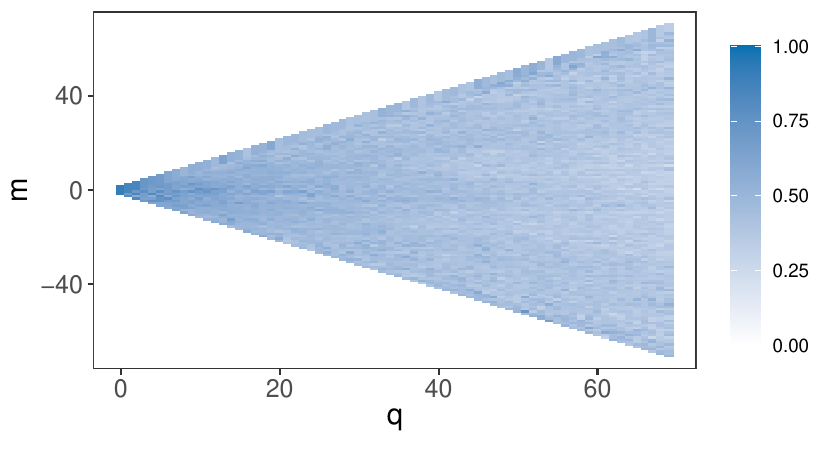}}
\caption{Evaluation of the stochastic component of monthly simulations. (a) shows boxplots of $\{\mathrm{BIC}_{*,t}^{(r)}(Q)\}_{r=1,\ldots,R; t=1,\ldots,T}$ for the monthly temperature against different values of $Q$, where $*$ takes ``l" and ``o". Points $\times$ show median of BIC values. Red Points $\times$ indicate the minimum BIC medians. (b) is the map of $\hat v(L_i,l_j)$ for the monthly data. (c) is the map of the estimated $\{(\phi_1)_q^m\}_{q=0,\ldots,Q_o-1; m=-q,\ldots,q}$ for the monthly aggregated simulations.}
\label{Fig:time_monthly}
\end{figure}
After obtaining the stochastic component, we apply Model~(5) to reduce its dimension. By BIC shown in Fig.~S\ref{fig:subfig:BIC_monthly}, we choose $Q_l=36$ and $Q_o=70$ for the monthly data. These values are very close to those of the annual data and remain unaffected by the increase in temporal resolution. Correspondingly, the estimate of $v(L_i,l_j)$ is also provided in Fig.~S\ref{fig:subfig:v2hat_monthly}, which has similar pattern and scale to Fig.~3(d). Note that aside from regions with high altitudes and complex topography, certain coastlines display larger $\hat v(L_i,l_j)$ values. It means that our low-rank approximation encounters difficulties in capturing intricate details along these coastlines. As the temporal resolution increases, the temperature fluctuations on these transition points become more pronounced. This leads to a lack of spatial continuity, posing a challenge for SHT, whereas has not much impact on the performance of emulations regarding to $\mathrm{I}_{\rm uq}$ and $\mathrm{WD}_T$ shown in following Figs.~S\ref{fig:subfig:Iuq_monthly_Tukey} and S\ref{fig:subfig:WD_monthly_Tukey}. Now, we assess the temporal dependence in the spectral domain using $(\tilde{s}_t^{(r)})_q^m$'s. 
% From Fig.~S\ref{fig:subfig:Bera_spectral}, only $42.6\%$ coefficients reject the Gaussianity and belong to $\mathcal{S}_{gh}$, although $\{Z_t^{(r)}(L_i,l_j)\}_{t=1,\ldots,T; r=1,\ldots,R}$ at most grid points are not Gaussian. It implies that the non-Gaussianity of monthly data is not severe. 
The proportions of $P=1,\ldots,5$ in the TGH autoregressive model are $87.3\%$, $8.8\%$, $3.4\%$, and $0.3\%$ and $0.2\%$, respectively. Therefore, we choose $P=1$ for the monthly data. The corresponding $\{(\hat\phi_1)_q^m\}_{q=0,\ldots,Q_o-1; m=-q,\ldots,q}$ are displayed the  in Fig.~S\ref{fig:subfig:TPhihat_monthly_Tukey}, which have an obvious trend compared with those for annual data in Fig.~4(a). That is, the lower the degree, the stronger the temporal correlation. The assessment of spatial dependence is easily performed by evaluating $\check{\mathbf{K}}$ with $(\check{s}_t^{(r)})_q^m$.

\begin{figure}[!b]
\centering
\subfigure[$\mathrm{I}_{\rm uq}(L_i,l_j)$ for SHT-SG(with TGH)]{
\label{fig:subfig:Iuq_monthly_Tukey}
\includegraphics[scale=0.5]{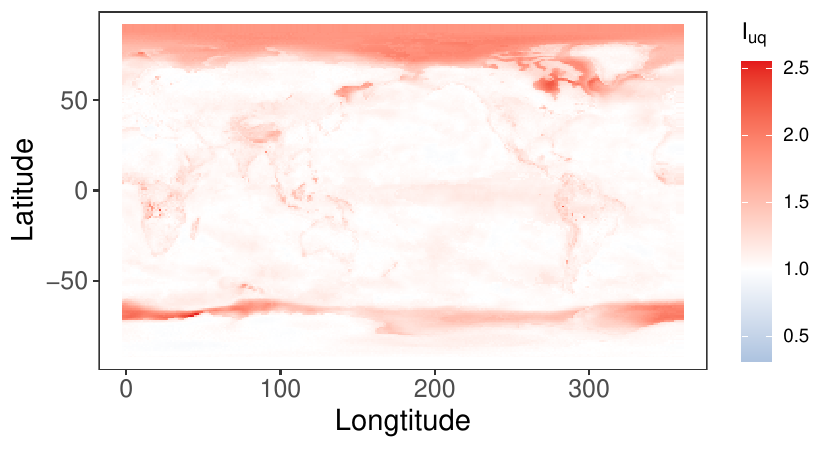}}
\subfigure[$\mathrm{I}_{\rm uq}(L_i,l_j)$ for HCBG-SG]{
\label{fig:subfig:Iuq_monthly_Huang}
\includegraphics[scale=0.5]{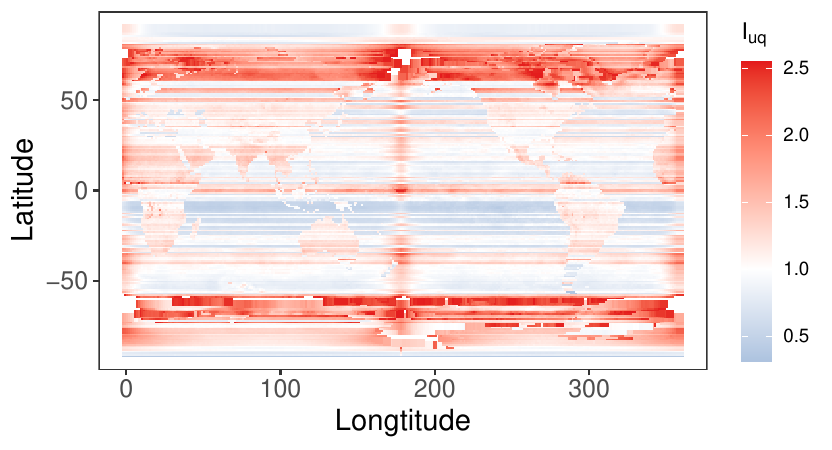}}\\
\subfigure[$\mathrm{WD}_{S}(L_i,l_j)$ for SHT-SG(with TGH)]{
\label{fig:subfig:WD_monthly_Tukey}
\includegraphics[scale=0.5]{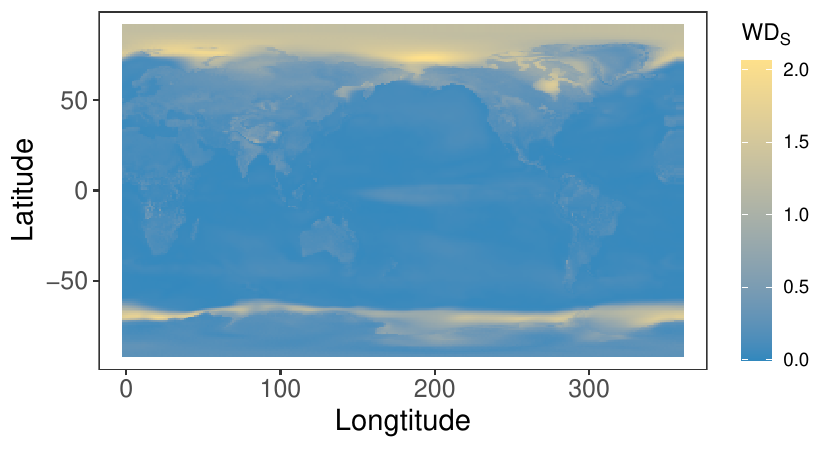}}
\subfigure[$\mathrm{WD}_{S}(L_i,l_j)$ for HCBG-SG]{
\label{fig:subfig:WD_monthly_Huang}
\includegraphics[scale=0.5]{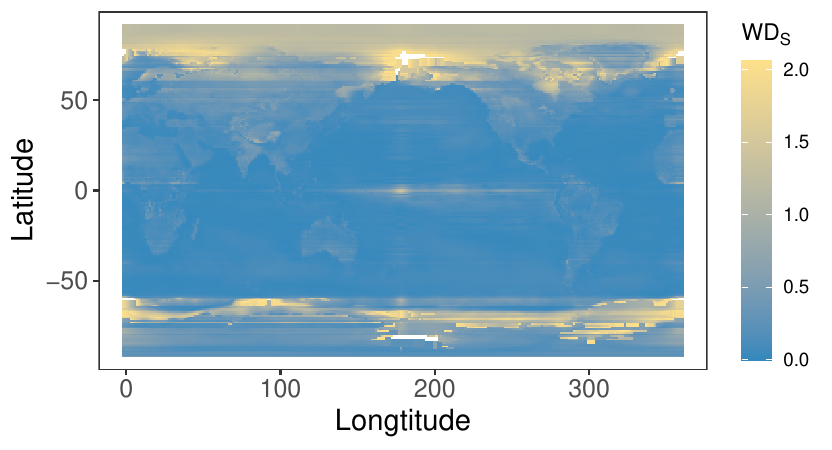}}\\
\subfigure[$\mathrm{I}_{\rm uq}(L_i,l_j)$ of SHT-SG]{
\label{fig:subfig:Box_Iuq_monthly}
\includegraphics[scale=0.5]{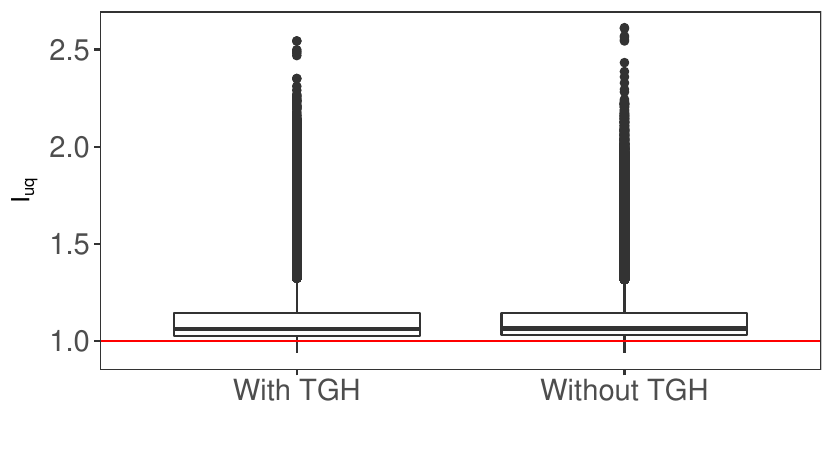}}
\subfigure[$\mathrm{WD}_{S}(L_i,l_j)$ of SHT-SG]{
\label{fig:subfig:Box_WD_monthly}
\includegraphics[scale=0.5]{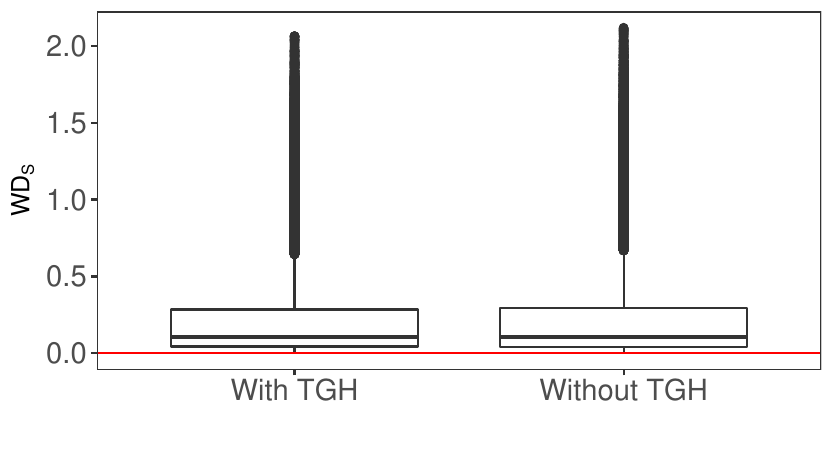}}\\
% \subfigure[$\{y_{t}^{(r)}(\rm GO)\}_{t,r}$ and $\{y_{SG,t}^{(r)}(\rm GO)\}_{t,r}$]{
% \label{fig:subfig:Hist1}
% \includegraphics[scale=0.5]{Fig_LENS2/Hist1.pdf}}
% \subfigure[$\{y_{555}^{(r)}(L_i,l_j)\}_{i,j,r}$ and $\{y_{SG,555}^{(r)}(L_i,l_j)\}_{i,j,r}$]{
% \label{fig:subfig:Hist2}
% \includegraphics[scale=0.5]{Fig_LENS2/Hist2.pdf}}
\caption{Performance assessment of $R'=7$ monthly emulations. (a) and (b) are maps of $\{\mathrm{I}_{\rm uq}(L_i,l_j)\}_{i=1,\ldots,I; j=1,\ldots,J}$ for SHT-SG(with TGH) and HCBG-SG, respectively. (c) and (d) are maps of $\{\mathrm{WD}_{S}(L_i,l_j)\}_{i=1,\ldots,I; j=1,\ldots,J}$ for SHT-SG(with TGH) and HCBG-SG, respectively. (e) and (f) are boxplots of $\{\mathrm{I}_{\rm uq}(L_i,l_j)\}_{i=1,\ldots,I; j=1,\ldots,J}$ and $\{\mathrm{WD}_{S}(L_i,l_j)\}_{i=1,\ldots,I; j=1,\ldots,J}$ for comparing the proposed SG with and without a modified TGH.}
\label{Fig:emulations_monthly}
\end{figure}

We generate $R'=7$ monthly emulations using the proposed SG with a modified TGH (referred to as SHT-SG(with TGH)). Additionally, we consider the HCBG-SG and the proposed SG without TGH (referred to as SHT-SG(without TGH)) as rivals. Fig.~\ref{Fig:emulations_monthly} illustrates their performance. In Figs.~S\ref{fig:subfig:Iuq_monthly_Tukey} and S\ref{fig:subfig:WD_monthly_Tukey}, $\mathrm{I}_{\rm uq}$ values (median: $1.060$) are close to $1$ and $\mathrm{WD}_S$ values (median: $0.106$) are close to $0$ at most grid points. This suggests that the proposed SG performs well for monthly simulations. Emulations on regions with high altitudes and complex topography exhibit higher variability. Note that we remove some extremely large $\mathrm{I}_{\rm uq}$ and $\mathrm{WD}_S$ from Figs.~S\ref{fig:subfig:Iuq_monthly_Huang} and S\ref{fig:subfig:WD_monthly_Huang} for better demonstration and comparison. Similar to the annual case, $\mathrm{I}_{\rm uq}$ and $\mathrm{WD}_S$ for HCBG-SG exhibit an obvious spatial discontinuity. The medians of $\mathrm{WD}_T$ for SHT-SG and HCBG-SG are $0.369$ and $0.454$, respectively.  Both SGs encounter challenges in the North Pole and Band regions due to the numerical instabilities. However, the proposed SG demonstrates greater robustness.
% Intuitively, emulations in Fig.~\ref{fig:subfig:Temporal_monthly} have similar variability and pattern to those in Fig.~\ref{fig:subfig:Mionthlytemporal}. Moreover, Figs.~S5(a) and S5(b) compare empirical distributions of emulations with those of simulations from different views, which can be measured using the Wasserstein distances $\mathrm{WD}_S$ and $\mathrm{WD}_T$, respectively. 
% The values of $\mathrm{WD}_T$ are also small, especially at time points in the middle. The medians of $\mathrm{WD}_T$ for emulations with and without TGH are $0.370$ and $0.369$, respectively. 
From Figs.~S\ref{fig:subfig:Box_Iuq_monthly} and S\ref{fig:subfig:Box_WD_monthly}, the performance of SHT-SG(without TGH) is comparable to that of SHT-SG(with TGH). On the one hand, as shown in Figs.~1 and \ref{Fig:SkewandKurt}, and discussed in Sections~2 and S2, skewness and heavy tailedness for the monthly temperature data may not be pronounced enough to significantly impact the performance of the SG without using the TGH. On the other hand, the incorporation of TGH may introduce additional uncertainty.

\begin{figure}[!b]
\centering
\subfigure[$\{y_{t}^{(r)}(\rm GO)\}_{t,r}$ and $\{y_{SG,t}^{(r)}(\rm GO)\}_{t,r}$]{
\label{fig:subfig:Hist1}
\includegraphics[scale=0.5]{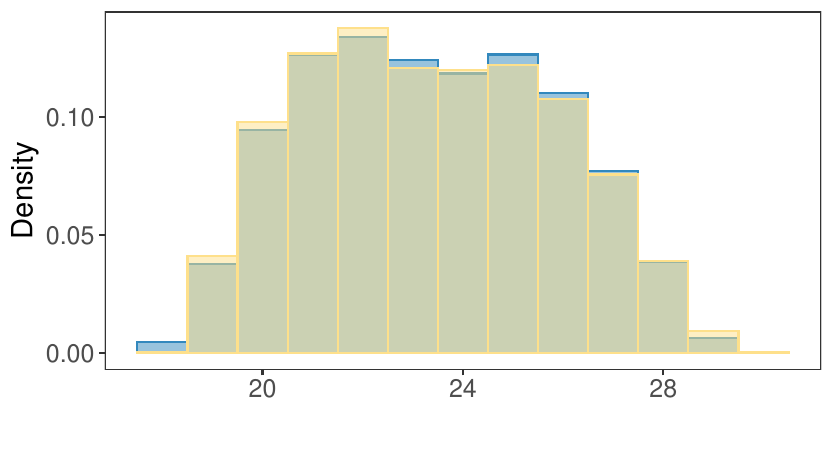}}
\subfigure[$\{y_{555}^{(r)}(L_i,l_j)\}_{i,j,r}$ and $\{y_{SG,555}^{(r)}(L_i,l_j)\}_{i,j,r}$]{
\label{fig:subfig:Hist2}
\includegraphics[scale=0.5]{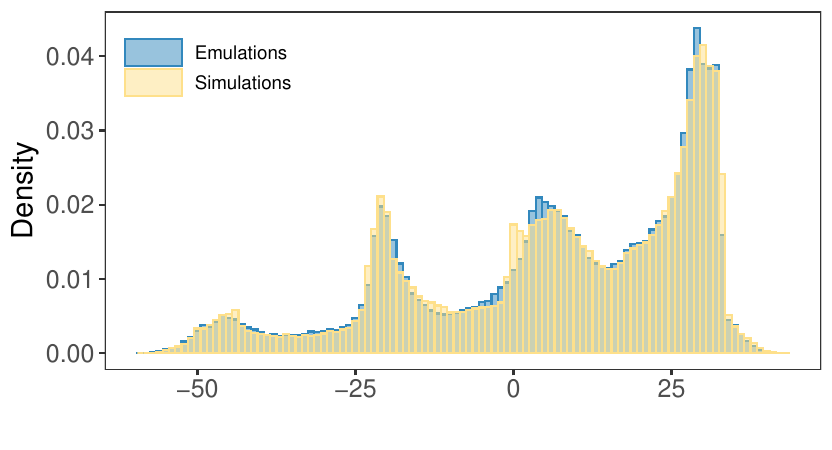}}\\
% \subfigure[$\mathrm{I}_{\rm uq}$]{
% \label{fig:subfig:Iuq_monthly_noTukey}
% \includegraphics[scale=0.5]{Fig_LENS2/Iuq_monthly_noTukey.pdf}}
% \subfigure[$\mathrm{WD}_{S}$]{
% \label{fig:subfig:WD_monthly_noTukey}
% \includegraphics[scale=0.5]{Fig_LENS2/WD_monthly_noTukey.pdf}}\\
\subfigure[Time series at GO]{
\label{fig:subfig:GenMonthlyTempO}
\includegraphics[scale=0.5]{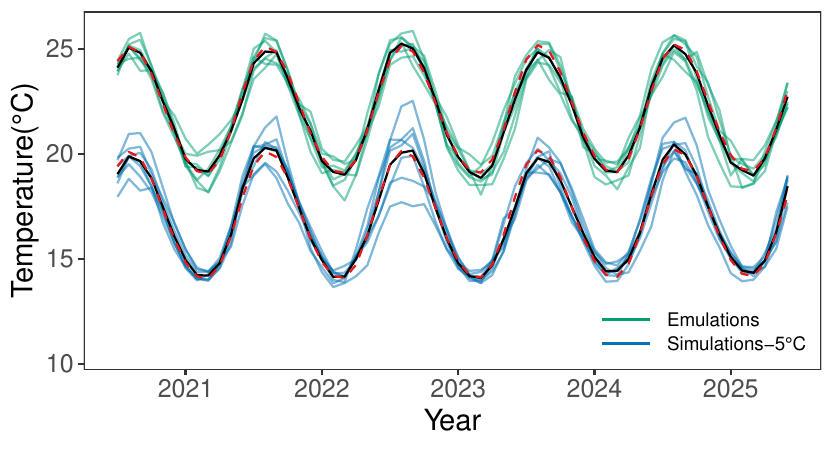}}
\subfigure[Time series at GL]{
\label{fig:subfig:GenMonthlyTempL}
\includegraphics[scale=0.5]{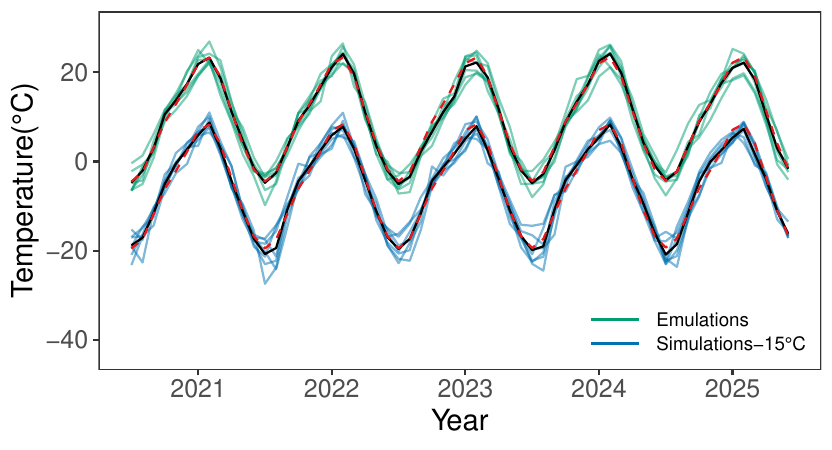}}\\
\caption{Comparison of some statistical characteristics between monthly emulations generated by SHT-SG(with TGH) and simulations. (a) displays the histogram of $\{y_t^{(r)}(\mathrm{GO})\}_{t=1,\ldots,T; r=1,\ldots,R}$, and compares it with that of $\{y_{SG,t}^{(r)}(\mathrm{GL})\}_{t=1,\ldots,T; r=1,\ldots,R}$. (b) displays histograms of $\{y_{555}^{(r)}(L_i,l_j)\}_{i=1,\ldots,I; j=1,\ldots,J; r=1,\ldots,R}$ and $\{y_{SG,555}^{(r)}(L_i,l_j)\}_{i=1,\ldots,I; j=1,\ldots,J; r=1,\ldots,R}$. (c) and (d) depict emulations and simulations at grid points Go and GL. The black solid curves are ensemble means for simulations and emulations. The red dashed curves are $\hat m_t$s.
% (c) and (d) are maps of $\{\mathrm{I}_{\rm uq}(L_i,l_j)\}_{i=1,\ldots,I; j=1,\ldots,J}$ and $\{\mathrm{WD}_{S}(L_i,l_j)\}_{i=1,\ldots,I; j=1,\ldots,J}$ for the monthly emulations generated without using TGH transformation, respectively. 
}
\label{Fig:emulations_monthly1}
\end{figure}
We further examine other statistical characteristics of our monthly emulations. Figs.~S\ref{fig:subfig:Hist1} and S\ref{fig:subfig:Hist2} compare their empirical distributions with those of simulations, which can be measured using the Wasserstein distances $\mathrm{WD}_S$ and $\mathrm{WD}_T$, respectively. We replicate the procedures outlined in Fig.~1(d) using the generated emulations. From Figs.~S\ref{fig:subfig:GenMonthlyTempO} and S\ref{fig:subfig:GenMonthlyTempL}, emulations can mimic simulations well, exhibiting similar patterns and larger variability at the peaks and valleys of the curves, despite the assumption that $\sigma(L_i,l_j)$ remains constant over time.

Another intriguing exploration involves annually aggregating the monthly emulations to obtain another set of annual emulations, denoted as $\{y_{MtoA,t}^{(r')}(L_i,l_j)\}_{i=1,\ldots,I; j=1,\ldots,J; t=1,\ldots,T; r'=1,\ldots,R}$. We compare these with the annual emulations directly generated in Section 4.1, denoted as $y_{SGA,t}^{(r)}(L_i,l_j)$ in this section. Fig.~\ref{Fig:emulations_monthly2} reveals that the aggregated emulations exhibit higher $\mathrm{I}_{\rm fit}$, $\mathrm{I}_{\rm uq}$, and $\mathrm{WD}_S$ values over land and two regions with numerical instabilities compared to the directly generated annual emulations. The poor performance of the annually aggregated emulations in the North Pole and Band regions is inherited from the monthly emulations and is a consequence of the inadequate model for data with numerical instabilities. Several factors contribute to the subpar performance on land. Firstly, the $\hat\sigma$ on land may be overestimated. On the one hand, it stems from an inadequate evaluation of $m_t$ on land, where the temporal dynamics are more complex. Although we select $K_M=3$ for all grid points, Fig.~S\ref{fig:subfig:Monthly_BIC} suggests that larger values of $K_M$ are needed for land points. On the other hand, it stems from an idealistic assumption of a constant $\sigma$ for monthly data. For example, Fig.~1(d) reveals larger standard errors at peaks and valleys of the time series curves. Secondly, additional procedures in the monthly SG, such as developing a modified TGH transformation, may introduce further errors. Finally, the larger $\hat\sigma$ on land may amplify both variability and potential errors in the stochastic component. These issues are not readily apparent when directly comparing monthly emulations with simulations, as monthly data exhibit larger variations that can obscure these issues. However, after aggregation, when the scale of variation of the data diminishes, these issues become more pronounced.
\begin{figure}[!h]
\centering
\subfigure[$\mathrm{I}_{\rm uq}(L_i,l_j)$ for $y_{MtoA,t}^{(r)}(L_i,l_j)$]{
\label{fig:subfig:Iuq_Monthly_to_Annual}
\includegraphics[scale=0.5]{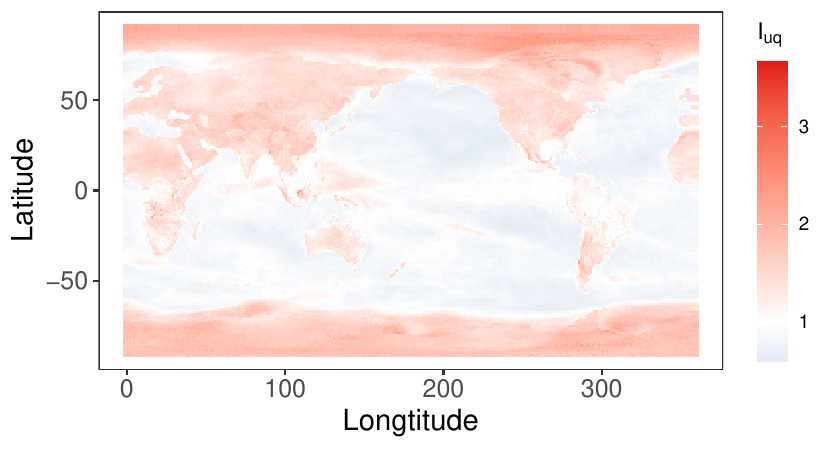}}
\subfigure[$\mathrm{I}_{\rm uq}(L_i,l_j)$ for $y_{SGA,t}^{(r)}(L_i,l_j)$]{
\label{fig:subfig:Iuq_Annual_B}
\includegraphics[scale=0.5]{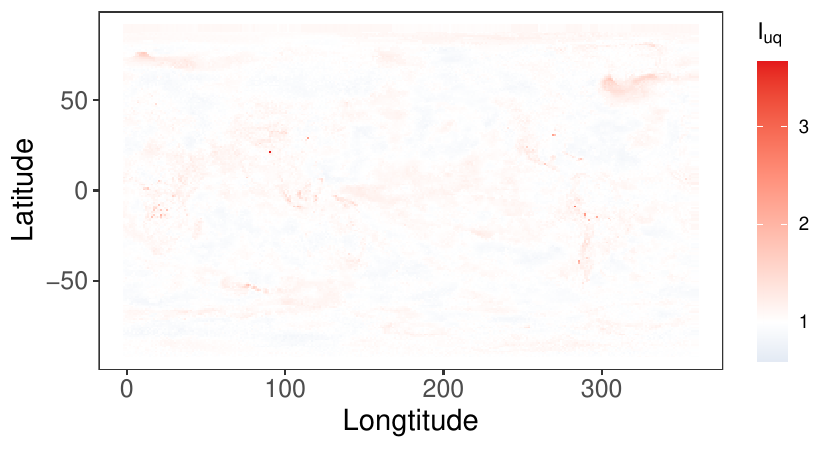}}\\
\subfigure[$\mathrm{WD}_{S}(L_i,l_j)$ for $y_{MtoA,t}^{(r)}(L_i,l_j)$]{
\label{fig:subfig:WD_Monthly_to_Annual}
\includegraphics[scale=0.5]{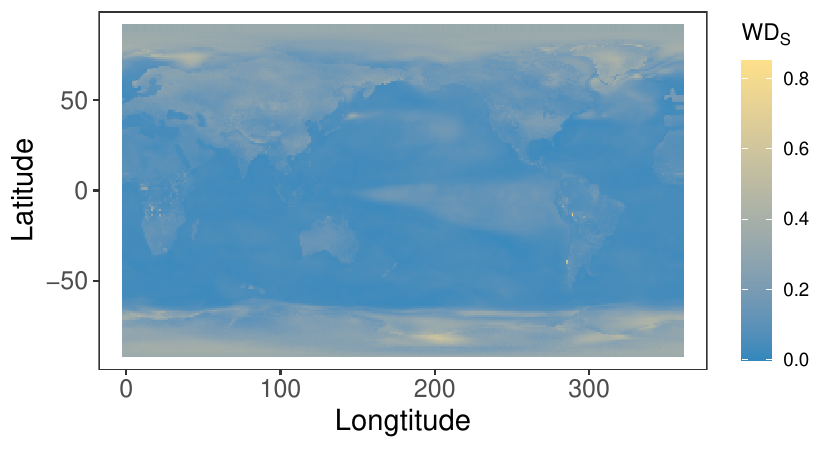}}
\subfigure[$\mathrm{WD}_{S}(L_i,l_j)$ for $y_{SGA,t}^{(r)}(L_i,l_j)$]{
\label{fig:subfig:WD_Annual_B}
\includegraphics[scale=0.5]{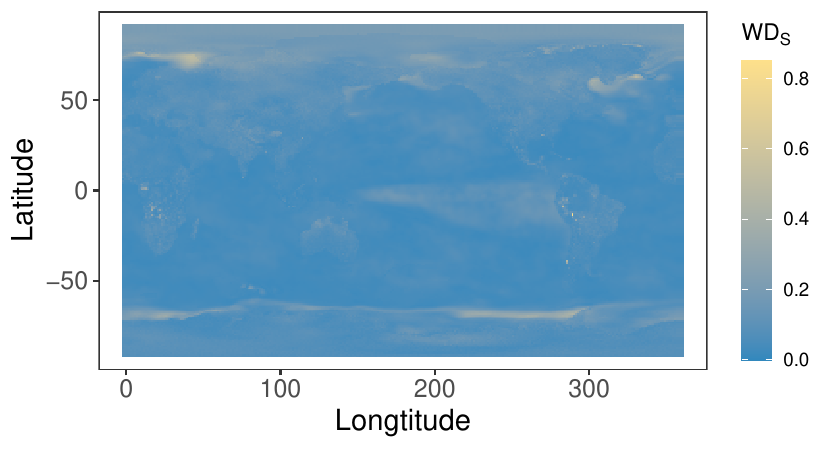}}\\
\subfigure[$\mathrm{I}_{\rm fit}(L_i,l_j)$ for $y_{MtoA,t}^{(r)}(L_i,l_j)$]{
\label{fig:subfig:Ifit_Monthly_to_Annual}
\includegraphics[scale=0.5]{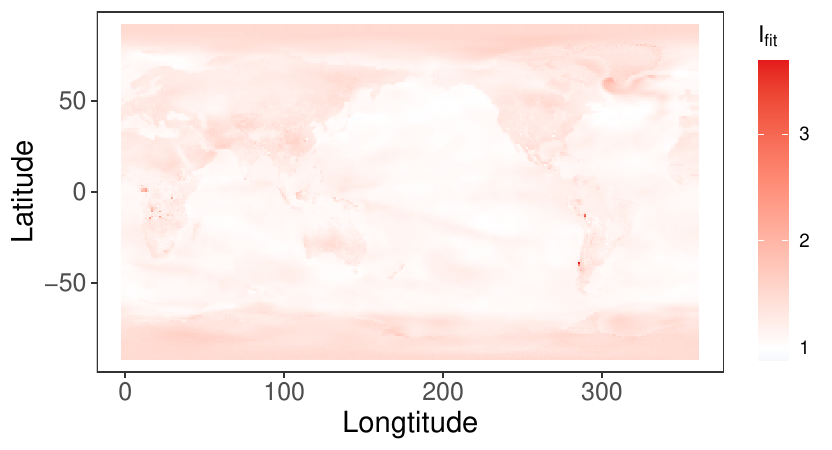}}
\subfigure[$K_M$]{
\label{fig:subfig:Monthly_BIC}
\includegraphics[scale=0.5]{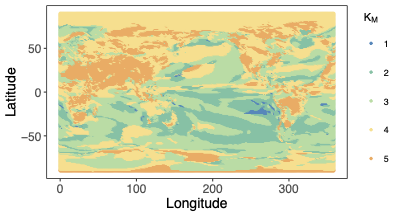}}\\
\caption{Performance assessment for the annually aggregated monthly emulations. (a), (c), and (e) show maps of  $\mathrm{I}_{\rm uq}$, $\mathrm{WD}_S$, and $\mathrm{I}_{\rm fit}$ values for the annually aggregated monthly emulations, where the $\mathrm{I}_{\rm fit}$ compares the ensemble mean of $\{y_{MtoA,t}^{(r)}(L_i,l_j)\}_{i=1,\ldots,I; j=1,\ldots,J; t=1,\ldots,T; r=1,\ldots,R}$ to that of simulations. (b) and (d) depict $\mathrm{I}_{\rm uq}$ and $\mathrm{WD}_S$ values for the annual emulations in Section~4.1, which have different color legends compared to Fig.~5, for clear illustration. (f) provides the $K_M$ chosen by BIC for each grid point.}
\label{Fig:emulations_monthly2}
\end{figure}

\subsection{Daily aggregated temperature}
\label{sec:subsec:supplement_daily}
\subsubsection{Development of the daily SG}
\begin{figure}[!b]
\centering
\subfigure[$|\hat m_{474}(L_i,l_j)-\Bar{y}_{474}(L_i,l_j)|$]{
\label{fig:subfig:goodfit_daily}
\includegraphics[scale=0.5]{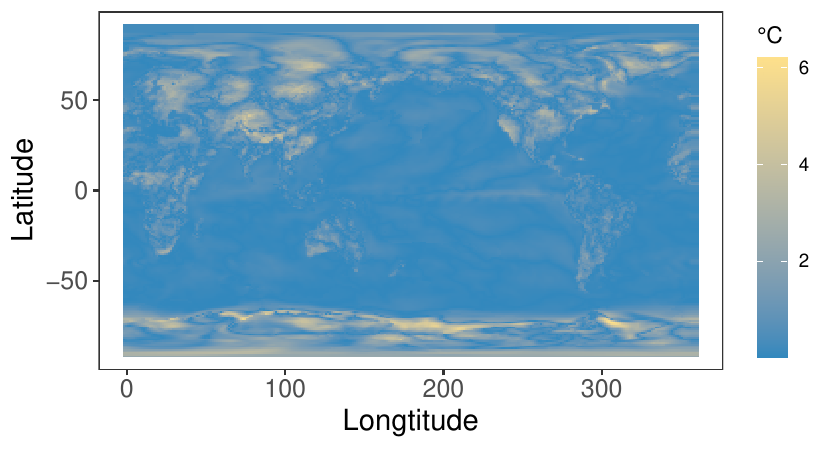}}
\subfigure[$\hat{\sigma}(L_i,l_j)$]{
\label{fig:subfig:Sig_daily}
\includegraphics[scale=0.5]{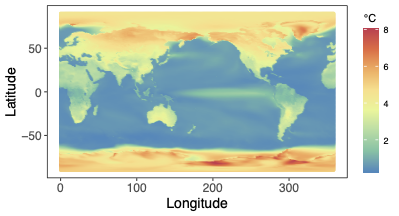}}\\
%\subfigure[Jarque-Bera test]{
%\label{fig:subfig:Bera_Daily}
%\includegraphics[scale=0.5]{Fig_LENS2/Bera_Daily.pdf}}\\
\subfigure[BIC for choosing $Q_l$ and $Q_o$]{
\label{fig:subfig:BIC_daily}
\includegraphics[scale=0.5]{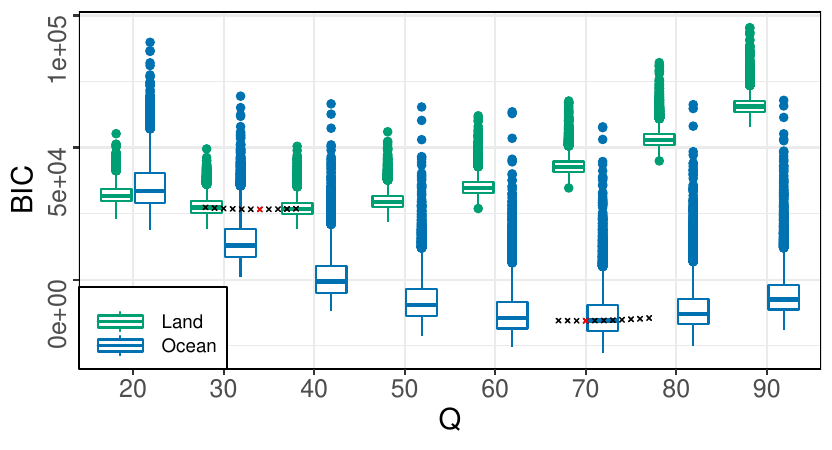}}
\subfigure[$\hat v(L_i,l_j)$]{
\label{fig:subfig:v2hat_daily}
\includegraphics[scale=0.5]{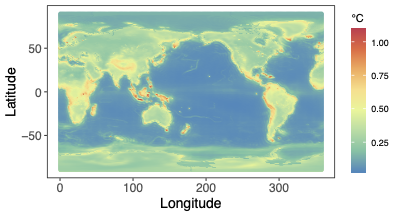}}\\
\subfigure[BIC for choosing $P$]{
\label{fig:subfig:BICp_Tukey}
\includegraphics[scale=0.5]{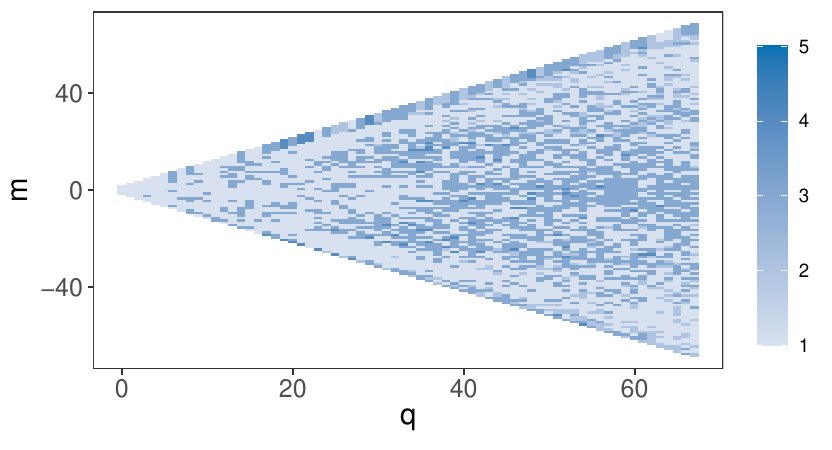}}
\subfigure[$(\hat\phi_1)_q^m$ for daily data]{
\label{fig:subfig:phihat_Tukey_daily}
\includegraphics[scale=0.5]{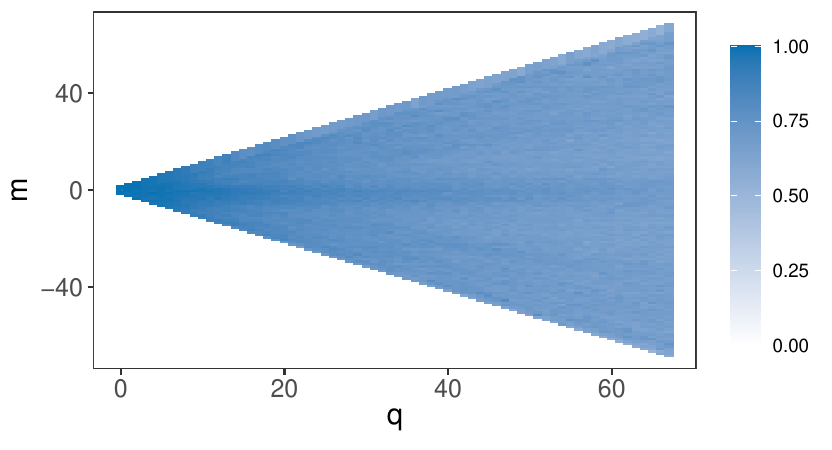}}\\
% \subfigure[$\mathrm{WD}_T$]{
% \label{fig:subfig:WDspace_daily}
% \includegraphics[scale=0.5]{Fig_LENS2/WDspace_daily.pdf}}\\
% \subfigure[$\mathrm{WD}_S$ with Tukey g-and-h]{
% \label{fig:subfig:WDtime_Tukey_daily}
% \includegraphics[scale=0.5]{Fig_LENS2/WDtime_Tukey_daily.pdf}}
% \subfigure[$\mathrm{WD}_S$ without Tukey g-and-h]{
% \label{fig:subfig:WDtime_noTukey_daily}
% \includegraphics[scale=0.5]{Fig_LENS2/WDtime_noTukey_daily.pdf}}
\caption{Inference process for the annual SG. (a) shows the gap between the estimated mean and ensemble mean of the daily aggregated temperature simulations for September 13, 2023. (b) is the map of the estimated standard deviation $\hat\sigma(L_i,l_j)$ for the daily aggregated temperature. (c) are boxplots of $\{\mathrm{BIC}_{*,t}^{(r)}(Q)\}_{r=1,\ldots,R; t=1,\ldots,T}$ for the daily temperature under different values of $Q$, where $*$ takes l and o. Points $\times$ show the median of BIC values. Red Points $\times$ indicate the minimum BIC medians. (d) is the map of $\hat v(L_i,l_j)$ for the daily data. (e) shows the order of $P$ selected by BIC in the TGH auto-regressive model. (f) is the map of the estimated $\{(\phi_1)_q^m\}_{q=0,\ldots,Q_o-1; m=-q,\ldots,q}$ for the daily aggregated simulations.}
\label{Fig:Supplement_daily}
\end{figure}
This subsection supplements the case study of daily aggregated temperature simulations. Now, assume that $y_t^{(r)}(L_i,l_j)$ is  the temperature at grid point $(L_i,l_j)$, day $t$ after year 2014, and ensemble $r$. In the deterministic component of daily temperature data, we choose $K_D=4$ using the BIC. Specifically, the proportions of grid points favoring $K_D=1$ through $K_D=5$ are $0.4\%$, $8.5\%$, $26.7\%$, $35.3\%$, and $29.1\%$, respectively. In Fig.~S\ref{fig:subfig:goodfit_daily}, we intuitively compare the difference between $\hat m_{474}$ and the ensemble mean at $t=474$, which is a randomly chosen time point. There is a larger gap on land. It may result in the larger variation and uncertainty of temperature on land. The map of $\{\hat\sigma(L_i,l_j)\}_{i=1,\ldots,I; j=1,\ldots,J; t=1,\ldots,T; r=1,\ldots,R}$ is illustrated in Fig.~S\ref{fig:subfig:Sig_daily}, which has similar pattern to those of annual and monthly data but with larger variation.

After removing the deterministic components, we investigate the stochastic component. The BIC in Fig.~S\ref{fig:subfig:BIC_daily} helps us to choose $Q_l=36$ and $Q_o=68$. Fig.~S\ref{fig:subfig:v2hat_daily} presents the corresponding $\hat v(L_i,l_j)$, which exhibits more pronounced larger values on regions with high altitudes and complex topography, and coastlines. We would see higher $\mathrm{I}_{\rm uq}$ and $\mathrm{WD}_S$ values on high-altitude regions in Fig.~6. However, the influence of non-continuous temperature at transition points is not obvious. 
% Jarque-Bera tests reveal that over $98\%$ of grid points reject the Gaussian assumption in the spatial domain. 
The non-Gaussianity in the spatial domain is carried over to the spectral domain through the SHT. Moreover, lower-degree coefficients exhibit more pronounced skewness and heavier tails. For example, the skewness and kurtosis of $\{(\tilde s_t^{(r)})_0^0\}_{t=1,\ldots,T; r=1,\ldots,R}$ is $0.18$ and $3.07$, respectively. However, after applying the TGH transformation, these statistics of $\{(\check{s}_t^{(r)})_0^0\}_{t=1,\ldots,T; r=1,\ldots,R}$ improve to $0.02$ and $3.02$, respectively. In Fig.~S\ref{fig:subfig:BICp_Tukey}, the proportions of coefficients choosing $P=1,\ldots,5$ are $55.5\%$, $7.4\%$, $35.5\%$, $1.6\%$, and $0.0\%$, respectively. Most coefficients at lower degree, which is more influential, opt for $P=1$. Therefore, $P=1$ is ultimately selected for the daily data. In Fig.~S\ref{fig:subfig:phihat_Tukey_daily}, the evaluation of $(\hat\phi_1)_q^m$ reveals a more structured and highly correlated pattern compared to the results in Fig.~S\ref{fig:subfig:TPhihat_monthly_Tukey} for the monthly data.

\subsubsection{Further illustration of daily simulations}
\begin{figure}[!t]
\centering
\subfigure[TGN and TGN']{
\label{fig:subfig:Demo_TGN_TGN1}
\includegraphics[scale=0.5]{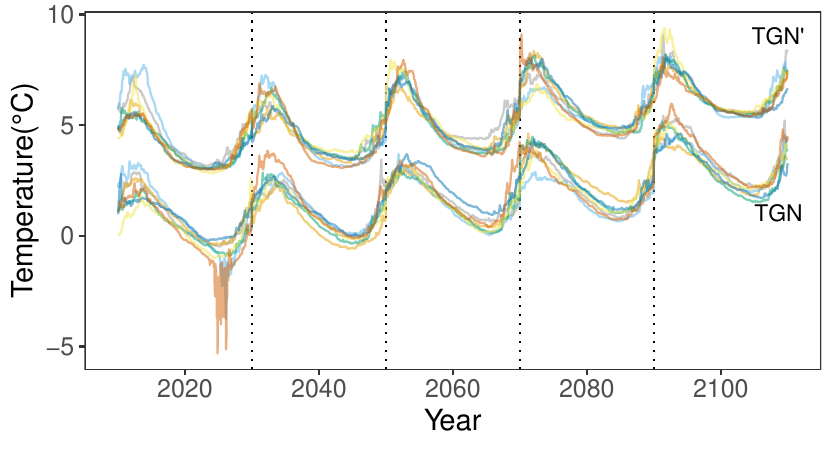}}
\subfigure[TGB and TGNP]{
\label{fig:subfig:Demo_TGB_TGNP}
\includegraphics[scale=0.5]{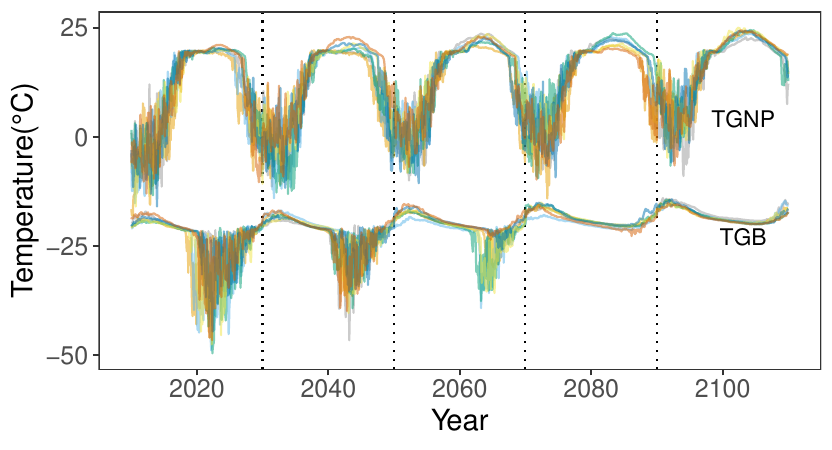}}\\
\caption{Detailed illustration of daily aggregated surface temperature simulations at four selected grid points. (a) shows simulations at grid points that are near to the Band region, where TGN=$(-56.07,10.00)$, and TGN'=$(-56.07,135.00)$. (b) shows (shifted) simulations at TGB=$(-64.55,10.00)$ in the Band region and TGNP=$(76.81,135.00)$ in the North Pole region. For better visualization, the simulations at TGB and TGNP are shifted below and above by $20^\circ$C, respectively.}
\label{Fig:Demo_daily_2}
\end{figure}

As a supplement to Fig.~6, Fig.~\ref{Fig:Demo_daily_2} provides more details about the daily aggregated surface temperature simulations at four selected grid points, where ensembles are represented by various colours. 

\end{document}